\documentclass{aa}
\usepackage{amsmath}
\usepackage[varg]{txfonts}

\def\lsim{\mathrel{\rlap{\lower4pt\hbox{\hskip1pt$\sim$}}
    \raise1pt\hbox{$<$}}}                
\def\gsim{\mathrel{\rlap{\lower4pt\hbox{\hskip1pt$\sim$}}
    \raise1pt\hbox{$>$}}}                

\def\plotonespecial#1{\centering \leavevmode
    \includegraphics[angle=90,width=1.0\columnwidth]{#1}}

\def\plotonesuperspecial#1{\centering \leavevmode
    \includegraphics[angle=90,width=1.4\columnwidth]{#1}}

\def\plotonekindaspecial#1{\centering \leavevmode
    \includegraphics[angle=0,width=1.6\columnwidth]{#1}}

\def\plottwo#1#2{\centering \leavevmode
    \includegraphics[angle=0,width=1.0\columnwidth]{#1} \hfil
    \includegraphics[angle=0,width=1.0\columnwidth]{#2}}

\def\plottwokindaspecial#1#2{\centering \leavevmode
    \includegraphics[angle=0,width=1.0\columnwidth]{#1} \hfil
    \includegraphics[angle=0,height=0.98\columnwidth]{#2}}

\def\plottwospecial#1#2{\centering \leavevmode
    \includegraphics[angle=90,width=1.0\columnwidth]{#1} \hfil
    \includegraphics[angle=90,width=1.0\columnwidth]{#2}}

\def\plottwokindaspecialrot#1#2{\centering \leavevmode
    \includegraphics[angle=90,width=1.0\columnwidth]{#1} \hfil
    \includegraphics[angle=90,height=0.71\columnwidth]{#2}}

\def\plottwoveryspecial#1#2{\centering \leavevmode
    \includegraphics[angle=0,width=1.0\columnwidth]{#1} \hfil
    \includegraphics[angle=0,width=1.0\columnwidth]{#2}}

\def\plotone#1{\centering \leavevmode
    \includegraphics[angle=0,width=1.0\columnwidth]{#1}}

\def\plotonetiny#1{\centering \leavevmode
    \includegraphics[angle=0,width=0.94\columnwidth]{#1}}

\def\eps@scaling{1.0}%
\newcommand\epsscale[1]{\def\eps@scaling{#1}}%

\begin{document}

\title{Hidden starbursts and active galactic nuclei at $0<z<4$ from the \emph{Herschel}\thanks{{\it Herschel} is an ESA space observatory with science instruments provided by European-led Principal Investigator consortia and with important participation from NASA.}-VVDS-CFHTLS-D1 field: \\ Inferences on coevolution and feedback\thanks{All data used for this study are available in electronic format the CDS via anonymous ftp to cdsarc.u-strasbg.fr (130.79.128.5) or via http://cdsweb.u-strasbg.fr/cgi-bin/qcat?J/A+A/}} 
\author{B.~C.\ Lemaux\inst{\ref{inst1}}\and E.\ Le Floc'h\inst{\ref{inst2}}\and O.\ Le F\`evre\inst{\ref{inst1}}\and O.\ Ilbert\inst{\ref{inst1}}\and L.\ Tresse\inst{\ref{inst1}}\and L.~M.\ Lubin\inst{\ref{inst3}}\and G.\ Zamorani\inst{\ref{inst4}}\and R.~R.\ Gal\inst{\ref{inst5}}\and P.\ Ciliegi\inst{\ref{inst4}}\and P.\ Cassata\inst{\ref{inst1},\ref{inst9}}\and D.~D.\ Kocevski\inst{\ref{inst6}}\and E.~J.\ McGrath\inst{\ref{inst7}}\and  S.\ Bardelli\inst{\ref{inst4}}\and E.\ Zucca\inst{\ref{inst4}}\and G.~K.\ Squires\inst{\ref{inst8}}}
\institute{Aix Marseille Universit\'e, CNRS, LAM (Laboratoire d'Astrophysique de Marseille) UMR 7326, 13388, Marseille, France \email{brian.lemaux@lam.fr}\label{inst1}
\and
Laboratoire AIM-Paris-Saclay, CEA/DSM/Irfu - CNRS - Universit\'e Paris Diderot, CE-Saclay, pt courrier 131, F-91191 Gif-sur-Yvette, France\label{inst2}
\and
Department of Physics, University of California, Davis, 1 Shields Avenue, Davis, CA 95616, USA\label{inst3}
\and
INAF - Osservatorio Astronomico di Bologna, via Ranzani 1, 40127 Bologna, Italy\label{inst4}
\and
University of Hawai'i, Institute for Astronomy, 2680 Woodlawn Drive, Honolulu, HI 96822, USA\label{inst5}
\and 
Department of Physics and Astronomy, University of Kentucky, Lexington, KY 40506-0055, USA\label{inst6}
\and
Department of Physics and Astronomy, Colby College, Waterville, ME 04901, USA\label{inst7}
\and
Spitzer Science Center, California Institute of Technology, M/S 220-6, 1200 E. California Blvd., Pasadena, CA 91125, USA\label{inst8}
\and
Instituto de Fisica y Astronom\'ia, Facultad de Ciencias, Universidad de Valpara\'iso, Gran Breta$\rm{\tilde{n}}$a 1111, Playa Ancha, Valpara\'iso Chile\label{inst9}
}

\date{Received November 19th, 2013 / Accepted September 17th, 2014} 
\abstract{We investigate of the properties of $\sim$2000 \emph{Herschel}/SPIRE far-infrared-selected galaxies from $0<z<4$ in the CFHTLS-D1 field. Using a combination 
of extensive spectroscopy from the VVDS and ORELSE surveys, deep multiwavelength imaging from CFHT, VLA, \emph{Spitzer}, \emph{XMM-Newton}, and \emph{Herschel}, and 
well-calibrated spectral energy distribution fitting, \emph{Herschel}-bright galaxies are compared to optically-selected galaxies at a variety of redshifts. 
\emph{Herschel}-selected galaxies are observed to span a range of stellar masses, colors, and absolute magnitudes equivalent to galaxies undetected in SPIRE. Though many 
\emph{Herschel} galaxies appear to be in transition, such galaxies are largely consistent with normal star-forming galaxies when rest-frame colors are utilized. 
The nature of the star-forming ``main sequence'' is studied and we warn against adopting this framework unless the main sequence is determined precisely.  
\emph{Herschel} galaxies at different total infrared luminosities ($L_{TIR}$) are compared. Bluer optical colors, larger nebular extinctions, and larger contributions from 
younger stellar populations are observed for galaxies with larger $L_{TIR}$, suggesting that low-$L_{TIR}$ galaxies are undergoing rejuvenated starbursts while galaxies 
with higher $L_{TIR}$ are forming a larger percentage of their stellar mass. A variety of methods are used to select powerful active galactic nuclei (AGN). Galaxies hosting all 
types of AGN are observed to be undergoing starbursts more commonly and vigorously than a matched sample of galaxies without powerful AGN and, additionally, the fraction of galaxies 
with an AGN increases with increasing star formation rate at all redshifts. At all redshifts ($0<z<4$) the most prodigious star-forming galaxies are found to contain the highest fraction 
of powerful AGN. For redshift bins that allow a comparison ($z>0.5$), the highest $L_{TIR}$ galaxies in a given redshift bin are unobserved by SPIRE at subsequently lower redshifts, 
a trend linked to downsizing. In conjunction with other results, this evidence is used to argue for prevalent AGN-driven quenching in starburst galaxies across cosmic time.} 
\titlerunning{Starbursts and AGN at $0<z<4$ in the CFHTLS-D1 Field}
\authorrunning{B.~C.\ Lemaux et al.}
\maketitle

\section{Introduction}

Questions related to the processes responsible for transforming galaxies from the large populations of 
low-mass, gas-dominated, blue star-forming galaxies found at high redshifts to the massive, quiescent, bulge-dominated 
galaxies observed in the local universe remain numerous. Gas depletion resulting from normal, continuous, or quasi-continuous star formation 
extended over a large fraction of the Hubble time, perhaps coupled with an initial phase of vigorous star formation, is a seemingly viable 
way of explaining the cessation of star-formation activity in galaxies and is invoked by several studies as the primary 
method of building up stellar mass in galaxies (e.g., Legrand et al.\ 2001; Noeske et al.\ 2007a, 2007b; James et al.\ 2008; Rodighiero et al.\ 2011; 
Gladders et al.\ 2013; Oesch et al.\ 2013). This scenario is not, however, the only method of transforming the gas-rich galaxies observed in 
the high-redshift universe (e.g., Daddi et al.\ 2010a) into those whose baryonic component is dominated by their stellar matter. In some cases, star formation is thought to 
proceed in a sporadic manner, particularly at high redshifts, with undisturbed, low-level, continuous star formation being the exception in galaxies rather 
than the primary mechanism of building stellar mass (e.g., Kolatt et al.\ 1999; Bell et al.\ 2005; Papovich et al.\ 2005, 2006; Nichols et al.\ 2012; Pacifici et al.\ 2013). 
Violent episodes of star formation, known as starbursts, can dramatically alter the stellar mass of galaxies with sufficient 
gas reservoirs during the relatively short period that constitutes the lifetime of a starburst (typically $\sim$100-500
Myr; Swinbank et al.\ 2006; Hopkins et al.\ 2008; McQuinn et al.\ 2009, 2010b; Wild et al.\ 2010; though perhaps up to $\sim$700 Myr for particular 
populations, e.g., Lapi et al.\ 2011; Gruppioni et al.\ 2013). Such events contribute 
significantly to the global star formation rate ($\mathcal{SFR}$) of galaxies both in the high-redshift universe (e.g., Somerville et al.\ 2001; Chapman et al.\ 2005; 
Erb et al.\ 2006; Magnelli et al.\ 2009) and locally (e.g., Brinchmann et al.\ 2004; Kauffmann et al.\ 2004; Lee et al.\ 2006, 2009) and 
are furthermore thought to be a typical phase of evolution undergone by massive quiescent galaxies observed at low redshifts early in their formation 
history (e.g., Juneau et al.\ 2005; Hickox et al.\ 2012).

While the operational definition of the term starburst changes significantly throughout the literature (see, e.g., Dressler et al.\ 1999 vs. 
Daddi et al.\ 2010b and the discussion in Lee et al.\ 2009 and references therein), typically the phenomenon is attributed to  
galaxies that are prodigious star formers, i.e., forming stars at rates of $\sim$100-1000 $\mathcal{M}_{\odot}$ yr$^{-1}$, though 
dwarf galaxies with substantially lower star formation rates have also fallen under the dominion of this term (see, e.g., McQuinn et al. 2010a).
Beyond this common thread, weaving a coherent picture of how starbursting galaxies are formed and how they evolve is difficult due to the large range of
properties observed in the population. Though galaxies undergoing a starburst 
event are commonly associated with a late-type morphology (i.e., spiral, irregular, or chaotic), 
a non-negligible fraction of starbursts are, surprisingly, associated with early-types 
(Mobasher et al.\ 2004; Poggianti et al.\ 2009, Abramson et al.\ 2013). Additionally, 
starbursts are observed in hosts that span a wide range of stellar masses (e.g., Lehnert \& Heckman 1996; Elbaz et al.\ 2011, Hilton et al.\ 2012; Ibar et al.\ 2013), color properties 
(e.g., Kartaltepe et al.\ 2010b; Kocevski et al.\ 2011a), and environments (e.g., Marcillac et al.\ 2008; Galazzi et al.\ 2009; Hwang et al.\ 2010a; Lemaux et al.\ 2012; Dressler et al.\ 2013). 
And though some starbursting populations show morphologies indicative of a recent merging or tidal interactions, others appear primarily as undisturbed, 
grand-design spirals, with little evidence of recent interactions with their surroundings (see, e.g., Ravindranath et al.\ 2006; Kocevski et al.\ 2011a; Tadhunter et al.\ 2011). 
Thus, it is likely that a variety of different processes are responsible for inducing starburst events in galaxies, though which mechanisms are 
responsible and how prevalent their effects are, remain open issues. 

Just as the relative importance of the processes that serve to induce starbursts in 
galaxies remain ambiguous, so too are the mechanisms responsible for the cessation of star-formation activity in 
galaxies. There are many processes capable of negatively feeding back on a galaxy once a 
starburst is underway, but what role each of these plays in frustrating or quenching star formation is not fully
understood. These processes include photoionization by hot stars (Terlevich \& Melnick 1985; Filippenko \& Terlevich 1992; Shields 1992), galactic shocks 
(Heckman 1980; Dopita \& Sutherland 1995; Veilleux et al.\ 1995), post-asymptotic giant branch stars (Binette et al.\ 1994; Taniguchi et al.\ 2000), 
and emission from an active galactic nucleus (AGN; Ferland \& Netzer 1983; Filippenko \& Halpern 1984; Ho et al.\ 1993; Filippenko et al.\ 2003; Kewley et al.\ 2006). 
The latter is of particular importance, as AGN feedback is often invoked as the primary culprit in transforming the morphologies of galaxies (e.g., Fan et al.\ 2008; Ascaso et al.\ 2011; 
Dubois et al.\ 2013), in color space (e.g., Springel et al.\ 2005; Georgakakis et al.\ 2008; Khalatyan et al.\ 2008; though see Aird et al.\ 2012 for an opposing view), and in either 
stimulating (i.e., positive feedback) or truncating (i.e., negative feedback) star-formation activity in their host galaxies.  

However, the observational connection between starbursts or galaxies that have recently undergone a starburst 
(i.e., post-starburst or K+A galaxies; Dressler \& Gunn 1983; Dressler \& Gunn 1992) and AGN activity 
is, to date, largely circumstantial. 
This is not due to to a lack of studies investigating this connection, nor to the quality of the data in those 
studies, but is rather due to the difficulty in proving a causal relation between the two phenomena. 
While AGNs, especially those of the X-ray variety, have been observed in hosts that exhibit transitory properties 
either in their broadband colors (e.g.,  Silverman et al.\ 2008, Hickox et al.\ 2009, Kocevski et al.\ 2009b, though 
see Silverman et al.\ 2009 for an alternative view) or in their spectral properties (e.g., Cid Fernandes et al.\ 2004; Yan et al.\ 2006; 
Kocevski et al.\ 2009b; Lemaux et al.\ 2010; Rumbaugh et al.\ 2012), determining timescales related to each phenomenon with precision 
in any particular galaxy remains extremely challenging. And it is this lack of precision which is fatal to directly proving causality (or lack thereof) 
between AGN activity and the cessation of star formation. Thus, many questions remain about the nature of the interaction
between the AGN and its host in such galaxies. Does the AGN play a role in causing this transitional phase or does the 
nuclei become active after the host has already shutdown its star formation? If feedback originating from an AGN is 
primarily responsible for quenching star formation in galaxies, why are many transition galaxies observed without a 
luminous AGN? And how efficient is the quenching provided by an AGN, i.e., on what timescale can the two phenomena be 
observed coevally? 

The difficulty of answering these questions is compounded by the confusion of separating starburst
and AGN phenomena observationally. For those studies which have the luxury of integrated spectra and the 
requisite wavelength coverage, the latter condition becoming increasingly difficult to satisfy at high redshifts,
starbursts can be separated from AGN, typically utilizing a BPT (Baldwin, Phillips, \& Terlevich 1981) emission line ratio 
diagnostic or some variation thereof. However, even with high-quality spectral information, attributing the 
fractional contribution of processes related to star formation and those related to AGN to the strength of 
the emission lines is challenging. While attempts at quantifying relative contributions have been made using 
photoionization models (e.g., Kewley et al.\ 2001), much ambiguity still remains. In the absence of spectra, 
separating, or quantifying the relative strengths of AGN and starbursts have been attempted through spectral 
energy distribution (SED) fitting (e.g., Marshall et al.\ 2007; Lonsdale et al.\ 2009; Pozzi et al.\ 2010; Johnson et al.\ 2013). 
This method, however, rapidly loses its effectiveness for galaxies 
whose SEDs are not dominated by one of the two phenomena, a difficulty which limits its usefulness for investigations
on the connection between AGN and starbursts. Another approach is to use passbands that are (spectrally) far removed from 
typical wavelengths that AGN, when present, dominate. Over the past decade this has typically been achieved using 
\emph{Spitzer} Multiband Imaging Photometer for \emph{Spitzer} (MIPS; Rieke et al.\ 2004) 24$\mu$m imaging to observe 
large populations of starbursting galaxies up to $z\sim1$ and beyond (e.g., Le Floc'h et al.\ 2005). Again, however, the 
difficulty of separating light originating from an AGN and that originating from a starburst persists. The SEDs of AGN
obscured by dust, a population that is, not surprisingly, preferentially associated with the dusty starbursts 
selected by such imaging, peaks at 10-30$\mu$m, roughly the center of the MIPS passband over a broad redshift range 
(i.e., $0<z<2$). Given that the reprocessed emission from HII regions in dusty starbursting galaxies peaks at 
50-150$\mu$m (e.g., Chary \& Elbaz 2001), in order to minimize contamination from emission originating from an AGN, redder 
observations are necessary.

\begin{figure*}
\plotonekindaspecial{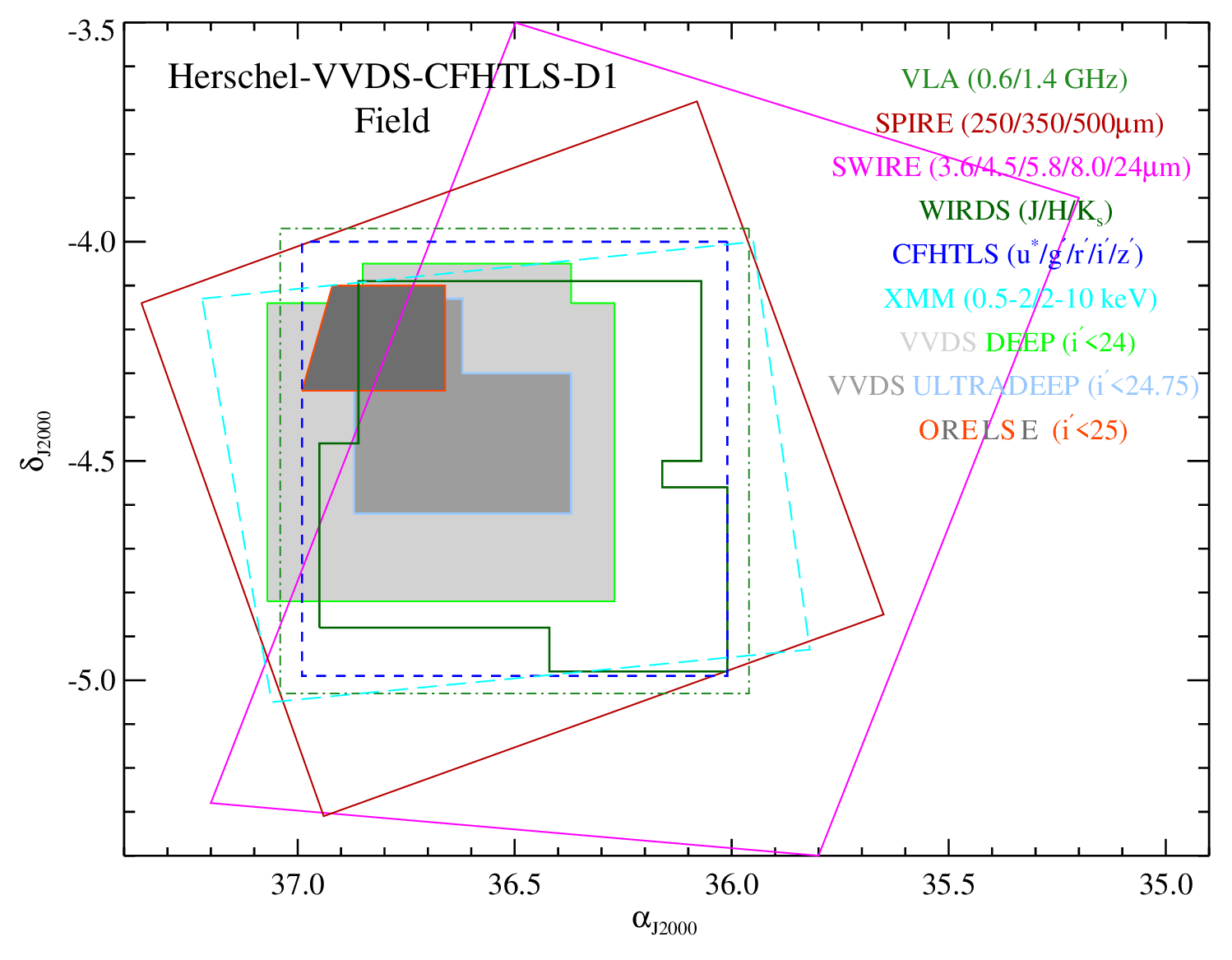}
\caption{Overview of the sky coverage of the observations available on the CFHTLS-D1 field. More details on each observation are
given in Table \ref{tab:D1overview}. Imaging data is represented by open regions bounded by the various lines. Spectroscopic observations are represented
by shaded regions. The overlap region that is used to select our final \emph{Herschel} sample is defined by the intersection of the CFHTLS
optical imaging (dashed blue line), SWIRE NIR/MIR imaging (solid magenta line), and the \emph{Herschel} far-infrared imaging (solid dark red line).
In many cases throughout the paper the sample is further restricted to those galaxies that are covered by WIRDS NIR imaging (solid dark
green line). The full extent of the SWIRE coverage is truncated here for clarity, though the coverage over the overlap region is accurately
represented.}
\label{fig:coverage}
\end{figure*}

The recently launched \emph{Herschel} Space Observatory provides, for the first time, the means to make 
these observations for statistical samples of starbursting galaxies in the local and distant universe. 
Already, observations from \emph{Herschel} have been used to great effect to investigate the relationship
between various types of AGN and their starbursting host galaxies. In galaxies with X-ray-selected AGN, 
numerous studies have found correlations between the strength of the starburst in galaxies, as determined by 
\emph{Herschel}, and the strength of the AGN they host (Shao et al.\ 2010; Harrison et al.\ 2012b; Rovilos et al.\ 2012; Santini et 
al.\ 2012; though for an alternative viewpoint see Page et al.\ 2012). The host galaxies of radio- and 
IR-selected AGN have also been the subjects of investigation with \emph{Herschel} (e.g., 
Hardcastle et al.\ 2012; Del Moro et al.\ 2013), though the connection between these AGNs and their 
host galaxies is less clear than in the case of X-ray-selected AGN. Additionally, large populations 
of starbursting galaxies without an active nucleus have been studied with \emph{Herschel}, resulting 
in insights into the relationship between stellar mass and the $\mathcal{SFR}$ in star-forming 
galaxies (e.g., Elbaz et al.\ 2011; Rodighiero et al.\ 2011; Hilton et al.\ 2012), specific star formation 
rates ($\mathcal{SSFR}$s) and dust mass (Smith et al.\ 2012a), various $\mathcal{SFR}$ indicators (Wuyts et al.\ 2011; 
Dom{\'{\i}}nguez S{\'a}nchez et al. et al.\ 2012), and a variety of other properties (e.g., Casey 2012a; Casey et al.\ 2012b; 
Roseboom et al.\ 2012). 

In this study we utilize wide-field \emph{Herschel}/SPIRE observations taken of the CFHTLS-D1 field. The deep multiband optical/NIR imaging, 
extensive coverage of both radio and X-ray imaging, and exhaustive spectroscopic campaigns by the VVDS and ORELSE surveys on this 
field allow a unique view on both the properties of \emph{Herschel}-selected galaxies and of the interplay between the 
star-forming properties of such galaxies and all flavors of AGN. In the first part of this paper we use all imaging and spectroscopic 
data on the field, along with the results of well-calibrated spectral energy distribution (SED) fitting, to broadly 
investigate the properties of galaxies detected by \emph{Herschel}. Comparisons are made both between SPIRE-detected and equivalent 
samples of SPIRE-undetected galaxies and between various subsets of SPIRE-detected galaxies. In the second part of the paper, we 
use every method at our disposal (IR, radio, X-ray, and spectra) to select AGN and investigate the properties of starbursting galaxies
who play host to such phenomena. The structure of the paper is as follows. In \S\ref{obsnred} the properties of the vast datasets available 
on the CFHTLS-D1 field are described, including the \emph{Herschel}/SPIRE imaging, and, to varying degrees, the reduction of these data 
are also described. In \S\ref{analysis} we describe various analyses which setup the framework of the sample and the subsequent interpretation
of its properties, including the match of SPIRE sources, the selection of AGN, and the SED fitting process. In \S\ref{fullSPIRE} 
the color, magnitude, stellar mass, and spectral properties of the full sample of $\sim$2000 SPIRE-matched galaxies is discussed.
The contents of \S\ref{AGNfull} is specifically devoted to those SPIRE-detected galaxies hosting AGN and to investigating both their prevalence and their 
properties relative to the full SPIRE population. Finally, \S\ref{conclusions} presents a summary of the results. Throughout this paper all 
magnitudes, including those in the IR, are presented in the AB system (Oke \& Gunn 1983; Fukugita et al.\ 1996). We adopt a standard, 
pre-\emph{Planck}, concordance $\Lambda$CDM cosmology with $H_{0}$ = 70 km s$^{-1}$, $\Omega_{\Lambda}$ = 0.73, and $\Omega_{M}$ = 0.27.

\section{Observations and reduction}
\label{obsnred}

Over the past decade and a half, the Canada-France-Hawai'i Telescope Legacy 
Survey (CFHTLS)\footnote{http://www.cfht.hawaii.edu/Science/CFHTLS/}-D1 field (also known as the VVDS-0226-04, the 02h XMM-LSS, or the \emph{Herschel}-VVDS 
field) has been the subject of exhaustive photometric and spectroscopic campaigns, both by
telescopes spanning the surface of the Earth and, more recently, those far above it. The observational 
advantages of this field are numerous. It is an equatorial field, allowing observations from telescopes 
located in both the southern and northern hemispheres. In addition, the field lies at a high (absolute) galactic 
latitude $(l,b)=(172.0,-58.1)$, resulting in low galactic extinction, and is reasonably far removed 
from the ecliptic, resulting in acceptable levels of zodiacal light. The field first 
became the subject of study in 1999 when it was targeted with the Canada-France-Hawai'i 12K wide-field 
imager (CFH12K; Cuillandre et al.\ 2000) to provide $BVRI$ photometry (McCracken et al.\ 2003; 
Le F{\`e}vre et al.\ 2004) for the initial phase of the 
VIMOS VLT Deep Survey (VVDS; Le F{\`e}vre et al.\ 2005)\footnote{All VIMOS VLT Deep Survey data are publicly available on
http://cesam.lam.fr/vvds}. Shortly thereafter, a subsection of the field was observed in multiple 
optical bands as part of the deep portion of the CFHTLS providing the basis for numerous future studies. Since the inception of the
CFHTLS imaging campaign in this field, CFHTLS-D1 has been targeted by a large number imaging and 
spectroscopic surveys at wavelengths that span nearly the entirety of the electromagnetic spectrum. 
In this section we describe the current data available in the CFHTLS-D1 field. For those readers
interested only in a brief overview of the data, Table \ref{tab:D1overview} and 
Fig. \ref{fig:coverage} outline the basic properties and coverage, respectively, of each dataset 
used in the paper.

\begin{table*}
\caption{CFTLS-D1 imaging and spectral data \label{tab:D1overview}}
\centering
\begin{tabular}{lccccc}
\hline \hline
Telescope/Instrument & Band & Survey & $\lambda_{c}$ & Limit\tablefootmark{a} & $N_{obj}$\tablefootmark{b}\\[0.5pt]
\hline
\emph{XMM}/EPIC & CD (2-10 keV) & XMM-LSS & 2$\times10^{-4}\mu$m & 3.7$\times$10$^{-15}$ \tablefootmark{c,d} & --- \\[4pt]
\emph{XMM}/EPIC & B (0.5-2 keV) & XMM-LSS & 1$\times10^{-3}\mu$m & 1.2$\times$10$^{-14}$ \tablefootmark{c,d}& --- \\[4pt] 
CFHT/MegaCam & $u^{\ast}$ & CFHTLS & 0.37$\mu$m& 26.8 & --- \\[4pt] 
CFHT/MegaCam & $g^{\prime}$ & CFHTLS & 0.49$\mu$m & 27.4 & --- \\[4pt] 
CFHT/MegaCam & $r^{\prime}$ & CFHTLS & 0.65$\mu$m & 27.1 & --- \\[4pt] 
CFHT/MegaCam & $i^{\prime}$ & CFHTLS & 0.77$\mu$m & 26.9 & --- \\[4pt] 
CFHT/MegaCam & $z^{\prime}$ & CFHTLS & 0.91$\mu$m& 25.7 & --- \\[4pt] 
CFHT/WIRCam & $J$ & WIRDS & 1.25$\mu$m & 24.2 & --- \\[4pt] 
CFHT/WIRCam & $H$ & WIRDS & 1.63$\mu$m & 24.1 & --- \\[4pt] 
CFHT/WIRCam & $K_{s}$ & WIRDS & 2.15$\mu$m & 24.0 & --- \\[4pt] 
\emph{Spitzer}/IRAC & $Ch1$ & SWIRE & 3.6$\mu$m & 22.8 (3 $\mu$Jy) & --- \\[4pt] 
\emph{Spitzer}/IRAC & $Ch2$ & SWIRE & 4.5$\mu$m & 22.1 (5 $\mu$Jy) & --- \\[4pt] 
\emph{Spitzer}/IRAC & $Ch3$ & SWIRE & 5.8$\mu$m & 20.2 (29 $\mu$Jy) & --- \\[4pt] 
\emph{Spitzer}/IRAC & $Ch4$ & SWIRE & 8.0$\mu$m & 20.1 (31 $\mu$Jy) & --- \\[4pt] 
\emph{Spitzer}/MIPS & $Ch1$ & SWIRE & 24$\mu$m & 18.5 (141 $\mu$Jy) & --- \\[4pt] 
\emph{Herschel}/SPIRE & $Ch1$ & HerMES & 250$\mu$m & 13.7\tablefootmark{c} (12 mJy) & --- \\[4pt] 
\emph{Herschel}/SPIRE & $Ch2$ & HerMES & 350$\mu$m & 13.6\tablefootmark{c} (13 mJy) & --- \\[4pt] 
\emph{Herschel}/SPIRE & $Ch3$ & HerMES & 500$\mu$m & 13.5\tablefootmark{c} (13 mJy) & --- \\[4pt] 
GMRT & Band-4 & VLA-VIRMOS & 49 cm (610 MHz)& 17.8 (275 $\mu$Jy) & --- \\[4pt] 
VLA/B-Array & $L-Band$ & VLA-VIRMOS & 21 cm (1.4 GHz) & 19.0 (94 $\mu$Jy) & --- \\[4pt]    
Keck II/DEIMOS & 1200 l mm$^{-1}$ & ORELSE & 0.79$\mu$m& $i^{\prime}<25$ & 123\tablefootmark{e} \\[4pt] 
VLT UT3/VIMOS & LR-red & VVDS Deep & 0.65$\mu$m & $i^{\prime}<24$ & 6889\tablefootmark{e} \\[4pt] 
VLT UT3/VIMOS & LR-blue/LR-red & VVDS Ultra-Deep & 0.65$\mu$m & $i^{\prime}<24.75$ & 692\tablefootmark{e} \\[4pt] 
\hline
\end{tabular}
\tablefoot{
\tablefoottext{a}{Corresponds to 5$\sigma$ point source completeness limits unless otherwise specified}
\tablefoottext{b}{Numbers given only for spectroscopic datasets}
\tablefoottext{c}{3$\sigma$ point source limit}
\tablefoottext{d}{In units of ergs s$^{-1}$ cm$^{-2}$}
\tablefoottext{e}{Only those spectra considered high quality and within the overlap region are included (see text for definitions)}
}
\end{table*}

\subsection{Optical and near-infrared imaging}
\label{optNIR}

In this paper the deep CFHTLS imaging was chosen instead of  
the original VVDS CFH12K imaging due to the increased depth and number of bands. The deep CFHTLS imaging 
covers 1 deg$^2$, centered on $[\alpha_{J2000},\delta_{J2000}]$=[02:26:00,-04:30:00], and consists of five bands, 
$u^{\ast}g^{\prime}r^{\prime}i^{\prime}z^{\prime}$, with 5$\sigma$ point source completeness limits (i.e., $\sigma_{m}=0.2$) 
of 26.8/27.4/27.1/26.1/25.7, respectively\footnote{These values differ, sometimes significantly, from
the official values given by the CFHTLS collaboration due to the differing definitions of ``completeness'' (see 
http://www.cfht.hawaii.edu/Science/CFHLS/).
In practice, however, the choice of definitions has few discernible consequences for our results, as the 
galaxies used for this study are cut at $i^{\prime}<25.5$, only slightly fainter than the official
CFHTLS completeness limit ($i^{\prime}=25.4$).}. Magnitudes were taken from the penultimate release of 
the CFHTLS data (T0006, Goranova et al.\ 2009) and corrected for Galactic extinction and reduction artifacts
using the method described in Ilbert et al.\ (2006). The pixel scale of the CFHTLS imaging is 0.186$\arcsec$/pixel and 
the average seeing for the five bands ranges between 0.7$\arcsec$-0.85$\arcsec$ over the entire area of the 
imaging used for this paper, with progressively better seeing in progressively redder bands. For further details 
on properties of the CFHTLS-D1 imaging and the reduction process see the CFHTLS TERAPIX 
website\footnote{http://terapix.iap.fr/rubrique.php?id\_rubrique=268}, Ilbert et al.\ (2006), 
and Bielby et al. (2012).

Imaging in the near infrared (NIR) of a subsection the CFHTLS-D1 field was taken from WIRCam (Puget et al.\ 
2004) mounted on the prime focus of the CFHT. This imaging was taken as part of the WIRCam Deep Survey 
(WIRDS; Bielby et al.\ 2012), a survey designed provide deep $JHK_{s}$ NIR imaging for a large portion
of all four CFHTLS fields. In the CFHTLS-D1 field, the one relevant to this paper, the WIRDS imaging
covers roughly $\sim75$\% of the area covered by the CFHTLS imaging, reaching 5$\sigma$ point source 
completeness limits of 24.2, 24.1, \& 24.0 in the $J$, $H$, \& $K_{s}$ bands, respectively. While the 
native pixel of WIRCam is 0.3$\arcsec$/pixel, imaging was performed with half-pixel dithers (also known
as micro-dithering due to the dither pattern resulting in moves of 9$\mu$m at the scale of the detector)
resulting in an effective pixel scale of 0.186$\arcsec$/pixel for the WIRDS imaging. This pixel scale
matches that of the CFHTLS imaging and, more importantly, is sufficient to Nyquist sample the 0.6$\arcsec$ 
average seeing achieved by WIRDS at Mauna Kea in the NIR. For a full characterization of the WIRDS 
survey, as well as full details of the WIRDS imaging properties and reduction see Bielby et al.\ (2012).

Additional NIR imaging spanning a portion of both the CFHTLS and WIRDS coverage was drawn from the 
\emph{Spitzer} Wide-Area InfraRed Extragalactic survey (SWIRE; Lonsdale et al.\ 2003). The SWIRE survey
has imaged 49 deg$^2$, of which the CFHTLS-D1 field is a part, with \emph{Spitzer} at 
3.6/4.5/5.8/8.0 $\mu$m from the \emph{Spitzer} InfraRed Array Camera (IRAC; Fazio et al.\ 2004) and at 
24$\mu$m from the Multiband Imaging Photometer for \emph{Spitzer} (MIPS; Rieke et al.\ 2004). All data were 
taken from the band-merged catalog of this field, released as part of SWIRE Data Release 2. While 
this release is deprecated, no changes were made to the CFHTLS-D1 SWIRE data (referred to as 
SWIRE-XMM-LSS by the SWIRE collaboration) in future data releases. The rough 
full-width half-maximum (FWHM) point spread functions (PSFs) of the 
IRAC images are 1.9$\arcsec$, 2.0$\arcsec$, 2.1$\arcsec$, \& 2.8$\arcsec$ at 3.6, 4.5, 5.8, \& 8.0 $\mu$m, 
respectively, while the FWHM PSF of the MIPS image is $\sim6.5\arcsec$. The data were reduced by the 
standard SWIRE pipeline (Surace et al.\ 2005)\footnote{For further details see 
http://swire.ipac.caltech.edu/ swire/astronomers/publications/SWIRE2\_doc\_083105.pdf}. Magnitudes 
were derived using a combination of aperture-corrected fixed-aperture measurements for 
fainter sources ($m_{3.6\mu\rm{m}}>19.5$) and variable diameter Kron apertures (Kron 1980; 
Bertin \& Arnouts 1996) for brighter sources. The SWIRE imaging reaches 5$\sigma$ point source limiting
magnitudes of 22.8, 22.1, 20.2, 20.1, \& 18.5 (3, 5, 21, 29, \& 141 $\mu$Jy) in the 3.6, 
4.5, 5.8, 8.0, \& 24$\mu$m bands, respectively\footnote{These values differ from the official SWIRE 
values due to the differing definitions of completeness (see http://swire.ipac.caltech.edu/swire/astronomers/program.html). 
This has no effect on our results.}.
For further details on the SWIRE data used in this study see Arnouts et al.\ (2007) and the footnote 
below. A schematic diagram showing the CFHTLS, WIRDS, \& SWIRE coverage for the CFHTLS-D1 field is 
shown in Fig. \ref{fig:coverage}. In this study we will be focusing exclusively
on the 0.8 deg$^2$ region where the available SPIRE, CFHTLS, and SWIRE observations overlap (i.e., the part of the
blue CFHTLS coverage square in Fig. \ref{fig:coverage} that is to the right 
of intersecting SWIRE magenta line). This area is referred to in the remainder of the paper as the ``overlap 
region''. 

\subsection{Spectroscopic data}
\label{spectra}

While the deep ground-based broadband optical/NIR data described above are used to estimate redshifts of 
all detected objects solely from the photometry itself (hereafter photo-$z$s, see \S\ref{SEDfitting}), 
it is important to complement these photo-$z$s with spectroscopic redshifts of a large subsample 
of these objects in order to check for biases, consistency, precision, and accuracy. While we 
did not, in this study, perform a dedicated spectroscopic survey of \emph{Herschel} 
counterparts (as was done in, e.g., the study of Casey et al.\ 2012b, 2012c), the CFHTLS-D1 field is 
the subject of extensive spectroscopic campaigns from two different surveys, both of which are
used in this study. 

The first survey, and the survey from which a large majority of our spectroscopy are drawn, 
is the VVDS (see Le F{\`e}vre et al.\ 2005, 2013a for details on the survey design and goals). The VVDS 
utilizes the Multi-Object Spectroscopy (MOS) 
mode of the VIsible MultiObject Spectrograph (VIMOS; Le F{\`e}vre et al.\ 2003) mounted on 
the Nasmyth Focus of the 8.2-m VLT/UT3 at Cerro Paranal. In the CFHTLS-D1 field, the VVDS survey 
was broken into two distinct phases: the VVDS Deep survey (hereafter VVDS-Deep), 
a spectroscopic survey of $\sim50$\% of the CFHTLS-D1 imaging coverage, magnitude
limited to $i^{\prime}<24$, and the VVDS Ultra-Deep survey (hereafter VVDS-Ultra-Deep), 
which covers a smaller portion of the field but targets galaxies to a magnitude limit of 
$i^{\prime}<24.75$. 

The two phases of the VVDS in this field used distinctly different spectroscopic setups and 
observing strategies. Briefly, VVDS-Deep observations were made with the LR-red grism, resulting
in a spectral resolution of $R=230$ for a slit width of 1$\arcsec$ and a typical 
wavelength coverage of $5500<\lambda<9350$\AA. Integration times were 16000s per pointing. 
VVDS-Ultradeep observations were taken for both the LR-blue and LR-red grisms again with 1$\arcsec$
slit widths, resulting in a spectral resolution of $R=230$ for both grisms. Due to the increased
faintness of the target population the integration times of the VVDS-Ultra-Deep observations 
were significantly longer, typically 65000s per pointing per grism. The observation of both grisms 
resulted in a typical wavelength coverage of $3600<\lambda<9350$\AA\ for the VVDS-Ultra-Deep 
spectra. For further details on the observation and reduction of the full VVDS dataset see 
Le F{\`e}vre et al.\ (2005), Cassata et al.\ (2011), and Le F{\`e}vre et al.\ (2013a).

In total, $\sim10000$ spectra of unique objects have been obtained in the CFHTLS-D1 field. In this 
study, only those VVDS spectra for which the redshift reliability has been determined to be in
excess of 87\% were used (see Le F{\`e}vre et al.\ 2013a), which, in VVDS nomenclature, corresponds to flags 2, 3, \& 
4\footnote{We also use those galaxies with flags 12, 13, \& 14, corresponding 
to broadline AGNs, as well as 22, 23, \& 24, corresponding to 
objects which were not targeted, but which appeared serendipitously on the slit in addition
to the target.} (see Le F{\`e}vre et al.\ 2005 and Ilbert et al.\ 2005 for more details on the VVDS flag system). Applying 
this reliability threshold results in 7352 and 722 spectra for the VVDS-Deep and VVDS-Ultra-Deep datasets, 
respectively, in the overlap region. Galaxies surrounding bright 
stars, for which the photometry was insufficient to perform accurate SED fitting (see 
\S\ref{SEDfitting}), were removed from this sample, resulting in a sample of 6889 and 692 high-quality
spectra in the overlap region for the VVDS-Deep and VVDS-Ultra-Deep samples, respectively.

Additional spectra were taken from the Observations of Redshift Evolution in Large Scale Environments
(ORELSE; Lubin et al.\ 2009) survey. The ORELSE survey is an ongoing multi-wavelength campaign mapping 
out the environmental effects on galaxy evolution in the large scale structures surrounding 20 known 
clusters at moderate redshift ($0.6 \leq z \leq 1.3$). One of these clusters, XLSS005 ($z\sim1.05$), 
lies in a portion of the CFHTLS-D1 field. ORELSE spectra were obtained with the DEep Imaging 
Multi-Object Spectrograph (DEIMOS; Faber et al.\ 2003) on the 10-m Keck II telescope.
Four slitmasks were observed of XLSS005 with an average integration time of $\sim10500$s and an
average seeing of $0.75\arcsec$. Observations were taken with the 1200 l mm$^{-1}$ grating 
tilted to a central wavelength of $\lambda_c =7900$\AA, yielding a typical wavelength coverage of
$6600-9200$\AA. All targets were observed with $1\arcsec$ slits, which results in a spectral
resolution of $R\sim5000$. Data were reduced with a modified version of the Interactive Data Language
(IDL) based \emph{spec2d} package and processed through visual inspection using the IDL-based \emph{zspec} 
tool (Davis et al.\ 2003; Newman et al.\ 2013). Only those spectra considered ``high quality'' (i.e., Q=-1,3,4) 
are used in this study (for an explanation of the quality codes see Gal et al.\ 2008), resulting in
317 high-quality spectroscopic redshifts (hereafter spec-$z$s) of which 123 lie in the survey area and
have sufficient photometric accuracy necessary for SED fitting. For further details on the reduction 
of ORELSE data see Lemaux et al.\ (2009). 

Accounting for duplicate observations, in total the VVDS and ORELSE datasets combine for high-quality 
spectra of $>7600$ unique objects in the overlap region. Spectroscopic coverage of both phases of the VVDS as well as ORELSE 
is represented as shaded regions in Fig. \ref{fig:coverage}. 


\subsection{Far-infrared imaging}
\label{FIR}

Imaging of the CFHTLS-D1 field at 250, 350, \& 500$\mu$m was provided by the Spectral and Photometric Imaging 
REceiver (SPIRE; Griffin et al.\ 2010) aboard the \emph{Herschel} space observatory (Pilbratt et al.\ 2010), 
taken as part of the \emph{Herschel} Multi-tiered Extragalactic Survey (HerMES; Oliver et al.\ 2012). HerMES is 
a massive imaging campaign with \emph{Herschel} designed to cover 340 deg$^2$ of the sky to varying depths 
with both SPIRE and the Photodetector Array Camera and Spectrometer (PACS; Poglitsch et al.\ 2010). HerMES observations
of the CFHTLS-D1 field (referred to as ``VVDS'' by the HerMES collaboration) were taken between the UTC 14th-15th of July 2010 
(Obs. IDs 1342201438-1342201444) and designed as ``level'' (i.e., tier) 4 observations, where the level system runs from 
1-7 and higher levels generally result in observation with increased area at the expense of depth. 
SPIRE observations were performed using Large Map mode, a mode which is described in \S3.2.1 of 
the SPIRE Observer's Manual\footnote{http://herschel.esac.esa.int/Docs/SPIRE/html/spire\_om.html}.
The CFHTLS-D1 field was imaged with SPIRE at a relatively deep exposure time/area; the roughly 2 deg$^2$ VVDS
area that is considered to be within the good part of the VVDS mosaic (i.e., $\Omega_{good}$, the area in
which the sampling is sufficiently deep relative to the mean sampling rate, see Oliver et
al.\ 2012) was imaged with a total integration time of $\sim$10.4 hr (essentially equivalent depth
as that of the Groth Strip and COSMOS HerMES, but shallower than the coverage in, e.g., the ECDFS and GOODS-N/S). The subsection 
of the $\sim$2 deg$^2$ area imaged by SPIRE that intersects the CFHTLS-D1 field is inclined slightly 
($\sim20^{\circ}$) to the CFHTLS coverage, but, despite this, spans the entirety of the CFHTLS imaging 
(see Fig. \ref{fig:coverage}). In addition, the entire $\sim20$ deg$^2$ 02h XMM-LSS field was imaged by both SPIRE
and PACS to moderate depth (level 6) and released as part of Data Release 1 by the HerMES collaboration. These images 
were, however, too shallow to consider for the sample presented in this paper. For more details on HerMES and the \emph{Herschel} 
observations relevant to this study see Oliver et al.\ (2012) and the SPIRE Observer's Manual. 

SPIRE observations of the CFHTLS-D1 field were obtained from the Herschel Data Archive (HDA). These observations consist of 14 scans which utilized 
the \emph{Herschel} Large Map mode (described previously) and the nominal SPIRE scan rate, with an equal number of scans in quasi-orthogonal directions so as to enable the 
removal of low-frequency drifts. Level 1 data products processed by the Herschel Interactive Processing Environment were retrieved from 
the HDA and combined using the package Scanamorphos (Roussel 2013), which is an IDL-based set of routines specifically 
designed to process observations obtained from bolometer arrays. These routines, by using the single scan overlap regions, allow for the subtraction of the 
thermal and pixel drifts as well as the removal of low-frequency noise. Subsequently, Scanamorphos combines the different scans to produce 
a single mosaic at each available wavelength. The resulting mosaics, which span $\sim2$ deg$^2$ each, have a pixel scale of 4.5, 6.25, and 9$\arcsec$ pix$^{-1}$ for
the 250, 350, and 500$\mu m$ bands, respectively.

Far-Infrared (FIR) source catalogs in the three SPIRE bands were obtained with standard PSF fitting using the Image Reduction and Analysis Facility (IRAF; Tody 1993)
\emph{daophot} package (Stetson 1987). First, the 24$\mu m$-detected sources taken from the MIPS-SWIRE coverage of the CFHTLS-D1 field were used to create a 
list of priors for positions of possible FIR sources and a PSF fitting was performed at the positions of the priors for each of the three SPIRE mosaics. However, 
because of the relatively shallow depth of the SWIRE MIPS CFHTLS-D1 observations, many sources detected with \emph{Herschel} lack counterparts at 24$\mu m$. We thus ran 
Sextractor on the residual image obtained after the first PSF fitting to search for any FIR sources that were not extracted with the 24$\mu m$ priors. This 
list of additional detections, comprising $\sim56$\% of our total SPIRE catalog, was merged with the list of FIR sources detected from the initial 24$\mu m$ prior 
input catalog to create a full list of FIR detections. A final round of PSF fitting was then performed over each of the initial science images, this time using the 
final list of positions as position priors. With the exception of a few rather bright and nearby sources, the residual images obtained after this procedure did not 
reveal any obvious excess of signal related to faint sources remaining after the PSF fitting, nor negative signal that could be attributed to PSF over-subtraction. 
The bright objects where the PSF fitting did not perform well were removed from the input list of prior sources and their flux density measurements were treated 
separately with aperture photometry.

The \emph{Herschel} data are calibrated in Jy/beam, so the flux density of each point source extracted by \emph{daophot} was derived from the central pixel value of the 
scaled PSF which fit the object. The flux density uncertainties provided by \emph{daophot} were verified against simulations that were also performed to estimate 
our source extraction completeness. In this simulation, modeled PSFs with flux densities ranging from 1$\sigma$ to 50$\sigma$ were randomly added to each mosaic and 
extracted with the real sources using the same PSF fitting technique as was used for building the \emph{Herschel} source catalogs. The distribution of the 
differences between the input and measured flux densities for these artificially added point sources yielded 1$\sigma$ uncertainties of 4.0, 4.7 and 4.8 mJy in the
250, 350 and 500$\mu m$ bands, respectively. In a given bin of flux density, the completeness of the source extraction was inferred as the fraction of 
recovered sources for which the measured flux density agrees within 50\% of that of the input PSF. This criterion led to 80\% completeness limits of 18.5, 16 and 
15 mJy in the 250, 350 and 500$\mu m$ bands, respectively\footnote{Though the PSF of the SPIRE images gets progressively larger with increasing wavelength, which would
ostensibly lead to a higher confusion limit, the density of real sources in these bands is considerably lower, which leads to the counterintuitive limits derived from 
the simulations}. These limits are slightly brighter than the 3$\sigma$ limits in each of the three \emph{Herschel} bands 
(see below and Table \ref{tab:D1overview}), meaning that, with these $3\sigma$ cuts we are not performing a complete census of these sources, but are rather detecting $\ga$70\% of 
such sources. The difference between these two numbers is largely driven by confusion, meaning that \emph{Herschel} sources in areas of high SPIRE density will
be underrepresented in our sample. Despite the high level of completeness for sources detected at $\ge3\sigma$ in at least one of the three SPIRE bands, this confusion
could induce some small level of bias to our sample if the SPIRE sources in areas of high real or projected density have appreciably different properties, either in the 
far infrared or in the other galaxy parameters used in this study. However, there is only one known cluster in the CFHTLS-D1 field 
(Valtchanov et al.\ 2004), which lies in an area of the sky that does not have a particularly high density of SPIRE sources. Thus, any sources missed due to this
confusion will be missed due to projection effects, and as such, there is no \emph{a priori} reason to believe such sources would not share the properties of
the $\sim$70\% of field galaxies recovered by imposing a significance threshold of $\ge3\sigma$. 

The PSFs of the SPIRE 250, 350, \& 500$\mu$m band images are 18$\arcsec$, 25$\arcsec$, \& 37$\arcsec$, respectively. This is a factor of $>$10 more than our 
complementary images in shorter wavelengths, a subject that is dealt with extensively in \S\ref{herschelmatch}. The SPIRE images reach measured 3$\sigma$ flux 
density limits of 12.0, 12.9, \& 13.2 mJy at 3$\sigma$ in the 250, 350, \& 500$\mu$m bands, respectively, (or, alternatively, $5\sigma$ point source completeness 
limits of 18.3, 21.8, \& 22.6 mJy, respectively). A total of 10431 SPIRE sources were extracted at all significances in the overlap region, of which 3843 were detected at $\ge3\sigma$ in 
at least one of the three SPIRE bands. This latter definition will be adopted widely throughout the paper to define the sample detected ``significantly'' in SPIRE. 

The above definition of significantly detected is more liberal than studies of \emph{Herschel}-selected samples (e.g., B{\'e}therman et al.\ 2012). As such, there is some concern that 
adopting such a limit would result in a higher number of spurious SPIRE sources. However, there are several reasons to believe this is not the case. Approximately half of the $>$$3\sigma$ SPIRE sample is
detected significantly in at least two SPIRE bands. Additionally, among the sample detected at $>$3$\sigma$ in at least one SPIRE band, the median S/N in the next most 
significant SPIRE band is $2.5\sigma$. Roughly two-thirds of the sample detected significantly in only a single SPIRE band is detected at $\ge$5$\sigma$ in the MIPS-SWIRE 
imaging (see \S\ref{herschelmatch}). Of the remaining objects, the median significance in the detection band is sufficiently high ($\sim3.7\sigma$) that 
only a small number of objects in our full SPIRE sample are likely due to stochastic fluctuations ($\sim$10 objects). The above arguments rely solely on the 
statistics of the MIR/FIR imaging. However, the sample we present here is not a \emph{Herschel}-selected sample, but rather an optical/NIR sample that is matched 
to \emph{Herschel}/SPIRE imaging (see \S\ref{herschelmatch}). Thus, we do not require only a $>$$3\sigma$ fluctuation in the SPIRE imaging, but a $>$$3\sigma$ fluctuation 
at specific points in the sky. If we instead consider the probability of a spurious $\ge$3$\sigma$ \emph{Herschel}/SPIRE source matching to an optical counterpart 
using either of the two matching methods detailed in \S\ref{herschelmatch}, the number of spurious sources drops essentially to zero. As such, the effect of 
spuriously-detected SPIRE sources is ignored for the remainder of the paper. 

\subsection{Other ancillary data}
\label{ancillary}

Integral to our study of coeval starbursts and AGN in this paper were the X-ray and radio imaging provided
by two different surveys. In this section we briefly describe the properties of this imaging. 

X-ray observations of the CFHTLS-D1 field were taken from the X-ray Multi-mirror Mission space telescope 
(\emph{XMM-Newton}; Jansen et al.\ 2001) as part of the XMM Medium Deep Survey (XMDS; Chiappetti et al.\ 2005). 
This survey, which lies at the periphery of the much larger XMM - Large Scale Structure survey (XMM-LSS; 
see Fig. 2 of Pierre et al.\ 2004 for rough survey geometries and details on survey motivation), utilized 
the MOS (Turner et al.\ 2001) and pn (Str$\ddot{\rm{u}}$der et al.\ 2001) cameras on the European Photon Imaging 
Camera (EPIC) aboard XMM to map out an area $\sim3$ deg$^2$ that is nearly coincident with the CFHTLS-D1 field. 
A majority of the survey area, and the entirety of the area used in this study, was imaged with a moderate 
integration time of 20-25 ks. This results in images with a $3\sigma$ point source flux limits of 3.7$\times$10$^{-15}$ and 
1.2$\times$10$^{-14}$ ergs s$^{-1}$ cm$^{-2}$ in the soft (0.5-2 keV) and hard (2-10 keV) bands, respectively.
This limit is equivalent to $L_{X,\;2-10 keV}\ga2.5\times10^{43}$ $h_{70}^{-1}$ ergs s$^{-1}$ at $z=1$, corresponding
to a moderately powerful Seyfert. 

X-ray fluxes were drawn from the catalog of Chiappetti et al.\ (2005), which provides fluxes and errors in the 
B (soft) and CD (hard) bands for 286 X-ray objects. The sources in this catalog are limited to those objects 
that satisfied a $4\sigma$ significance cut in one of the EPIC bands. Additionally, the area coverage of the 
XMDS sources are limited to the XMDS overlap with the original VVDS imaging, resulting in the pattern shown 
in Fig. \ref{fig:coverage}. Duplicate entries from this catalog were removed and sources were re-matched
to our own band-merged optical/NIR photometric catalog (see \S\ref{ancillarymatching} and Appendix A). For 
further details on the observations and reduction of the XMM data we refer to Chiappetti et al.\ (2005). 

Radio observations of the CFHTLS-D1 field were taken at 1.4 GHz and 610 MHz from the Very Large Array (VLA, B-Array) 
and the Giant Metrewave Radio Telescope (GMRT), respectively, as part of the VLT-VIRMOS survey (Bondi et al.\ 2003). 
Details of the observations and reduction of the VLA data are given in Ciliegi et al.\ (2005) and those of the GMRT 
data are given in Bondi et al.\ (2007). An updated band-merged version of the optical-radio catalog provided in 
Ciliegi et al.\ (2005) was generated. This updated catalog contains fluxes and errors in both 1.4 GHz and 
610 MHz for all 1054 objects that meet a significance threshold of $5\sigma$ in the VLA imaging. A maximum liklihood 
matching procedure was used to match the radio objects with optical counterparts (see Ciliegi et al.\ 2005), which 
we have in turn matched to our updated band-merged optical/NIR photometric catalog (see \S\ref{ancillarymatching} and
Appendix A). The observations reach $5\sigma$ point source completeness limits of 275 and 94 $\mu$Jy beam$^{-1}$ in 
the GMRT and VLA images, respectively. The PSF in both images is roughly $6\arcsec$. For further details on the 
observation and processing of the radio data see Ciliegi et al.\ (2005) and Bondi et al.\ (2007). 

\section{Analysis}
\label{analysis}

In this study we combine datasets from across the full electromagnetic spectrum, combining data from telescopes 
that have PSFs that span nearly two orders of magnitude, ranging from $0.6\arcsec$ for our ground-based NIR imaging to 
$37\arcsec$ in the reddest \emph{Herschel} band. In this section we discuss the process of matching and otherwise
combining these data into a coherent, self-consistent dataset. 

\subsection{Near-infrared, x-ray, and radio imaging} 
\label{ancillarymatching}

The near-infrared, X-ray, and Radio imaging catalogs used in this study were drawn from a variety of other
studies and adapted to our own purposes. A combination of methods was used to match these catalogs to 
optical sources in the CFHTLS images and to correct observed flux densities (or power densities in the 
case of radio) to rest-frame values for each source. The literature from which these catalogs 
were drawn as well as the various methods used to refine and adapt these datasets for use in this study
is discussed in Appendix A.

\subsection{Herschel counterparts}
\label{herschelmatch}

Two distinct methods were used to determine optical/NIR counterparts to those objects detected by 
\emph{Herschel}. While these methods take dramatically different forms and are used to select 
significantly different populations, we will show later that the two methods are complementary and 
consistent. Our initial method of matching utilized SWIRE/MIPS 24$\mu$m data. Because there is 
a strong physical motivation for expecting a positive correlation between 24$\mu$m flux and flux in the 
FIR, those objects detected in 24$\mu$m were used as a ``parent'' population to extract SPIRE 
sources at the position of the MIPS detection. This step was done during 
the extraction process described in section \S\ref{FIR}. Thus, any 
objects extracted through this methodology, by definition, had a counterpart in the MIPS data. Because
the MIPS data was matched to the SWIRE/IRAC data, which was in turn matched to our optical CFHTLS 
photometric catalog, all \emph{Herschel} objects extracted in this manner, except those MIPS detections 
that had no IRAC counterpart (comprising only $15$\% of the MIPS sample) for which no optical match was made or
objects for which a sufficiently robust match between the IRAC and CFHTLS optical catalog could not be made, 
necessarily had a optical counterpart. Of the 2146 \emph{Herschel} detections in the overlap region that were
detected at $\ge3\sigma$ in at least one of the three SPIRE bands and matched to a counterpart, 1682 (78.4\%) 
were determined through this method.

Because a second set of extractions was performed on the residual \emph{Herschel} mosaic subsequent 
to extracting all sources with 24$\mu$m parents, a large population of SPIRE detections 
were without optical counterparts after the extraction process. Due to the shallowness 
of the SWIRE imaging in 24$\mu$m relative to other similar fields (e.g., COSMOS), this 
24$\mu$m-``orphaned'' \emph{Herschel} population comprised a large fraction of the SPIRE sample over the full SPIRE 
mosaic ($\sim40$\% of significantly detected objects), though some are likely galaxies which
have low $f_{24\mu m}/f_{SPIRE}$ ratios (Magdis et al.\ 2011). In order to prevent severely limiting our sample
size, a second round of matching was performed on the orphaned SPIRE sample. While there is an 
obvious reason to expect galaxies that are intrinsically brighter at $24\mu$m to be, on average, 
intrinsically brighter in the FIR, there exists no other sufficiently deep dataset in the CFHTLS-D1 
field where a similar correlation is expected\footnote{While there is a well-established correlation 
between the radio and the FIR, the VLA/GMRT data was not deep enough to perform this type of matching.}. 
In the absence of an astrophysical motivation, we chose to adopt a statistical approach relying on
the SWIRE/IRAC photometry and a well-established methodology of matching sources between images of 
significantly different PSFs (Richter 1975; Sutherland \& Sanders 1992). This method, described 
in detail in Kocevski et al.\ (2009a) and Rumbaugh et al.\ (2012), uses a likelihood radio statistic of the form

\begin{equation}
\rm{LR}_{i,j} = \frac{\emph{w}_{i}e^{-r^2_{i,j}/2\sigma^2_j}}{\sigma^2_j},
\label{eqn:LR}
\end{equation}

\noindent where the indices $i$ and $j$ correspond to the $i^{th}$ IRAC and $j^{th}$ SPIRE object, respectively, $\sigma_j$
is the positional error of the SPIRE object (we assumed that the error in the NIR position is negligible), 
and $w_{i}$ is defined as $1/\sqrt{n(<m_i)}$, the square root of the inverse of the number density of objects in the IRAC image 
brighter than the $i^{th}$ IRAC source, following the methodology of Rumbaugh et al.\ (2012). Positional errors 
for SPIRE objects depended on flux density, a dependence determined from simulations to take the form

\begin{equation}
\sigma_j=7e^{-0.13f_{\nu,j}}+2,
\label{eqn:poserr}
\end{equation}

\noindent where $f_{\nu,j}$ is the flux density of the $j^{th}$ SPIRE source in mJy and $\sigma_j$ is 
defined to be in units of $\arcsec$. The behavior and magnitude resulting from the above formula as a function
of flux density are similar to those found in similarly deep fields by the HerMES collaboration (Smith et al.\ 2012b). 
Monte Carlo simulations were performed in an identical manner to 
the one described in Rumbaugh et al.\ (2012) and identical probability cuts were applied. For those SPIRE 
objects which were matched to more than one IRAC counterpart, the counterpart with the higher probability
was selected as the genuine match if the primary match met the conditions given in Rumbaugh et al.\ (2012). 
All SPIRE objects with multiple IRAC counterparts that did not meet these conditions were removed from 
our sample. In total, 464 24$\mu$m-undetected counterparts of significantly detected SPIRE sources were determined 
using this method, increasing our sample sample size by nearly a factor of 1.3 over the sample containing only those 
SPIRE objects with 24$\mu$m parents. While this is a sizeable increase over the original sample, the fraction 
of orphaned SPIRE objects matched to an optical counterpart was only $\sim31$\%, much lower than that fraction
for the SPIRE-detected objects with 24$\mu$m parents. This low fraction is as a result of two effects: the enormous PSF 
of SPIRE, which makes it incredibly difficult to properly identify which object (galaxy or star) the SPIRE source 
belongs to, and the conservativeness of the probabilistic matching scheme. For the latter point, the parameters of the 
scheme and the associated cutoffs were set sufficiently high to ensure matches are almost certainly real at
the expense of lower likelihood matches (see discussion in Rumbaugh et al.\ 2012). 

These two methods differ wildly in their approach, which causes some concern for induced differential bias 
in our full matched SPIRE sample. Consistency between the two methods was checked by blindly running the 
probabilistic matching technique on the full SPIRE sample and applying the same probability threshold as 
was applied for the orphaned matching. This catalog was then compared to the catalog of SPIRE objects 
with $24\mu$m parents. Remarkable consistency was found between the two methods, as $>99$\% of the 
objects determined by using $24\mu$m priors that had IRAC counterparts were recovered through the full 
probabilistic process. While this is perhaps to be expected given that the $24\mu$m position for a
given galaxy with an IRAC counterpart was, by virtue of the method used to generate the SWIRE catalog,
forced to be at the same position as centroid of the object in the IRAC 3.6$\mu$m band, the observed
consistency of the two methodologies lends credence both to the choice of the weighting scheme used
in the probabilistic method and the adopted reliability thresholds. In addition, since we do not 
use any galaxies in this paper that do not have an optical counterpart, i.e., galaxies that have
a 24$\mu$m and a SPIRE detection, for which we cannot test the consistency of the two methods, this
test shows that the sample used in this paper would have been largely unaffected if we had instead chosen
to solely adopt the probabilistic method for determining the optical counterparts to SPIRE detected
sources from the onset. Thus, for the remainder of the paper no distinction is made between orphaned SPIRE objects and those with 
24$\mu$m parents. Between the two methods, 2146 of the 3843 SPIRE sources detected at $\ge3\sigma$ in at least one of the three SPIRE 
bands were matched to an optical counterpart in the overlap region. Summarized in Table \ref{tab:SPIREtotals} are
the number of objects determined at each stage in the process using the two methods. 

A final note is necessary. Very recent results with the Atacama Large Millimeter/submillimeter Array (ALMA)
have found that a large fraction ($\sim35$\%) of sub-mm sources are resolved into 
multiple components when the resolution of the instrument used is increased dramatically (Hodge et al.\ 2013). 
The resolution of SPIRE, even in the shortest wavelength bandpass, is prohibitively large such that we cannot
determine the pervasiveness of this phenomenon for the current sample, but given that the selection methods used
in Hodge et al.\ (2013) are similar to our own, it is likely that a non-negligible fraction of the SPIRE-detected sources
in our sample are in fact the composite emission from multiple optical/NIR sources. Though such an effect confuses 
the interpretation of our sample, it is, by in large, only necessary for our study that this effect induces no differential bias, 
as all the comparisons made in this study are either internal or to other sub-mm-selected surveys. Over the dynamic range
probed by Hodge et al.\ (2013), the number of sub-mm sources found to have multiple counterparts, sources which were originally obseved 
with a similar PSF to the SPIRE 250$\mu$m channel, was broadly representative in flux density of the sample as a whole. Since many of the
results in this paper are predicated upon comparisons either between the properties of galaxies selected by their brightness in SPIRE 
or comparisons of the SPIRE brightnesses of different populations, the observed trend in Hodge et al.\ (2013) is encouraging. However, because ALMA 
observations have just begun, the number of sources studied in Hodge et al.\ (2013) was relatively small, and future work will certainly be necessary to fully 
characterize this effect as it relates to \emph{Herschel} studies. Without further ability to determine the true number of SPIRE 
sources in our sample that result from two or more optical/NIR sources, we take refuge in trend observed from preliminary ALMA results and 
ignore the effects of multiple SPIRE counterparts for the remainder of the paper. 

\begin{table}
\caption{Summary of \emph{Herschel}/SPIRE detections in the overlap region\label{tab:SPIREtotals}}
\centering
\begin{tabular}{lcc}
\hline \hline
Sample & 24$\mu$m Prior & 24$\mu$m Orphaned\tablefootmark{a}\\[0.5pt]
\hline
Total SPIRE Sources\tablefootmark{b} & 2366 & 1477  \\[4pt]
SPIRE Counterparts\tablefootmark{b} &  1682 & 464 \\[4pt]
Final SPIRE Sample\tablefootmark{c} & 1367  & 386 \\
\hline
\end{tabular}
\tablefoot{
\tablefoottext{a}{Refers to those galaxies detected in SPIRE with no 24$\mu$m counterpart}
\tablefoottext{b}{Numbers for each sample refer only to those galaxies in the overlap region detected at $\ge3\sigma$ in at least one of the three SPIRE bands}
\tablefoottext{c}{Refers to the final SPIRE sample satisfying the criteria above and with reliable redshifts and well-measured $L_{TIR}$, see \S\ref{SEDfitting} and \S\ref{TIRnSFRs}}
}
\end{table}

\subsection{Spectral energy distribution fitting and redshift distribution} 
\label{SEDfitting}

Many of the results in this study are not directly reliant on the observed properties of the SPIRE sample, 
but rather on secondary physical quantities derived from those properties. In order to link observed properties
to these secondary physical quantities we performed spectral energy distribution (SED) fitting on the entire SPIRE 
sample using the code Le Phare\footnote{http://cfht.hawaii.edu/$\sim$arnouts/LEPHARE/lephare.html} (Arnouts et al.\ 1999; 
Ilbert et al.\ 2006) in two steps. In this section the initial step of this process, used to derive all quantities 
for our sample except total infrared luminosities (TIRs) and $\mathcal{SFR}$s, is described and discussed.

The initial SED-fitting was performed solely on our ground-based optical and NIR photometry (i.e., CFHTLS+WIRDS) 
using the trained $\chi^2$ methodology similar to that described in Ilbert et al.\ (2006). 
\emph{Spitzer} IRAC and MIPS magnitudes were not used at this stage as those data were significantly shallower than the CFHTLS/WIRDS 
imaging. The method used for deriving photo-$z$s and almost all other physical parameters associated with our sample is identical to the one
described extensively in Bielby et al.\ (2012) for the CFHTLS-D1 field. As such, we describe it only briefly here. 

For the photo-$z$ fitting process, a combination of templates from Polletta et al.\ (2007) and Ilbert et al.\ (2009) were used. The final photo-$z$s were
drawn from the full probability distribution functions (PDFs) using a simple median as in Ilbert et al.\ (2010). For the CFHTLS-D1 field the 
reliability and precision of the derived photo-$z$s were checked against the full VVDS spectroscopic sample (flags 3 and 4 only). Since the 
photo-$z$s in this field were, by in large, trained on the VVDS spectral data, it is not suprising that the two sets of redshifts show 
excellent agreement. The normalized absolute median deviation (NMAD; Hoaglin et al.\ 1983), defined in Ilbert et al.\ (2013), is 
$\sigma_{\Delta z/(1+z_s)}\sim0.029$ (or, using a more traditional metric, a median $\Delta z/(1+z_s)=0.017$) for galaxies with both a reliable spec-$z$ and photo-$z$ 
(see Bielby et al.\ 2012). The catastrophic failure rate, defined as the percentage of galaxies with $|z_p-z_s|/(1+z_s) > 0.15$ (see Ilbert et al.\ 2013), 
is also quite low, roughly accounting for only 2.2\% of the sample. Stellar masses and other physical parameters were derived using a stellar population synthesis 
models from Bruzual \& Charlot (2003) and the methodology given in Ilbert et al.\ (2010) and Bielby et al.\ (2012). A rough range of parameters used in this
fitting, e.g., $\tau$, $E(B-V)_s$, metallicity, is given in Table 1 of Ilbert et al.\ (2010), with exceptions given in the text. As in Ilbert et 
al.\ (2010), we adopt a universal Chabrier initial mass function (IMF; Chabrier 2003) along with a Calzetti et al. (2000) reddening law. 

It is extremely important for this study to note that, unlike what is observed in Casey et al.\ (2012c; hereafter C12) and discussed extensively therein, the
disagreement between the spec-$z$s and photo-$z$s of our SPIRE-detected sample is not vastly larger than that of the full photometric sample. Applying the
same metrics to our SPIRE-detected sample with reliable photo-$z$s and spec-$z$s, we find only a marginal increase in the 
NMAD ($\sigma_{\Delta z/(1+z_s)}\sim0.047$ or $\Delta z/(1+z)= 0.031$) and the catastrophic failure rate ($\eta=4.3$\%) of the SPIRE-selected 
sample to the magnitude limit imposed for reliable photo-$z$s ($i^{\prime}<25.5$, see Ilbert et al.\ 2010). The former metric is nearly 
an order of magnitude lower than that of C12. While it is tempting to speculate on the source of 
this discrepancy, the photo-$z$s derived in C12 are drawn from a variety of different methodologies using fields 
that have wildly different photometric coverage and depths. This confuses our ability to discern the source of imprecision in 
their methodology and makes even a qualitative comparison between the two datasets a daunting task. Therefore, we simply 
state here what is important for our sample: that the precision or catastrophic failure rate of the photo-$z$s of our SPIRE-selected 
sample does not differ appreciably from the full photometric sample. This is integral to our study, as it not only gives us 
confidence in the derived photo-$z$s of those SPIRE-selected galaxies that do not have spec-$z$s, but allows us to make internal 
comparisons between SPIRE-selected and SPIRE-undetected galaxies without fear of induced bias. 

As is discussed extensively in the literature (e.g., Vanzella et al.\ 2001; Coe et al.\ 2006), self-consistent magnitude measurements from band 
to band are extremely important to the SED fitting process. While the rewards of analyzing secondary quantities derived from SED fitting 
can be significant -- stellar mass and rest-frame colors allow one to break degeneracies not possible when considering only 
observed-frame colors (e.g., Bell et al.\ 2004; Bundy et al.\ 2006; Lemaux et al.\ 2012; Brada{\v c} et al.\ 2013; Hern{\'a}n-Caballero et al.\ 2013; 
Ryan et al.\ 2013) -- there are also many potential hazards and subtleties associated with this type 
of analysis. However, the primary comparisons made in this study are internal. Thus, biases that introduce themselves during 
the process of fitting for these secondary quantities which affect the entire population are of less concern than those 
that affect populations differentially. To mitigate the latter, we treat the SPIRE-selected sample in an identical manner 
during the SED-fitting process as the full photometric sample. 

\begin{figure*}
\epsscale{1.03}
\plottwo{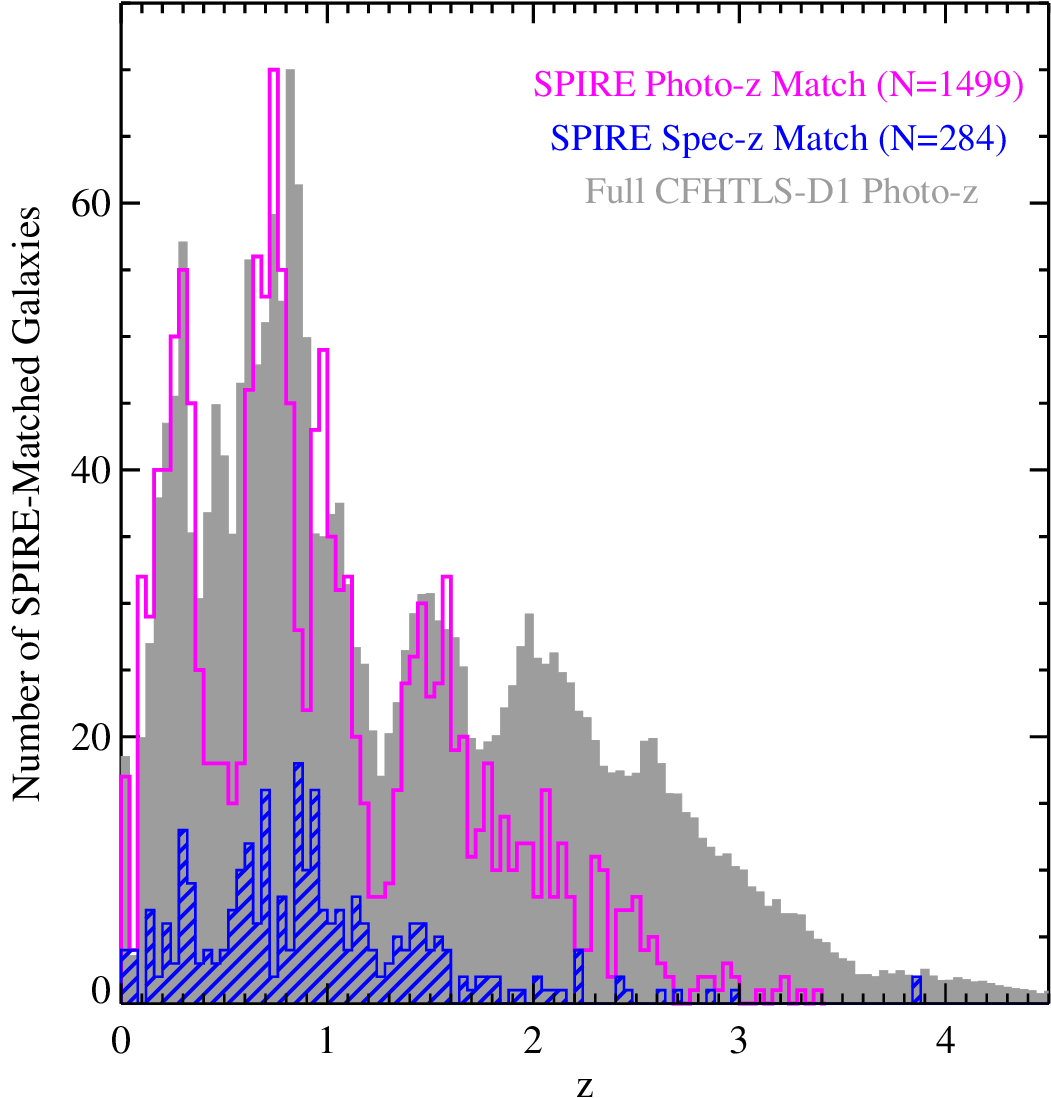}{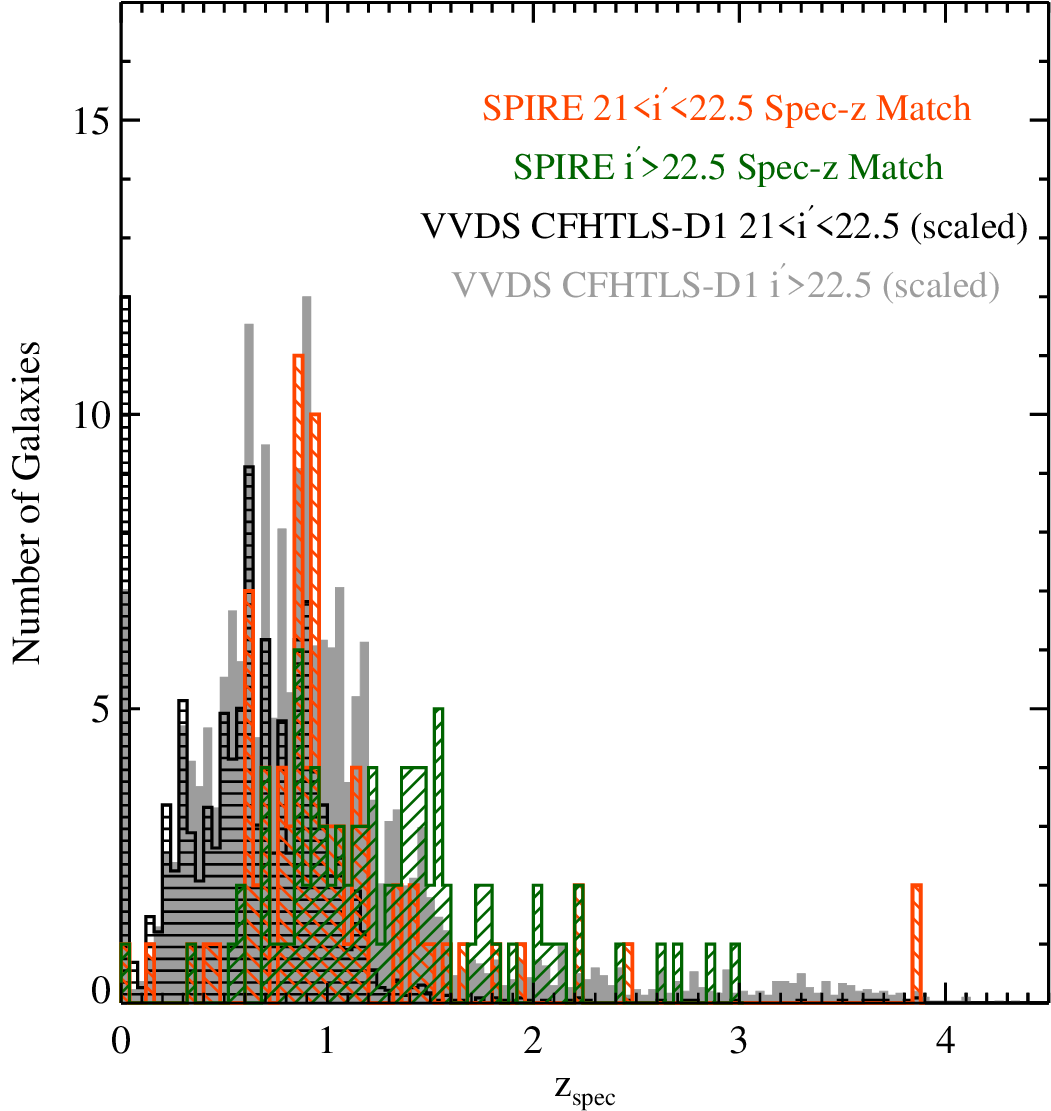}
\caption{\emph{Left:}Photometric (open magenta histogram) and spectroscopic (dashed blue histogram) redshift distribution of the 1783 \emph{Herschel}/SPIRE-matched
sources in our sample. Only those objects which have reliable redshifts (see text) and are detected at $\ge3\sigma$ in at least one of the three
SPIRE bands are plotted. The various peaks in the photometric redshift distribution are not real but are rather an artifact of the method
used to calculate the photometric redshifts as can be seen in the photometric redshift distribution of the full CFHTLS-D1 sample ($i^{\prime}<25.5$)
plotted as the gray filled histogram. \emph{Right:} Spectroscopic redshift histogram of two samples of SPIRE-detected galaxies at different optical brightness 
(orange and green hashed histograms). Also plotted is the spectroscopic redshift distribution of the full VVDS sample in this field for  
the same optical magnitude bins (black hashed and gray filled histograms). Despite median magnitudes that differ by $\sim1.5$ in the $i^{\prime}$ band, the redshift 
distributions of the two SPIRE-detected samples appear largely similar, with an excess of galaxies at higher redshift in the (optically) fainter sample that is dwarfed by 
the excess observed in the full optical sample.} 
\label{fig:zhist}
\end{figure*}

For the initial SED-fitting process, a process which resulted in our derived photo-$z$s, rest-frame colors, 
and stellar mass measurements, aperture magnitudes of 2$\arcsec$ were used for all input photometric measurements. While this choice 
can result in increased global bias relative to measurements using MAG\_AUTO (Kron 1980; Bertin \& Arnouts 1996) especially for lower 
redshift galaxies, the PSF across the 8-band CFHTLS/WIRDS imaging was essentially constant, meaning that there is no induced bias 
in our fitting process from differential loss of light at a given redshift. In addition, at the conclusion of the fitting process, 
quantities which are heavily dependent on aperture corrections, such as stellar mass and absolute magnitudes, were scaled by the 
difference\footnote{More specifically, the quotient of the flux densities} between the aperture and MAG\_AUTO magnitudes using the appropriate 
band (e.g., the median of the $JHK$ bands for stellar mass). Using this methodology allows for the determination of the best-fit model 
through the SED fitting process with a set of self-consistent magnitude measurements that are minimally effected by blending, while still 
allowing for a recovery of quantities which require a full account of the light of a galaxy. This statement is given credence by the photo-$z$s 
derived using aperture magnitudes for the full CFHTLS-D1, which show considerably better agreement with the full VVDS spectral training set 
than did those derived using, e.g., MAG\_AUTO.

A subset of several hundred SED fits, both for SPIRE-detected objects and those undetected in SPIRE, were examined by eye. 
While this was by no means a definitive test of bias, it was comforting that we found no major differences between the best-fit templates
in the two subsamples relative to their photometric data points. This is reflected in similarity of the $\chi_{\nu}^2$ distribution of the 
SPIRE-selected sample relative to the full photometric sample. Both distributions are to a close approximation log normal, characterized
by a mean of $\langle \log(\chi_{\nu}^2) \rangle$ of 0.07 and -0.24 for the SPIRE-detected and full CFHTLS samples, respectively. Both 
values translate to statistically acceptable fits to the data and, though the SPIRE-detected objects have, on average, higher $\chi_{\nu}^2$ 
values, the difference between the mean of the two samples deviates by $<1\sigma$. With no \emph{a priori} knowledge of the physical characteristics
of the SPIRE-selected galaxies, we cannot go further in testing for differential bias induced on the physical parameters during the SED-fitting 
process. However, the consistency of the analysis used and the similarity of the primary statistic used to determine the quality of
the SED-fitting process strongly suggests that no differential bias was induced during this process. Of the 2146 galaxies in the overlap region 
matched to a counterpart and detected at $\ge3\sigma$ in at least one of the three SPIRE bands, $\sim$1800 had sufficiently precise photometry to 
perform this portion of the SED fitting. The spectral and photometric redshift distribution of these galaxies is shown in Fig. \ref{fig:zhist}. 

There are a few interesting notes to be made about the redshift distribution of SPIRE-matched galaxies. The first regards the photometric redshift
distribution, which is not a smooth function, but rather varies appears with various peaks and troughs. While these peaks and troughs are mimicked to 
some degree by the spectroscopic redshift distribution, they are not necessarily a reflection of the underlying population. The method and statistical 
analysis used to calculate photometric redshifts, a method which is widely used, tend to siphon galaxies that are close in redshift
into relatively narrow photometric redshift bins. The effect can also be seen in full CFHTLS-D1 photo-$z$ sample for galaxies brighter than 
$i^{\prime}<25.5$ plotted as the gray shaded histogram in the background of Fig. \ref{fig:zhist}. The cause of this effect has to do with 
the narrowness of the various continuum breaks which are used as the primary discriminators of photometric redshift and the wideness of the 
broadband filters used to constrain the photometric redshift. This effect is mitigated to some extent when medium- or narrow-band filters are 
observed in addition to the broadband filters (see, e.g., Ilbert et al.\ 2009), a luxury that is not available
for the CFHTLS-D1 field. However, the mean error on the photometric redshift of a given galaxy remains small (see \S\ref{SEDfitting}) and, further, 
this effect is expiated by the use of large redshift bins for our analysis. 

The second note of interest regards the spectroscopic redshift distribution,
which appears qualitatively similar to the redshift distribution found in C12. At first glance this is perhaps not surprising; both samples
are probing the redshift distribution of SPIRE-detected sources, albeit in different fields. However, the nature of the two surveys, and specifically the 
spectroscopy of the two surveys, is considerably different such that this similarity did not necessarily need to exist. As mentioned in \S\ref{spectra}, 
the study of C12 presented a systematic spectroscopic campaign of SPIRE-selected sources. In our paper, the spectroscopy was 
performed blindly with respect to SPIRE, in that no knowledge of the SPIRE-detected sources were used to target objects. While there are some additional
criteria imposed on ORELSE spectroscopic targets, the vast majority of the spectroscopy in this paper is magnitude limited in the $i^{\prime}$ band, as
opposed to the study of C12 in which galaxies were magnitude limited in the 250$\mu$m band\footnote{Though the $i^{\prime}$ magnitude
obviously has an important role in this study as well, as SPIRE counterparts were observed using optical spectroscopy.}. The similarity of the 
two spectroscopic redshift distributions then has two main consequences. The first consequence is comforting for this study. The spectroscopy of the CFHTLS-D1 
field and the method of matching between optical and SPIRE counterparts reveals no bias in redshift with respect to the underlying 
population presented in C12, or only insomuch as their spectroscopic sample was biased with respect to the underlying population revealed 
by their photometric redshifts (though see the discussion earlier in the section regarding these measurements). The second consequence is of more 
interest astrophysically. While the (observed frame) wavelength coverage of the two samples is roughly similar, the spectroscopic data in the CFHTLS-D1 
field probe much deeper than those of C12. The number of spectroscopically confirmed objects in C12 begins to drop dramatically at $i^{\prime}\sim22.5$, 
whereas such objects are routinely observed in both phases of VVDS (see Le F{\`e}vre et al.\ 2013a) and ORELSE (see Lubin et al.\ 2009). The broad 
similarity between the two distributions strongly suggest that the redshift distribution of SPIRE-detected sources does not vary rapidly
with decreasing optical brightness. Plotted in the right panel of Fig. \ref{fig:zhist} are the spectroscopic redshift distributions of SPIRE counterparts
in the CFHTLS-D1 field separated in two bins of optical brightness. The spectroscopic redshift distribution of galaxies with 21$<i^{\prime}<$22.5 appears
remarkably similar to galaxies with $i^{\prime}>22.5$ given that the two samples differ in their median $i^{\prime}$ magnitude by $\sim1.5$ magnitudes.
Only a slight increase in the number of galaxies at higher redshifts is observed in the histogram of the fainter sample relative to the brighter sample, a trend not shared by 
galaxies in general (see, e.g., Le F{\`e}vre et al.\ 2013a; Newman et al. 2013). This lack of evolution to higher redshifts is further quantified by inspection of the spectroscopic 
redshift distribution of the full VVDS sample in the CFHTLS-D1 plotted for the same two optical magnitude bins in the right panel of Fig. \ref{fig:zhist}.
While the redshift distribution of SPIRE-detected counterparts is observed to peak at higher redshifts relative to the full VVDS sample for both the magnitude bins,
the number of spectroscopically confirmed galaxies at $z>1.5$ increases only marginally, a factor of $\sim2.5$ from the optically bright 
to the optically faint SPIRE sample. For the full VVDS sample, a sample which serves (largely) as the parent sample from which
the SPIRE-counterparts are drawn and, as such, has the same observational biases, a factor of $>20$ increase is observed in the number of $z>1.5$ galaxies 
over the same change in optical magnitudes (see also Le F{\`e}vre et al.\ 2013b). In other words, it appears that large statistical samples of higher redshift ($z>1.5$) SPIRE-detected galaxies cannot
simply be obtained by spectroscopically targeting galaxies at fainter optical/UV magnitudes. This phenomenon is perhaps due to the complex relationship that SPIRE-detected 
galaxies have with their optical/UV magnitudes, as dust can significantly modulate the latter at a given redshift, an effect which we will show later to be related, 
in turn, to the infrared luminosity of a given SPIRE-detected galaxy. 

 
\subsection{Total infrared luminosities and $\mathcal{SFR}$s}
\label{TIRnSFRs}

The second stage of the SED fitting process involved the full photometric dataset available for the CFHTLS-D1 field and was performed only on
those galaxies detected with SPIRE. The goal for this stage was to derive total infrared (TIR) luminosities ($8-1000\mu m$), which allows us to 
calculate infrared-derived $\mathcal{SFR}$s for the SPIRE-selected sample. In principle, we would be better served by calculating 
the true global $\mathcal{SFR}$, a quantity which combines the ultraviolet (UV) light produced by young stars unobscured by dust 
and the component which is re-radiated at IR wavelengths by interaction with dust particles. It has been shown, however,
that even in broad surveys of galaxies selected by (observed-frame) optical wavelengths, calculations of $\mathcal{SFR}$s from UV indicators 
lead to deficiencies of a factor of $\sim3-10$ relative to the true, dust-corrected $\mathcal{SFR}$ (e.g., Cucciati et al.\ 2012).
As the SPIRE sample represents, by construction, the extreme of such galaxies in terms of their dust content, these values 
represent lower limits for the sample presented here. Empirical evidence of this using a combination of SPIRE imaging and 
spectral properties is given later in this paper (see \S\ref{AtoK}). In addition, the SEDs and $\mathcal{SFR}$s of FIR-selected samples of galaxies
appear dominated by the FIR portion of their spectrum (e.g., Elbaz et al.\ 2011; Rodighiero et al.\ 2010b, 2011; Smith et al.\ 2012a; 
Burgarella et al.\ 2013; Heinis et al.\ 2013). Also, as we will show later (see \S\ref{AtoK}), the $\mathcal{SFR}$s of the SPIRE-detected
galaxies in our sample determined from TIR luminosities are, on average, vastly higher (i.e., a factor of $\ge20$) than those derived from 
the $\lambda$3727\AA\ [OII] UV emission feature. Thus, the contribution of UV light to the $\mathcal{SFR}$ is ignored and for the remainder of the 
paper the IR-derived $\mathcal{SFR}$ of the SPIRE-selected sample is equated with the global $\mathcal{SFR}$.


\begin{figure*}
\epsscale{1.02}
\plottwo{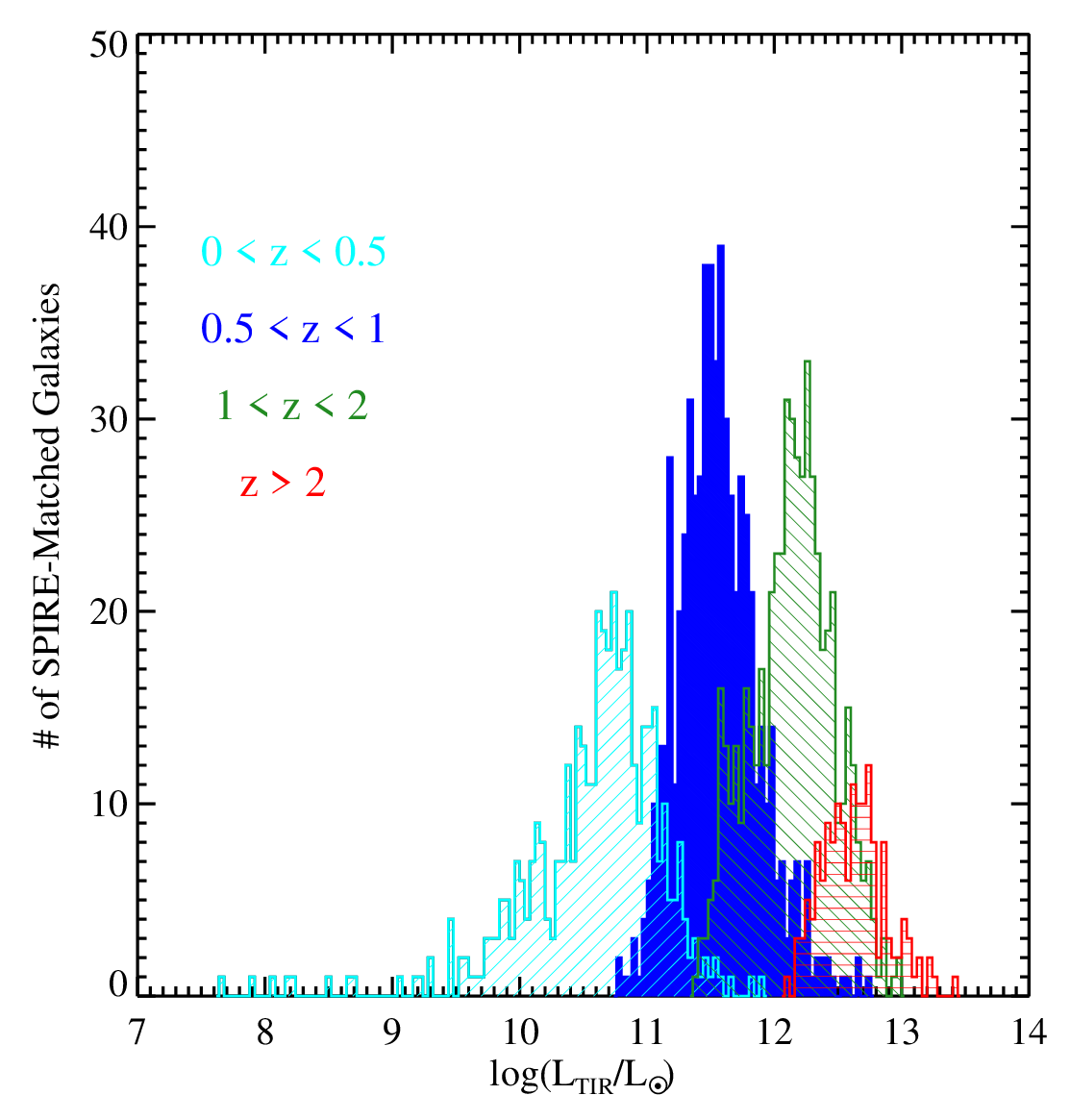}{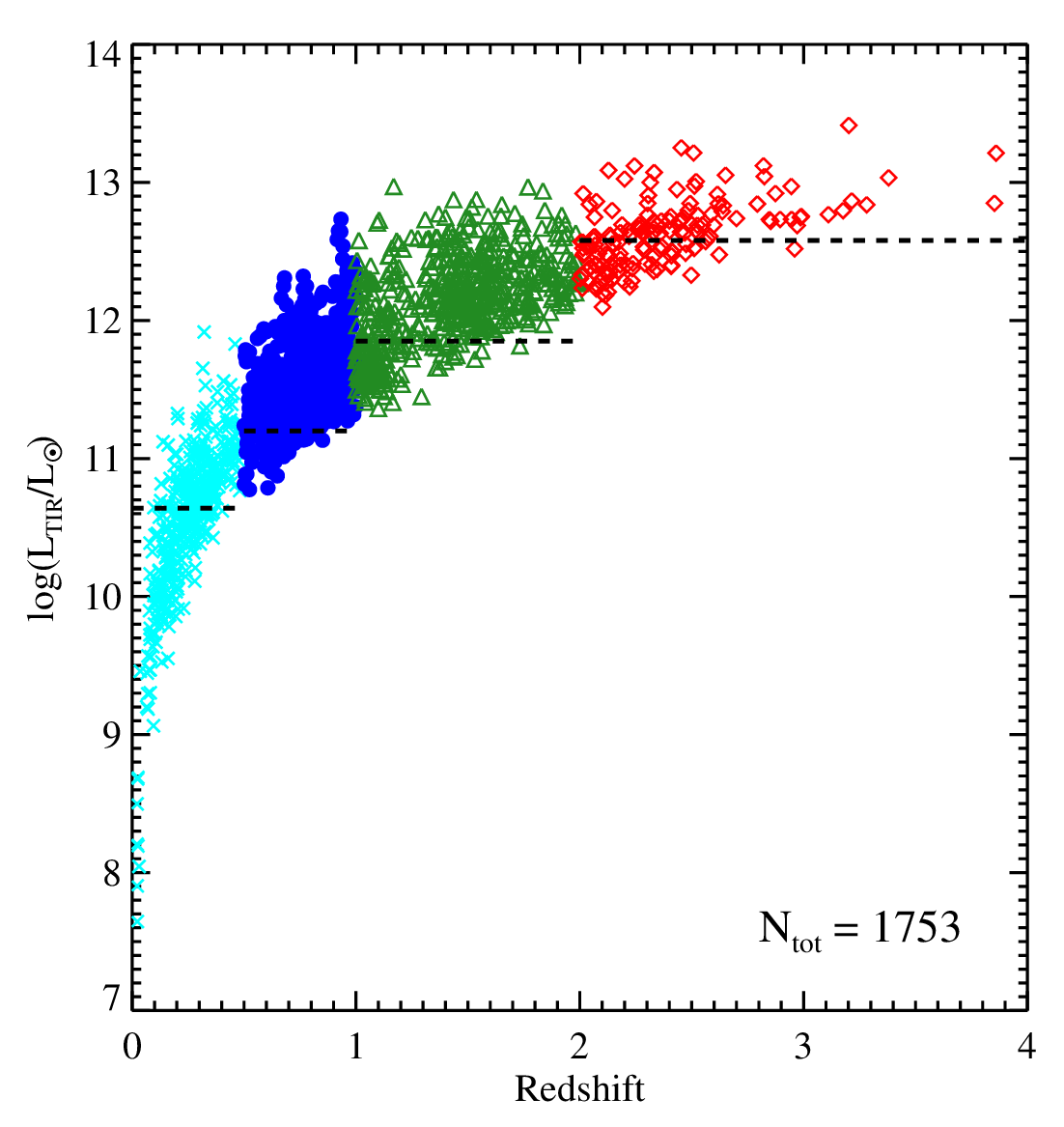}
\caption{\emph{Left:} Full total infrared (TIR) luminosity distribution for the 1753 galaxies in the final \emph{Herschel} sample. Galaxies are binned
as a function of (photometric or spectroscopic) redshift using the nominal redshift bins that are used thoughout most of the analysis in
this study. Though the dynamic range of SPIRE observations is small, the large redshift baseline of this study allows us to study galaxies
that span a factor of 10000 in TIR luminosity. \emph{Right:} Similar to the left panel except the dynamic range of the \emph{Herschel} sample is shown
at each redshift. \emph{Herschel}-detected galaxies are color coded in an identical manner to that of the left panel: galaxies with $0 < z < 0.5$ are
represented by cyan Xs, galaxies with $0.5 < z < 1$ are represented by filled blue circles, galaxies with $1 < z < 2$ are represented by dark
green open triangles, and galaxies with $2 < z < 4$ are represented by open red diamonds. The TIR limits used to contruct a volume-limited
sample for each redshift bin are denoted by the dashed horizontal lines.}
\label{fig:totlum}
\end{figure*}

An additional complication to the calculation of the $\mathcal{SFR}$ seemingly arises when discussing those SPIRE-detected galaxies that host a 
variety of active nuclei. Traditionally, separating emission coming from star-formation processes from that originating from AGNs has been 
extremely challenging. Even with high-quality spectroscopic information, which contains traditionally robust $\mathcal{SFR}$ indicators, one has 
to rely on modeling (e.g., Kewley et al.\ 2001) or statistical methods (e.g., Kauffmann et al. 2003; Vogt et al.\ 2014) to determine the fractional 
contribution of each process to the strength of a given recombination or forbidden emission line. One of the greatest virtues 
of \emph{Herschel}/SPIRE observations, however, is precisely the absence of this ambiguity. As has been shown for galaxies 
hosting X-ray-bright AGN (Mullaney et al.\ 2012; Johnson et al.\ 2013), spectroscopically confirmed samples of type-1 and type-2 AGN (Hatziminaoglou 
et al.\ 2010), those galaxies with IR-selected Compton-thick AGN (Sajina et al.\ 2012), and a variety of other types of AGN (Rowan-Robinson 2000)
the SEDs of galaxies at the rest-frame wavelengths probed by the SPIRE observations, even for the highest redshifts considered in this 
paper ($z\sim4$), are dominated by re-emitted IR photons originating from the cold dust component commonly associated with circumnuclear 
bursts in IR-bright galaxies rather than the warm dust component traditionally associated with the AGN torus. We 
can thus safely ignore the contributions of these types of AGN in the SPIRE bands. 

The remaining population to be considered is then those galaxies hosting radio AGN. If we consider the 
simplest possible scenario, i.e., that the SED generated purely by the mechanism powering the radio AGN is governed by the average 
spectral slope observed for galaxies hosting a radio AGN ($\alpha=0.7$), the power density at THz frequencies (the frequency of 
the SPIRE bands) is $\sim$1\% of that at 1.4 GHz. This translates to $<10$\% of the TIR luminosities for all but the least 
luminous SPIRE sources in our full sample and is essentially negligible at redshifts $z>0.5$. However, as is briefly discussed 
in the next section, radio emission originating from an AGN can exhibit a wide variety of spectral slopes, ranging from an 
extremely steep slope ($\alpha>1$) to an inverted spectrum ($\alpha<0$) in the case of optically thick synchrotron emission. 
In cases of the latter, it is possible that a large fraction of the SPIRE luminosity of the radio AGN hosts in our sample is 
dominated by self-absorbed synchrotron emission (see Gao et al.\ 2013 for a explanation of this phenomenon and Corbel et al.\ 2013 and
L{\'o}pez-Caniego et al.\ 2013 for astrophysical manifestations of it). On a global scale, however, this phenomenon seems to be an exception; the source
density of such objects is extremely low (e.g., L{\'o}pez-Caniego et al.\ 2013; Mocanu et al.\ 2013). In our sample, the median 
FIR colors ($m_{250\mu m}$-$m_{350\mu m}$ and $m_{350\mu m}$-$m_{500\mu m}$) of the SPIRE-detected radio AGN hosts in our sample, 
or indeed the hosts of any type of AGN, are statistically indistinguishable from those hosting dormant nuclei. This is not likely to be the case if 
the mechanism driving the FIR emission of AGN hosts was disparate from that of the SPIRE population with dormant nuclei 
(though see discussion in Hill \& Shanks 2012 and references therein for a possible alternative viewpoint). Another line 
of evidence comes from a more complex treatment of local ULIRGS involving multi-component SED fitting by Vega et al.\ (2008).
In that study it was found that the bandpass spanning the rest-frame wavelengths 40$\mu m$-500$\mu m$ was the least sensitive 
to the presence of a radio AGN, such that only $\sim5$\% of their sample needed AGN contributions to the TIR luminosity in 
excess of 10\% to properly reproduce the SEDs of their sample. These wavelengths are essentially identical to the rest-frame 
wavelengths probed by SPIRE in this study. While it is unclear whether the properties of these lower redshift ULIRG radio AGN
hosts are directly comparable to those in the redshift range of the galaxies presented in this study, the consistent picture painted by 
the various lines of evidence explored here allows us to confidently ignore the presence of any radio AGN, or indeed the 
presence of any other AGN, when deriving the $\mathcal{SFR}$ of our sample from the SPIRE bands.

Because we now incorporate the SWIRE and SPIRE photometry in this stage of the SED fitting process, we move from using aperture magnitudes 
(as were used in the previous section) to MAG\_AUTO for our CFHTLS/WIRDS photometry. At this stage of the fitting, we are attempting \emph{only} 
to determine the TIR $\mathcal{SFR}$ of our sample, the other physical parameters being set by the previous SED-fitting process. In order to determine
this quantity, the inclusion of the SWIRE and SPIRE photometry is of paramount importance. This photometry is, however,
aperture corrected, and thus it is necessary to use CFHTLS/WIRDS magnitudes which are corrected in a similar manner.
Though in practice we find that the choice of MAG\_AUTO over aperture magnitudes has only a small global impact on
the derived $\mathcal{SFR}$ (i.e., $\sim$5\% mean offset) as the $\mathcal{SFR}$s are mostly dependent on the observed SWIRE/SPIRE magnitudes, 
we prefer to adopt the $\mathcal{SFR}$s derived from magnitudes which are self-consistent across the full range of wavelengths used by
the SED-fitting process. The SED fitting was performed in the following manner. The photo-$z$ derived in the previous 
section or, where available, the spec-$z$, was used as a prior and simultaneous fitting of the rest-frame UV/optical and 
IR portions of the SED of each SPIRE-selected galaxy was 
performed in a method similar to the one described in the previous section. The cutoff between the two portions of the 
SED was set to 7$\mu$m in the rest-frame. This cutoff was chosen to place most of the polycyclic aromatic 
hydrocarbon (PAH) features observed in the typical spectrum of dusty starburst galaxies (see, e.g., Chary \& Elbaz 2001) 
redward of the cutoff while minimizing contamination by stellar continua. The optical templates used were identical 
to those used for fitting the full photometric sample. 

To these optical templates, added at this stage of the SED-fitting process were 64 different templates provided by 
Dale \& Helou (2002) to simultaneously constrain the IR part of the spectrum. We chose these templates because 
the models incorporate a variety of different dust components at different temperatures for each model and 
because the templates are calibrated at SPIRE wavelengths using the Submillimetre Common-User Bolometer Array 
(SCUBA; Holland et al.\ 1999). These TIR luminosities were translated into $\mathcal{SFR}$s using the conversion of Kennicutt 
(1998), which is based on the models of Leitherer \& Heckman (1995) and assumes a Salpeter (1955) initial mass 
faction (IMF)

\begin{equation}
\mathcal{SFR}(L_{TIR}) = 1.75\times10^{-10}\frac{L_{TIR}}{L_{\odot}} ~~ \mathcal{M}_{\odot}~ yr^{-1}, 
\label{eqn:SFRTIR}
\end{equation}
 
\noindent where $L_{\odot}$ is the solar luminosity and is equal to $L_{\odot}=3.84\times10^{33}$ ergs s$^{-1}$. 
Employing instead a Chabrier (2003) IMF results in a $\sim60$\% reduction in the TIR-derived $\mathcal{SFR}$ (e.g., Magnelli et al.\ 2013), 
though the absolute value of the $\mathcal{SFR}$ is not as important for this study as relative bias. While relative bias may enter
through a varying IMF (e.g., Swinbank et al.\ 2008), without an ability to constrain the IMF for individual 
galaxies we adopt a Salpeter IMF for all galaxies to consistently compare with
other surveys. Other suites of IR templates were experimented with to calculate TIR luminosities from our MIPS/SPIRE 
photometry. However, we found that the templates of Dale \& Helou (2002) yielded the best results both in visual 
inspection of the fits and in comparing the template-derived TIR luminosities with luminosities derived from 
directly translating SPIRE flux densities in individual bands. In addition, the models of Dale \& Helou (2002)
showed the most internal consistency when comparing the 24$\mu$m-derived TIR luminosity with the TIR luminosity 
derived from MIPS+SPIRE for lower redshift ($z<1$) galaxies. 

It has been suggested that, given the range of model parameter space and the small number of photometric constraints, 
a functional fit to the FIR photometry is perhaps a more robust methodology of deriving physical parameters from FIR
observations of galaxies (for a detailed discussion see Casey 2012a). This may be especially true for complex quantities
like the dust temperature or dust mass, where standard models from the literature may fail to capture the subtleties
of the physical conditions of sub-mm bright galaxies. Unfortunately, galaxies in our sample do not always have the number of 
photometric datapoints in the IR necessary to perform a functional fit to the data. However, we are concerned here 
only with $L_{TIR}$ of the SPIRE-detected galaxies, a quantity which is typically measured to within $\sim$10\% 
(at most) between the two methodologies (Casey 2012a). Because we have no way of verifying the differences between 
the two methods for each of the galaxies in the full SPIRE sample, and, more importantly, because the main 
results presented in this study are insensitive to this level of precision on the overall normalization of
$L_{TIR}$, we chose to adopt the SED-fitting process for the $L_{TIR}$ measurements of the full SPIRE sample.


At the conclusion of this fitting process, a final SPIRE sample of 1753 of the 1783 galaxies selected in the previous section
were selected by imposing that each galaxy, $i)$ be detected at $\ge3\sigma$ in at least one of the three SPIRE bands, $ii)$ have a well-defined 
redshift (a high-quality spectral redshift or a photo-$z$ with $i^{\prime}<25.5$) between $0<z<4$, and $iii)$ 
have sufficiently accurate photometry to perform both the initial SED-fitting (from which we derive stellar masses, 
etc.) and to perform the fitting discussed here. The median redshift of the finaal SPIRE sample is $\langle z\rangle =0.85$, identical to that 
of the 1783 galaxies selected in the previous section. In Fig. \ref{fig:totlum} we show the SED-derived TIR luminosities 
as a function of redshift for our final SPIRE sample. 

%

\subsection{Active galactic nuclei selection \& classification}
\label{AGNselection}

In this section we briefly discuss the various AGN populations used for this study. The general philosophy
adopted is to select AGN samples where purity is maximized without regard to completeness. This choice was made so that we 
need to make no allowance for ambiguity when discussing these populations relative to full SPIRE-selected sample. While 
it is true that we will miss, by virtue of these selections, those AGN which are less powerful (i.e., primarily lower 
radio power density populations), hybrid galaxies (e.g., Kewley et al.\ 2001; Davies et al.\ 2014) where the existence of 
the AGN cannot be proven definitively, and other exotic AGN or AGN-like populations which are discussed later in this section, 
it is enough for this study that the galaxies we do select here are definitively the hosts of powerful AGN. Four different selection 
techniques were used in this study to search for the presence of AGNs amongst the SPIRE sample. These were radio, 
NIR, X-ray, and spectral selection. 

For readers interested only in the basic properties of each AGN population selected by the various methods used in this paper, 
these are summarized in Table \ref{tab:AGNtotals}. There are many subtleties associated with the selection of each type of AGN, 
subtleties which induce a variety of consequences during the selection process. As a large portion of this study is predicated upon 
properly quantifying and contextualizing these consequences, a detailed discussion of each AGN selection method used for this
study as well as possible biases introduced by these selections follows in this section. 

\begin{figure*}
\plottwoveryspecial{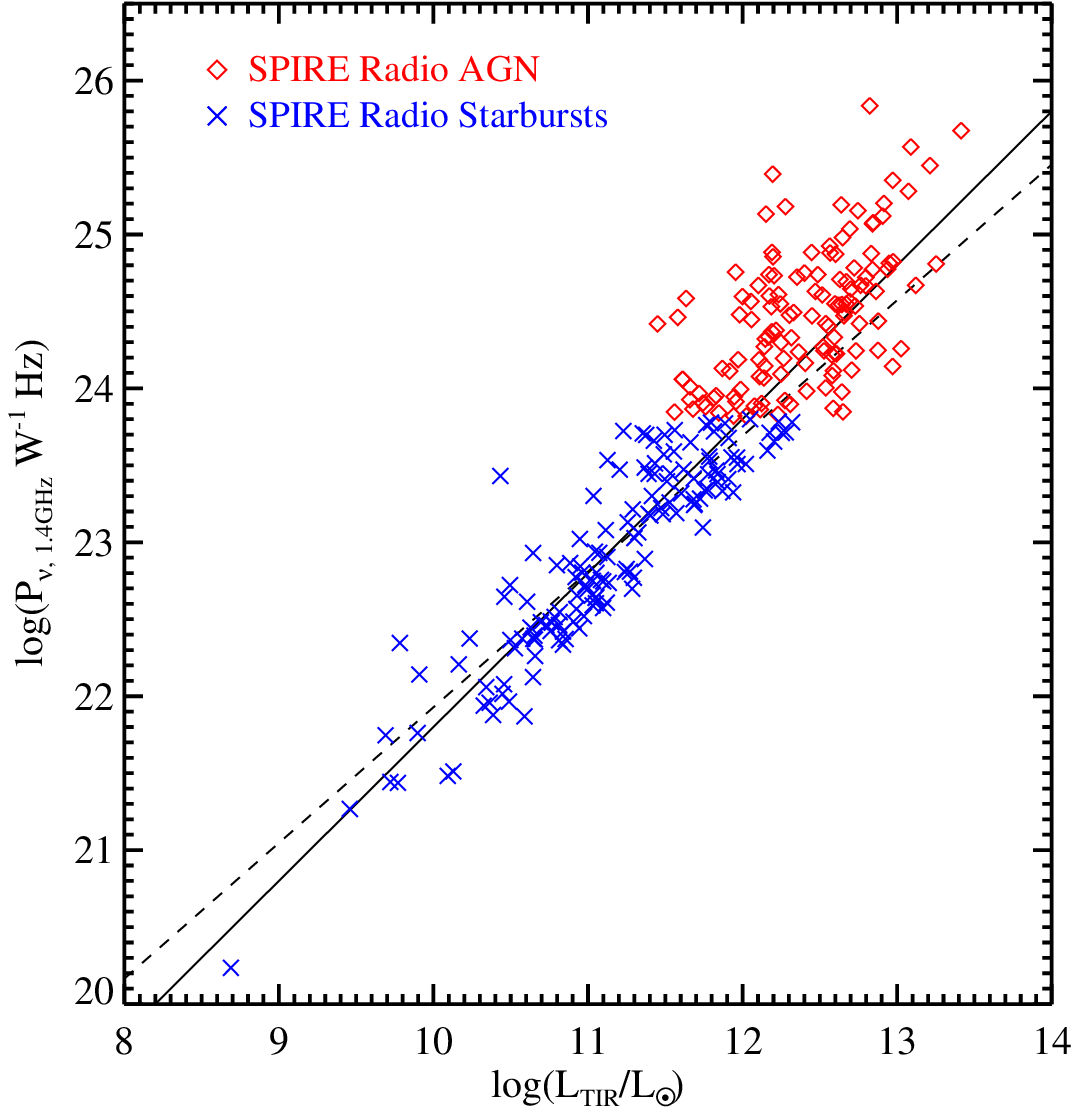}{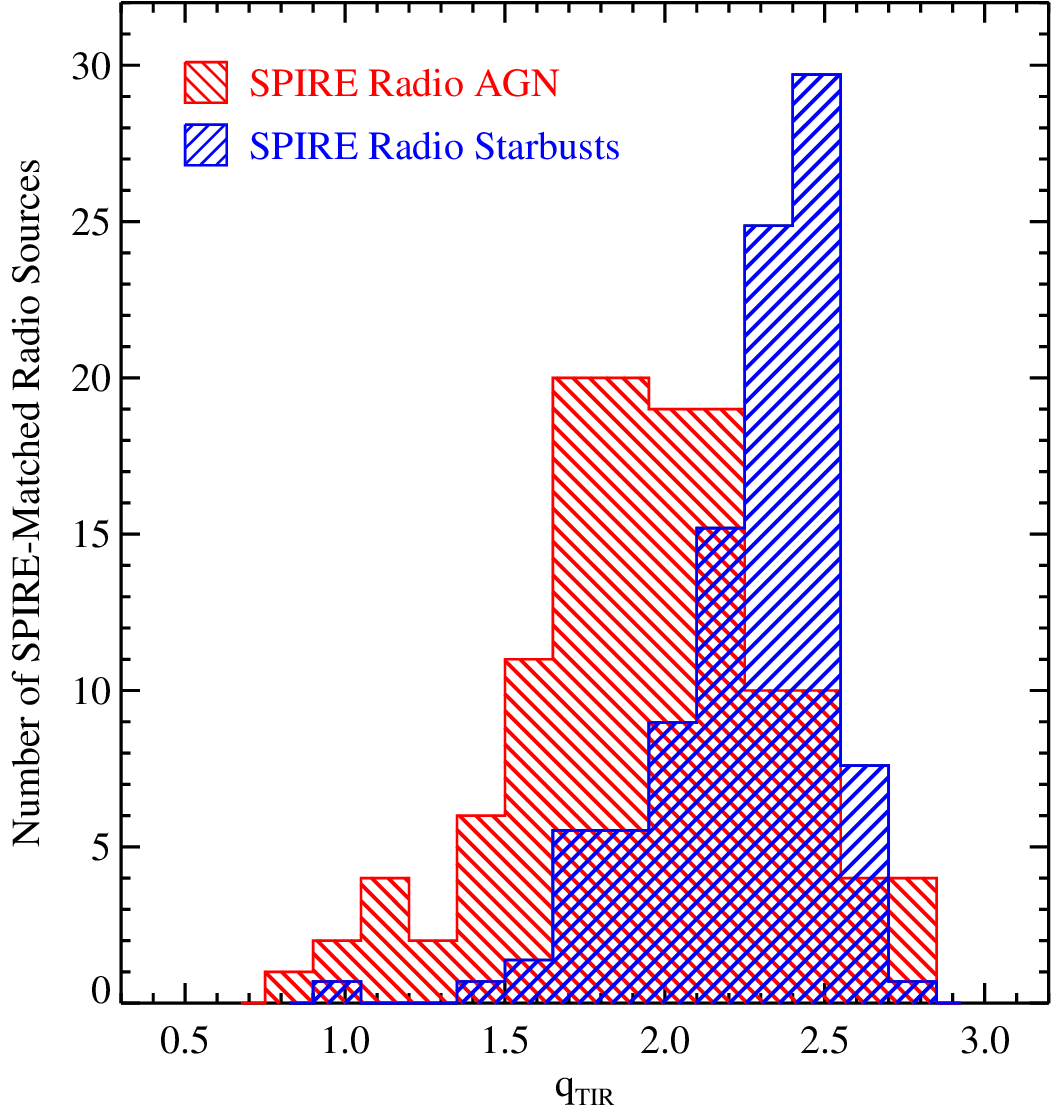}
\caption{\emph{Left:} Total infrared (TIR) luminosity plotted against the rest-frame power density at 1.4 GHz of all galaxies with reliable
redshifts detected significantly in both the VLA and \emph{Herschel} imaging. Galaxies which we have defined as hosting radio AGN are represented by 
open red diamonds and galaxies whose radio emission likely originates either solely from star-forming processes or some combination of star 
formation and AGN activity are shown as blue Xs. The dashed line denotes our formal best fit relation for SPIRE radio starbursts. 
The solid line is the best-fit relation from Del Moro et al. (2013) for ``radio normal'' galaxies. \emph{Right:} Histogram of $q_{TIR}$ for the SPIRE Radio 
AGN and SPIRE Radio starburst samples. The two histograms are normalized to each other. Though there is significant overlap between the two samples, 
in large part due to SPIRE radio starbursts that likely host lower luminosity AGN, the two distributions are inconsistent with being
drawn from the same parent sample.}
\label{fig:qAGNvsnon}
\end{figure*}

Initial investigations of the non-thermal emission from a radio AGN hinted that such emission exhibited a steeper 
spectral slope ($\alpha$) than that originating from star-forming regions (Heckman et al.\ 1983). However, more 
recent work has found that radio emission originating from an AGN shows a wide variety of spectral indices 
(e.g., Ho \& Ulvestad 2001; Clemens et al.\ 2008; Ibar et al.\ 2010). Because there is imaging in the radio of the 
CFHTLS-D1 field in two bands, we initially attempted to select radio AGNs based on a spectral slope 
($\alpha$) criterion alone. However, the large uncertainties in $\alpha$, primarily resulting from the 
shallowness of our 610 MHz imaging, made a selection for individual galaxies impossible. This selection 
was also not possible using a statistical treatment, as there was no observed correlation between radio power 
density and $\alpha$ in our data: the average $\alpha$ for sources with $\log(P_{\nu, 1.4GHz})>23$ was 
identical within the uncertainty to that of lower power density sources. As is also noted in Condon (1992), 
it is possible to use the strength of optical recombination lines (or their proxies) in concert with 
radio power densities to determine the dominant source of radiation. Again, however, this selection was 
not possible with our data as not all of our SPIRE-detected sample was targeted for spectroscopy. In the 
absence of a more evolved selection method, we resigned ourselves to a brute force power density cut, where 
any source with a rest-frame $\log(P_{\nu, 1.4GHz})>23.8$ was considered an AGN. This is consistent with 
maximal radio output of normal galaxies (Condon 1992; Kauffmann et al.\ 2008; Del Moro et al.\ 2013) and consistent 
with the selection used by other studies (e.g., Hickox et al.\ 2009) including other studies of SPIRE-selected galaxies 
(Hardcastle et al.\ 2012).  

%
 
To test the robustness of this cut we calculated $q_{TIR}$ for all radio sources matched to a high significance ($>3\sigma$) SPIRE detection, a
ratio which broadly characterizes the amount of light in the TIR band ($8-1000\mu m$) relative to that of the rest-frame 1.4 GHz radio band. 
This metric is used in some studies to discriminate between star-forming galaxies and those hosting AGN through the excess of radio emission in 
AGN-dominated galaxies relative to what is expected solely from processes related to star formation. Because a wide variety of observations 
are used especially to proxy the amount of light in the TIR band, a wide variety of definitions of this ratio are used, and each has its own subtleties 
with respect to selecting AGN phenomena (see discussion in Del Moro et al.\ 2012). Here, we define $q_{TIR}$ as

\begin{equation}
q_{TIR} = \log(\frac{3.939\times10^{26} L_{TIR}}{3.75\times10^{12} ~ L_{\odot}} ) - \log( \frac{P_{\nu, 1.4 GHz}}{\rm{W} ~ \rm{Hz^{-1}}}).
\label{eqn:qTIR}
\end{equation}

\noindent Because there is a strong correlation between the luminosity in the FIR bands and the luminosity 
in the TIR band, both in the models we use and in general in star-forming galaxies (see, e.g., Kennicutt 1998), we chose instead to calculate 
$q_{TIR}$ using the latter. In addition, because of the wide redshift baseline of our sample and the wide variety of $k-$corrections (or effective $k-$corrections
applied from the SED-fitting process) employed in this study, $q_{TIR}$ was calculated with the observed and modeled rest-frame $L_{TIR}$ and 
$P_{\nu, 1.4 GHz}$ values for all galaxies rather than observed-frame fluxes and flux densities.
While these choices will result in some absolute offset in $q_{TIR}$ relative to measurements made in other studies (i.e., Del Moro et al.\ 2013), we are 
only interested here in the relative difference between the value of $q_{TIR}$ in galaxies we have selected as hosting an AGN and those 
we consider as normal SPIRE-selected starbursts. We broke up the SPIRE-radio matched sample into radio AGN 
($\log(P_{\nu, 1.4GHz})>23.8$) and those galaxies likely either to be dominated by star formation or have some combination of 
star formation and lower level AGN activity ($\log(P_{\nu, 1.4GHz})\le23.8$). The resulting $q_{TIR}$ histograms are presented in 
Fig. \ref{fig:qAGNvsnon}. Visually, the two distributions appear to be quite 
different, with radio AGNs being biased, as expected, to lower values of $q_{TIR}$ (i.e., higher radio flux density for a given 
TIR flux density). To confirm this difference, we performed a Kolmogorov-Smirnov (KS) test on the two distributions, which resulted in a rejection of the 
null hypothesis that the two populations were drawn from the same parent population at a $>>99.99$\% confidence level (CL). 
The significance of this difference remains essentially unchanged if we instead chose $\log(P_{\nu, 1.4GHz})\le23.0$ for our 
star-forming criterion (i.e., the power density limit at which galaxies are likely dominated in the radio by their star formation; Condon 1992). 

However, the sample of galaxies studied here is special relative to many other studies of radio AGN host galaxies. By initially selecting galaxies 
on their $L_{TIR}$ over a large redshift range, we are probing galaxies whose $\mathcal{SFR}$s well exceed even the most vigorous ULIRGs in the local universe 
on which this radio selection was initially calibrated. There is, therefore, some concern that a radio AGN host sample selected on a power density cut 
based on the observed properties of local galaxies may not be sufficiently pure for this study. There are several reasons, however, to believe this 
is not the case. The statistical difference between the distributions of the $q_{TIR}$ values of the galaxies selected as radio AGN hosts and those
galaxies observed at lower power densities strongly suggested the two populations had different mechanisms powering their radio emission. 
This will be further demonstrated in the companion paper to this study (Lemaux et al.\ 2015) where we will remove the expected $P_{\nu, 1.4GHz}$, as derived from
$L_{TIR}$ and $q_{TIR}$, from the observed ($k$-corrected) $P_{\nu, 1.4GHz}$ of galaxies with $\log(P_{\nu, 1.4GHz})>23.8$, resulting in a residual power
density of $\log(P_{\nu, 1.4GHz})>23$ for $\sim90$\% of the sample. This excess can only be attributed to either an evolution in the $q_{TIR}$ parameter
as a function of redshift or $L_{TIR}$ (see Chapman et al.\ 2010 and Bourne et al.\ 2011 for a discussion on the former) or an AGN component. Conversely, for sources 
significantly detected in the radio but with $23<\log(P_{\nu, 1.4GHz})<23.8$, such an excess is \emph{not} observed in $\sim90$\% of the sample. In addition, if we instead 
select radio AGN hosts by a radio-excess criteria using the methodologies of Del Moro et al.\ (2013) none of the main results of this paper
are substantially changed. Because of these reasons and because the $q_{TIR}$ measurements of individual sources are considerably less precise 
than $\log(P_{\nu, 1.4GHz})$ measurements and, furthermore, because adopting a cut based on the former limits both internal comparisons and comparisons
to the literature which can be made without fear of differential bias, we choose to retain the definition of radio AGN hosts as those galaxies fulfilling the criterion 
$\log(P_{\nu, 1.4GHz})>23.8$.

Radio starburst galaxies are then defined as any galaxy detected at radio wavelengths at $\log(P_{\nu, 1.4GHz})\le23.8$, of which
251 lie within in the overlap region. This population will almost certainly contain a considerable number of lower luminosity 
AGN and hybrid galaxies. Similarly, the method of selecting X-ray AGN and IR AGN described later in this section will also 
leave open the possibility of lower level AGN activity in the non-AGN sample (mostly due to the shallowness of the 
\emph{XMM} and SWIRE imaging). However, as stated earlier, the methods of selecting AGN presented here are designed to 
maximize purity of galaxies hosting powerful AGN. When we compare, in the latter sections of this paper, galaxies 
hosting powerful AGN to those that, to the limit of our data and selection methods, do not host a powerful AGN, any contamination 
in the latter sample from low-level AGN activity will likely only serve to decrease the observed differences between the 
two samples. As a result of the previous arguments regarding the purity of the radio AGN host sample, despite the large redshift baseline of this sample ($0<z<4$), 
the redshift evolution of the radio power density (e.g., Condon 1989, 2002; Dunlop et al.\ 2003; Smol{\v c}i{\'c} et al.\ 2009a) 
has little effect on our results. Or, rather, this evolution will simply result in more radio AGN being selected at high redshifts 
as the redshift distribution of the $\sim$10\% of the radio AGN selected hosts whose $L_{TIR}$ subtracted $P_{\nu, 1.4GHz}$ is
less than $\log(P_{\nu, 1.4GHz})=23$ (see above) is indistinguishable from that of the remaining 90\% of the sample. Of the 567 
radio sources in the HerMES-SWIRE-CFHTLS overlap region that were matched to optical counterparts with well-measured redshifts, 
316 satisfied the radio AGN criterion.

\begin{figure}
\plotonespecial{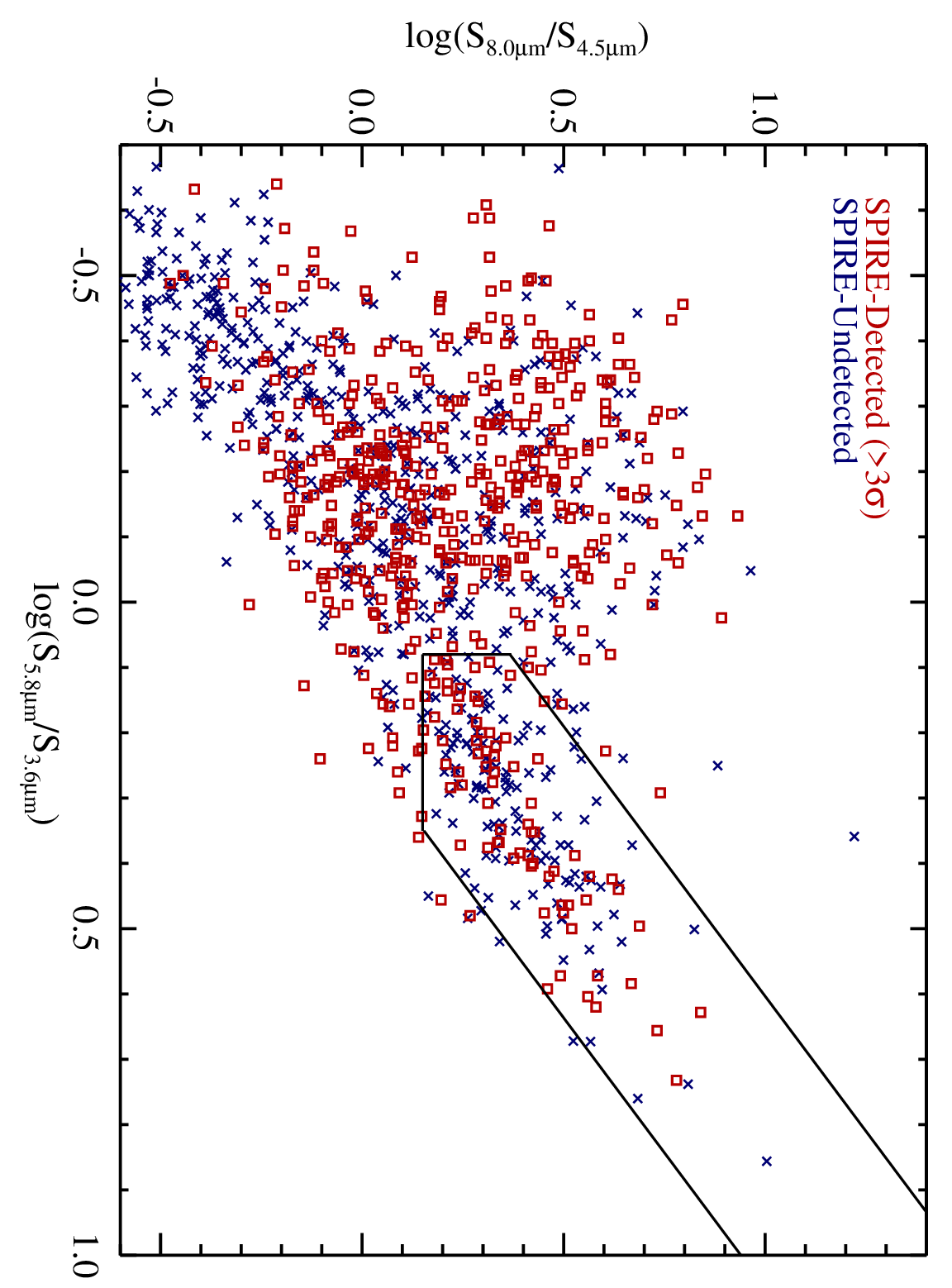}
\caption{\emph{Spitzer}/IRAC NIR color-color diagram for all galaxies in the overlap region (see \S\ref{optNIR}) with $>5\sigma$ detections 
in all four IRAC bands. Both galaxies which are undetected in \emph{Herschel}/SPIRE and galaxies which are marginally detected 
in \emph{Herschel}/SPIRE ($<3\sigma$) are plotted as blue Xs. Galaxies detected at $\ge3\sigma$ in at least one of the three SPIRE bands are shown as 
open red squares. The solid black lines are adopted from Donley et al. (2012) and denote the region used to select IR-AGN. While a large
fraction of galaxies with detections in all four IRAC bands are also detected in \emph{Herschel} (because of our selection method) many 
galaxies, including many in the AGN region, are not. The depth of the SWIRE imaging appears sufficient to probe AGNs with
NIR color properties spanning the entirety of the selection box.}
\label{fig:IRACcolorcolor}
\end{figure}

The selection of AGN obscured by dust has been the topic of much discussion since the advent of the \emph{Spitzer} space 
telescope. The selection of such AGN are particularly important to this study, as the host galaxies that we are selecting 
with SPIRE are necessarily dusty. While pervasive dust in star-forming regions of host galaxies does not necessarily have 
a bearing on the dust properties of its AGN, it is reasonable to assume, given that the most violent starbursting events 
are circumnuclear (e.g., Lehnert \& Heckman 1996; Kennicutt 1998; Kormendy \& Kennicutt 2004; Haan et al.\ 2013), that obscured AGN may be more prominent 
than other AGNs in a FIR-selected sample of galaxies. Indeed, this assumption has been recently given observational support (Juneau et al.\ 2013). A variety of 
different methodology have been used to select obscured AGN, mostly relying on various features in the NIR originating 
from re-radiated emission of the dust-enshrouded nucleus (Lacy et al.\ 2004; Stern et al.\ 2005; Donley et al.\ 2008; 
Kirkpatrick et al.\ 2012). In this paper we adopt the hybrid criteria of Donley et al.\ (2012) because of the high purity 
and completeness demonstrated for samples obtained using their methodology. In Fig. \ref{fig:IRACcolorcolor} we illustrate 
this selection applied to our sample, plotting the IRAC flux density ratios for all galaxies detected in the overlap 
region at $>5\sigma$ in all four IRAC bands along with the criteria given by Donley et al.\ (2012). Of the 1107 galaxies 
in the overlap region detected at $>5\sigma$ in all four IRAC bands and matched to a optical counterpart with a 
well-defined redshift, 172 were selected as IR-AGN using this method.

Selection of AGNs in the X-ray was considerably easier than selection in the radio or NIR. Because of the shallowness of the 
\emph{XMM} coverage in the CFHTLS-D1 field and the relatively low volume probed at low redshifts ($z<0.1$), nearly 
all ($\sim95$\%) of the objects detected in our X-ray catalog and matched to an optical counterpart with a well-measured 
redshift had X-ray luminosities in excess of what can be produced by normal star formation ($L_{X,\;2-10 keV} > 10^{42}$ ergs s$^{-1}$; Ranalli et al.\ 2003). 
Thus, any object detected in our X-ray catalog and matched to a SPIRE source, with the exception of only a few sources, was considered a X-ray bright 
AGN. It should be noted that this selection process is highly incomplete as the X-ray data are only complete to $L_{X,\;2-10 keV} > 10^{42}$
until redshift $z\sim0.3$. In addition, it has been shown that X-ray weak or X-ray obscured AGNs selected by a 
combination of rest-frame optical line ratio diagnostics and rest-frame colors (Yan et al.\ 2011) or stellar mass (Juneau et al.\ 2013) 
constitute a non-negligible fraction, perhaps as much as $\sim50$\%, of the total AGN population in exactly the type of 
infrared bright starbursts studied in this paper. Our selection is also incomplete with respect to those AGN, or AGN-like phenomena, whose presence is
inferred by other rest-frame optical indicators (e.g., Yan et al.\ 2006; Lemaux et al.\ 2010). However, such objects should be less 
prevalent in our sample, as such phenomena are typically associated with quiescent red-sequence galaxies, a galaxy type 
that SPIRE observations are blind to. The lack of a full spectral sample and the large redshift baseline of our sample 
precludes the possibility of making similar selections for our sample. While we mitigate the effects of this completeness 
somewhat by also selecting AGN in the infrared and radio, all percentages quoted in this paper for the X-ray AGN population 
are hard lower limits which likely severely underestimate the true number of high-excitation, high-accretion-efficiency AGNs 
in our sample. Because of this, and because of the shallowness of the \emph{XMM} data in the CFHTLS-D1 field relative to studies
dedicated to the interplay between X-ray AGN and their \emph{Herschel}-detected hosts (e.g., Harrison et al.\ 2012b),
we do not attempt later in this paper to study the X-ray AGN population alone, but rather as part of a larger ensemble comprised
of all AGN types. 

\begin{figure*}
\plottwokindaspecial{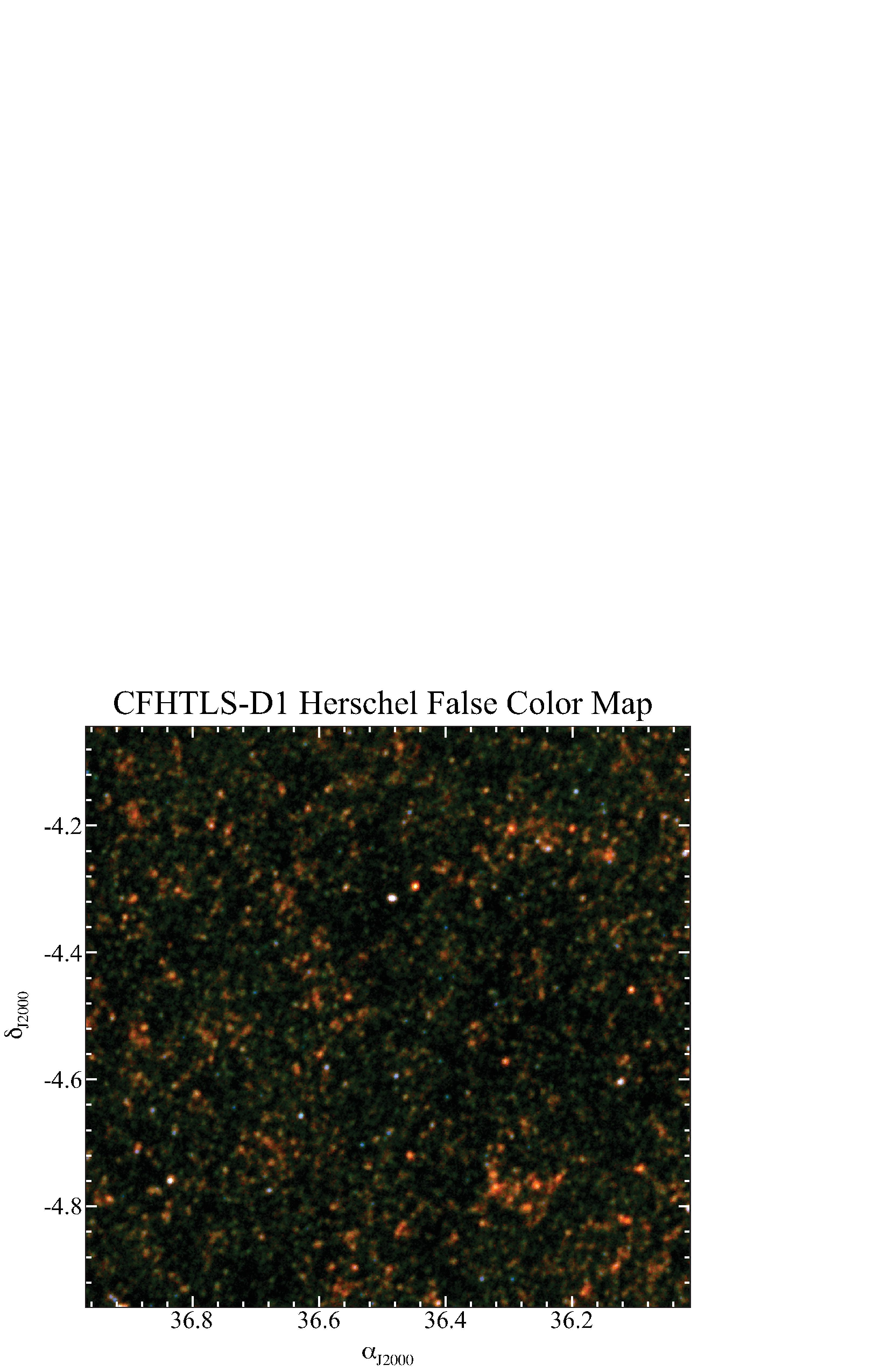}{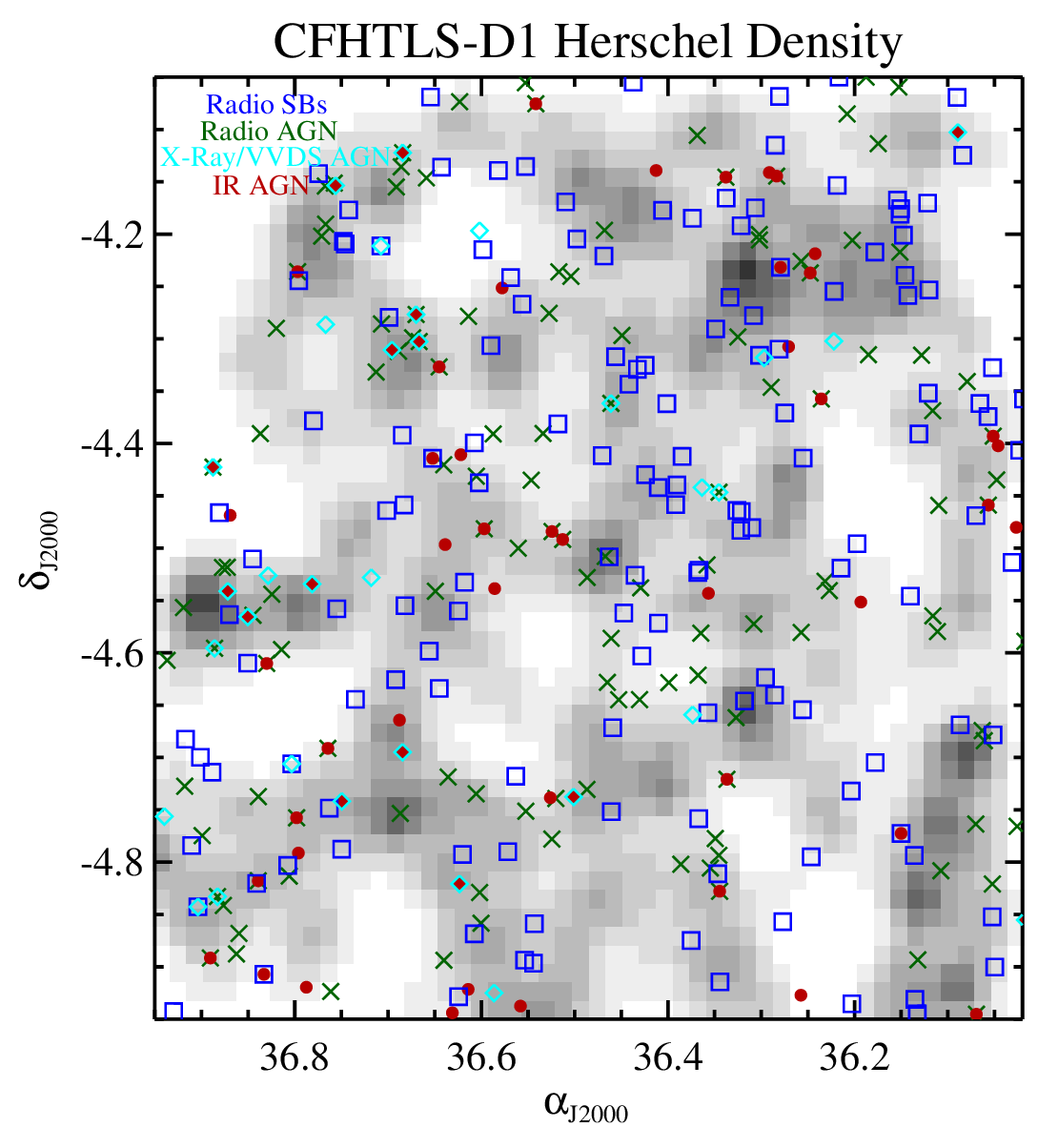}
\caption{\emph{Left:} False color image of the \emph{Herschel}/SPIRE mosaic in the CFHTLS-D1 overlap region (see \S\ref{optNIR}). The SPIRE
500$\mu$m, 350$\mu$m, and 250$\mu$m images served as the RGB channels, respectively. \emph{Right:} 
Sky distribution of multiwavelength galaxies also detected with SPIRE plotted against the backdrop of a density map of
the 1753 galaxies in our final SPIRE sample. The spatial density of the full SPIRE sample is a convolution of the true spatial distribution (left panel) 
and a number of metholodological and observational effects. Bright stars, which prevent both spectroscopy and 
high-quality photometric redshifts, account for the presence of most of the voids seen in the background density. This includes voids around areas of otherwise high densities 
of SPIRE-detected sources, such as the grouping seen in the left panel at $[\alpha_{J2000},\delta_{J2000}]\sim[36.28,-4.76]$.} 
\label{fig:densityplotwAGN}
\end{figure*}

Unobscured type-1 AGN, a population intimately related to the X-ray bright AGN population selected by \emph{XMM}, were drawn 
from those SPIRE-matched galaxies which were targeted for spectroscopy either by VVDS or ORELSE. These AGN are characterized 
by their broadline emission features (typically with $\sigma\ga1000$ km s$^{-1}$) and strong emission of those features which
can result only from ionization by extremely high energy photons. In such objects, the AGN continuum emission also tended to 
dominate the stellar continuum in these objects, suggesting significant levels of activity. For both the ORELSE and VVDS 
surveys these galaxies were noted in the visual inspection process and were selected based on these notes. The spectrum 
for each AGN selected in this manner was re-inspected visually to confirm both the presence of an AGN and the quality of the 
redshift measurement. In total, 130 X-ray and 77 type-1 AGN were identified in the overlap region. In the left panel of Fig. 
\ref{fig:densityplotwAGN} we plot a false color image of the \emph{Herschel}/SPIRE mosaic in the overlap region. The mosaic was 
generated with linear scaling and made use of the python-based APLpy package (http://aplpy.github.com). The false color mosaic is 
contrasted by the right panel of Fig. \ref{fig:densityplotwAGN} in which we plot the spatial distribution of all radio starbursts 
and all AGN selected in this section detected at a significance of $\ge3\sigma$ in at least one of the three SPIRE bands against the backdrop
of a density map composed from the 1753 SPIRE-detected galaxies fulfilling the criteria of the previous section. 
The density map was created using the methodology of Gutermuth et al. (2005), with the parameter dictating the number of nearest neighbors, 
$N$, set to $N=8$. Curiously, the AGN hosts appear to largely avoid the regions of highest SPIRE-detected galaxy density, especially in the 
case of radio AGN, an effect which cannot be caused by observational or selection issues. The environments of radio AGN in particular will be 
discussed in detail in a companion paper (Lemaux et al. 2015). Summarized in
Table \ref{tab:AGNtotals} are the total number of objects in the overlap region detected in each of the observations
relevant for AGN selection and matched to an optical counterpart along with the total number of these objects which were 
selected as AGN hosts.

\begin{table}
\caption{Multiwavelength objects in the HerMES/SWIRE/CHFTLS overlap region\label{tab:AGNtotals}}
\centering
\begin{tabular}{lcc}
\hline \hline
Sample & $N_{tot}$\tablefootmark{a} & $N_{AGN}$ \\[0.5pt]
\hline 
Radio & 567 & 316 \\[4pt]
IRAC & 28394 & 172 \\[4pt]
4-Channel IRAC & 1107 & 172 \\[4pt]
VVDS Type-1 AGN & 77 & 77 \\[4pt]
X-ray & 137 & 130 \\[4pt]
SPIRE\tablefootmark{b} & 1753 & 181\tablefootmark{c} \\ 
\hline
\end{tabular}

\tablefoot{
\tablefoottext{a}{Brighter than the $5\sigma$ completeness limit and matched to an optical counterpart with a well-defined redshift}
\tablefoottext{b}{Brighter than the $3\sigma$ completeness limit and matched to an optical counterpart with a well-defined redshift}
\tablefoottext{c}{These numbers reflect the number of unique galaxies in which an AGN is hosted, i.e.,  AGNs detected using multiple diagnostics are counted only once}
}
\end{table}

It is necessary at this point to make a final mention of subtleties regarding the SED-fitting process, this time for the 
AGN populations discussed in this section. Depending on the strength of the AGN and the dust properties of both the host 
and the region surrounding the nucleus, an active nucleus can dominate the stellar and nebular components of a host galaxy 
at a variety of wavelengths. This is particularly a problem for unobscured X-ray emitting AGN, quasars, and type-1 AGN, where the UV portion of 
the SED can almost entirely originate from the AGN (Vanden Berk et al.\ 2001; Gavignaud et al.\ 2006; Polletta et al.\ 2007). Obscured AGN can also present 
a problem, as the very features that allow their selection in the NIR can confuse ones ability to measure quantities such as stellar mass. We tested 
the SED fits of the parent samples of all the AGN (i.e., not only those detected by SPIRE) by comparing the photo-$z$s and 
spec-$z$s of those galaxies with reliable measurements for both and by comparing the $\chi_{\nu}^2$ value of the fits of 
the host galaxies of the various AGN with the full photometric sample. For radio- and IR-selected AGN the results were 
encouraging, no increase in the NMAD or the catastrophic outlier rate was observed for the host galaxies of those populations 
relative to the full photometric sample, nor was the $\chi_{\nu}^2$ statistic appreciably higher. However, this was not the 
case for the X-ray AGN and the type-1 AGN where both statistics were significantly higher. To alleviate this, we fit 
hybrid galaxy/QSO templates following the methodology of Salvato et al.\ (2009) using the templates of Polletta et al. 
(2007). This method resulted in a reduction of the catastrophic outlier rate of the full X-ray AGN host population with 
high-quality spectra from $\ga50$\% to $\sim$6\% and a reduction of the NMAD to $\sigma_{\Delta z/(1+z_s)}\sim0.15$.
While both of these values are considerably higher than that of the full photometric sample, our large redshift bins 
and the fact that galaxies hosting X-ray AGN make up a small fraction of our overall SPIRE-detected AGN host 
sample ($\sim10$\%), mean that this lack of precision is of limited consequence for our results. Because decomposing 
the stellar light from the light originating from the AGN is inexact, physical parameters other than redshift, such 
as stellar mass, are estimated for such galaxies with questionable accuracy. This, however, is of less consequence, as
the stellar mass of such galaxies is largely considered only as part of the full SPIRE population which is comprised of only 
$\sim1$\% of galaxies hosting X-ray AGN. For type-1 hosts, marked improvement in photo-$z$ reliability or accuracy was 
not observed upon the inclusion of the hybrid galaxy/QSO templates, as five out of the ten type-1 hosts detected at $\ge3\sigma$ 
in one of the three SPIRE bands remained as catastrophic outliers. The type-1 hosts which are catastrophic outliers, 
however, make up an even smaller fraction of our full SPIRE and full AGN samples than the X-ray AGN hosts (0.2\% and
2.7\%, respectively). Furthermore, since the $L_{TIR}$ of these galaxies, which is the quantity most relevant
to the analyses involving AGN host galaxies in this study, was calculated using spectroscopic redshifts when 
available, redshifts which the type-1 AGN hosts have by definition, these galaxies are retained as part of the  
full sample. 

\section{The properties of star-forming galaxies detected by SPIRE}
\label{fullSPIRE}

We break the remainder of the paper into two parts. In this section we discuss the properties of the 1753 \emph{Herschel}
bright galaxies selected in the previous sections making no differentiation between those galaxies hosting AGN and 
those with dormant nuclei. Because the former is a small fraction of our sample ($\sim10$\%), and because the 
fraction of AGN which may have their optical continua dominated by an AGN is an even smaller fraction of our 
sample (i.e., type-1 AGN and perhaps some X-ray AGN, $\sim$2\%), the results in the first part of this section 
will be broadly applicable to normal dusty star-forming or starbursting galaxies, with a 
minimal contamination to our sample of stellar masses, rest-frame colors, or magnitudes by galaxies hosting an AGN. 
In the subsequent section we will divide our sample into two distinct sub-groups, those SPIRE-detected galaxies 
with and without an AGN, and compare the properties of each subsample.

\subsection{Color, magnitudes, and stellar masses of the full SPIRE sample}
\label{colormagnSM}

\begin{figure*}
\plottwospecial{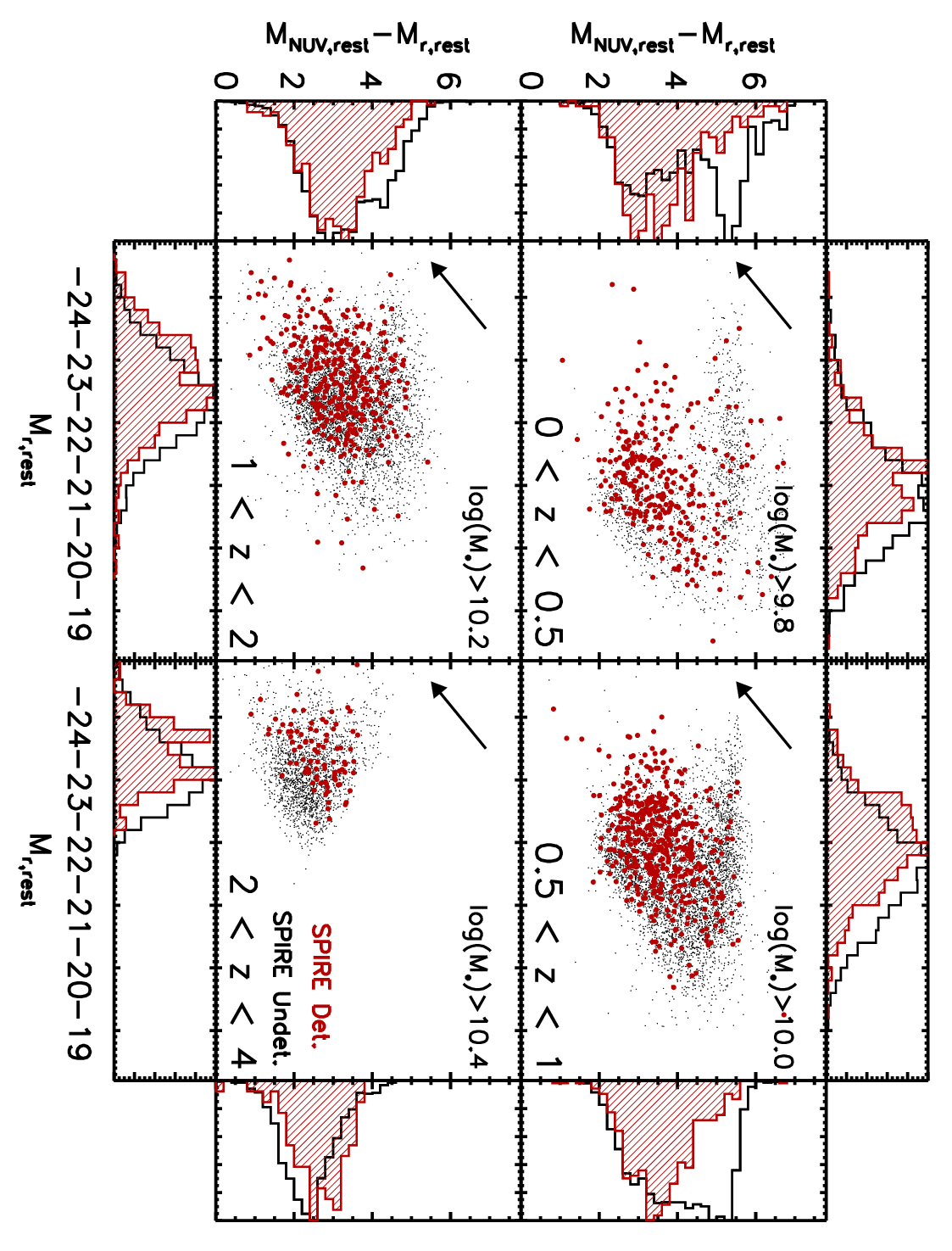}{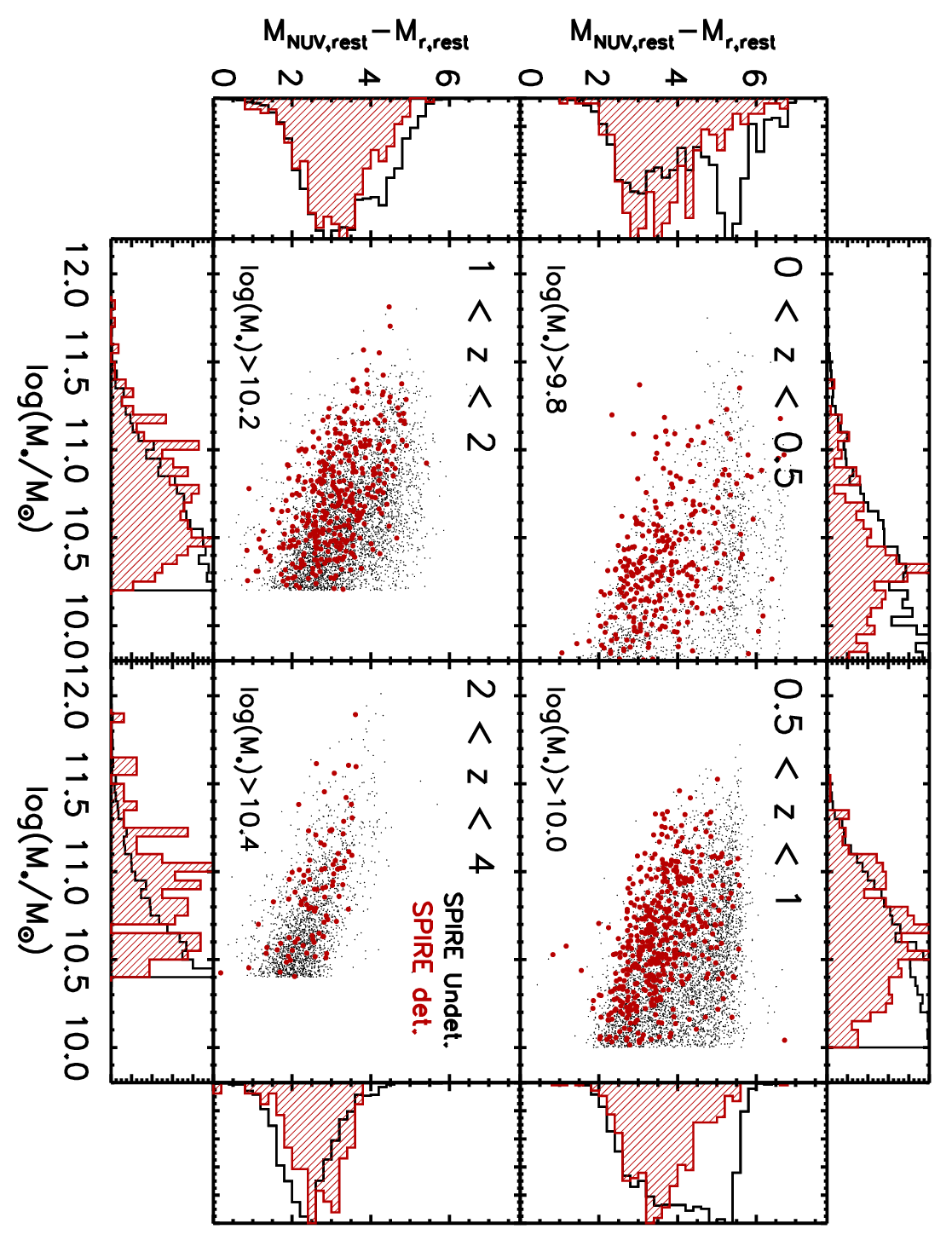}
\caption{Rest-frame $M_{NUV}-M_{r}$ color magnitude (\emph{left}) and color stellar mass (\emph{right}) diagrams of the full SPIRE sample (filled red
circles and dashed red histograms) and galaxies undetected in SPIRE (small black points and open black histograms). Galaxies are binned
into four different redshift bins and histograms of each sample are normalized (such that the maximal value is unity) for each redshift
bin. Only those galaxies detected at high significance in CFHT-WIRDS $K_s$-band imaging are shown in both panels. In each panel the sample is
restricted to those galaxies with stellar masses above the stellar mass limit for each redshift range (given in each panel). The arrow denotes
the change in color and (where appropriate) absolute magnitude for an extinction value of $E_{s}(B-V)\sim0.3$ using a Calzetti et al. (2000)
extinction law. The arrow is vertical in the right panel because the measured extinction values are already incorporated into
the SED-fitting process. The SPIRE-detected sample spans the full range of absolute magnitudes, stellar masses, and, surprisingly, colors
in each redshift bin.}
\label{fig:initialCMDnCSMD}
\end{figure*}

We begin the investigation of the full SPIRE sample\footnote{For the remainder of the paper, the phrase ``full SPIRE 
sample'' refers to the 1753 galaxies satisfying all the criteria given in \S\ref{TIRnSFRs}} by plotting in 
Fig. \ref{fig:initialCMDnCSMD} rest-frame $NUV-r^{\prime}$ color-magnitude diagram (CMD) and color-stellar-mass 
diagram (CSMD) of the full SPIRE sample in four different redshift bins. Rest-frame colors and magnitudes are 
derived from our SED-fitting process described in detail in \S\ref{SEDfitting}. Also plotted in Fig. 
\ref{fig:initialCMDnCSMD} are those galaxies in each redshift bin that went undetected in SPIRE but which fulfil 
all the other criteria of the full SPIRE sample (i.e., having a well-measured redshift between $0<z<4$ and sufficient 
photometry to perform SED fitting). For galaxies plotted in both the CMDs and CSMDs we additionally require galaxies be detected 
in the $K_s$ band at a magnitude brighter than the completeness limit of the CFHT-WIRDS imaging ($K_s<24$) and have a SED-derived stellar mass 
that is higher than the stellar mass completeness limit for each bin (see Fig. \ref{fig:initialCMDnCSMD}).

What is immediately apparent from the CMD is that the SPIRE-detected galaxies span a full range of colors and 
magnitudes, ranging from the brightest blue galaxies to the faintest red galaxies in each redshift bin. This is 
also true on inspection of the CSMD: the full SPIRE sample essentially spans the entire phase space encompassed 
by the SPIRE-undetected galaxies. The fraction of SPIRE-detected galaxies in the blue cloud (defined 
as roughly $NUV-r^{\prime}\la4$)\footnote{The definition of blue cloud, green valley, and red sequence changes 
slightly in each redshift bin. The numbers quoted here are an average of the cuts used across all bins. In practice,
the fractions quoted in this section are largely insensitive to the exact choice of color ranges of each of these
three regions} is quite high in every redshift bin (70\% on average), a consequence of the high rate of 
star formation in these galaxies. However, a non-negligible fraction (12.5\% on average) of galaxies in the full 
SPIRE sample have colors consistent with the red sequence (defined as $NUV-r^{\prime}\ga4.5$) for redshifts 
where the red sequence is clearly separable from the locus of star-forming galaxies (i.e., $0<z<2$). The remaining
galaxies in the full SPIRE sample (17.5\% on average) have intermediate colors, lying in a region 
of phase space commonly referred to as the green valley. This portion of phase space is typically thought to 
house a transitory galaxy population, comprised both of galaxies that were previously on the red sequence but 
are undergoing low-level rejuvenated star formation and those galaxies that have finished their star formation and 
are slowly transitioning onto the red sequence (Silverman et al.\ 2008; Kocevski et al.\ 2009b; Vergani et al.\ 2010; 
Mendez et al.\ 2011; Fang et al.\ 2012; Gon{\c c}alves et al.\ 2012). While the number of galaxies  
located in the green valley is a sizeable fraction of the full SPIRE sample, this fraction only
slightly exceeds the same quantity for all optical galaxies ($\sim15$\%). Still, it is surprising that galaxies
selected to be vigorously forming stars would (seemingly) be equally likely to be in a transition phase as 
galaxies which are either quiescent or undergoing low-level star formation. 

\begin{figure}
\epsscale{1.8}
\plotonespecial{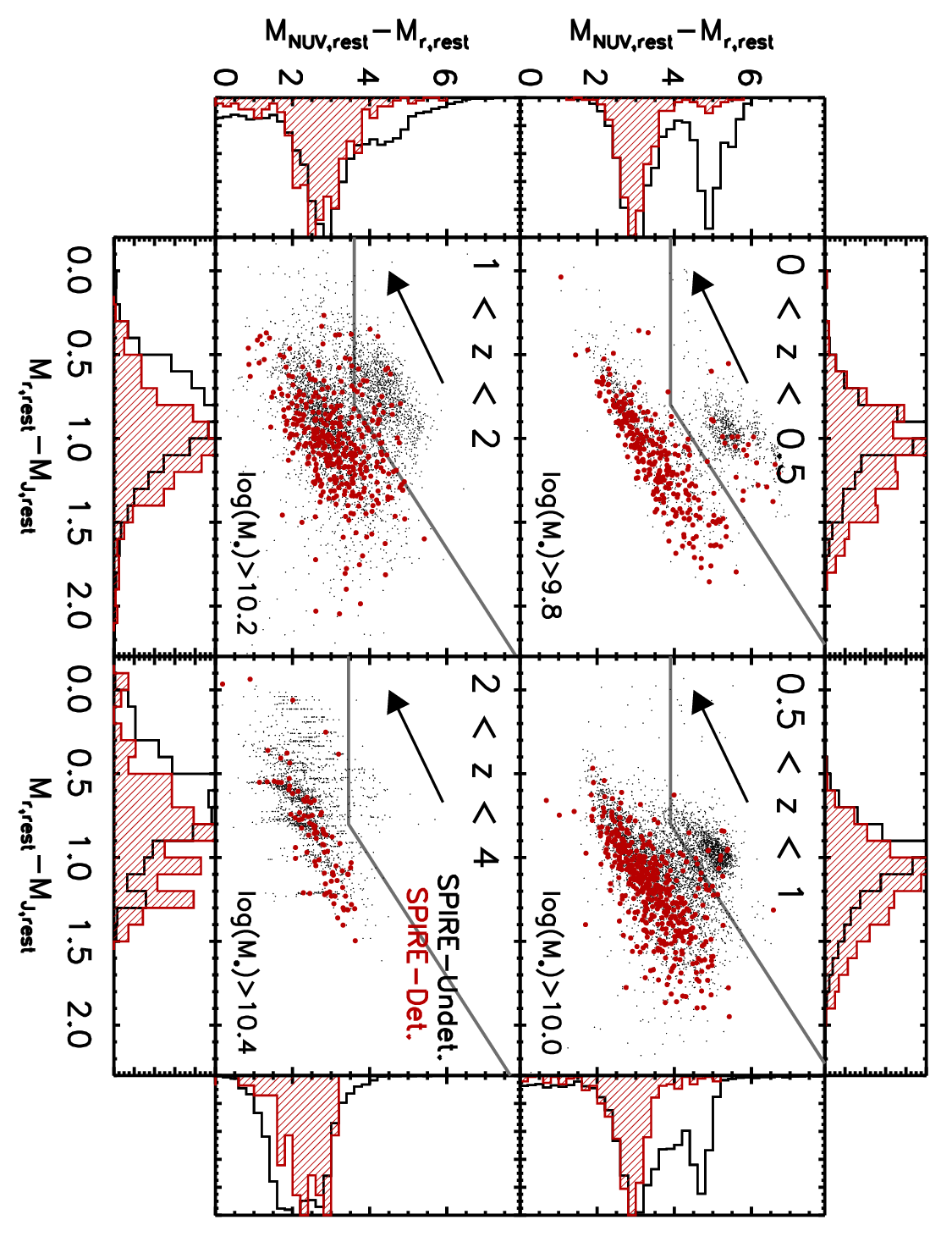}
\caption{Rest-frame $M_{NUV}-M_{r}$ vs. $M_r-M_J$ color-color diagram for both the full SPIRE sample and galaxies undetected in the SPIRE imaging. 
The gray lines in each panel divide quiescent galaxies (above the lines) from star-forming galaxies (below the line) and is determined using a 
modified version of the methodology employed by Ilbert et al. (2013). The meanings of colors and symbols are nearly identical to Fig. \ref{fig:initialCMDnCSMD}. The
one exception is the histogram plotted along the ordinate in each panel, in which galaxies are projected into an ``effective color'' defined as 
the offset from the dividing line. This representation of the sample serves to largely mitigate extinction effects, and as such, a large fraction 
($\sim$95\%) of the full SPIRE sample are now in the phase space coincident with normal star-forming galaxies.}
\label{fig:initialCCD}
\end{figure}

The prevalence of infrared-selected starburst galaxies in the green valley has been observed for a \emph{Spitzer}/MIPS 
survey of 70$\mu m$ sources in the COSMOS field (Kartaltepe et al.\ 2010b). However, as is discussed extensively in that study,
the high number of dusty starbursts in the green valley 
can be due to either properties intrinsic to the stellar populations of these galaxies or to dust extinction effects. 
In the case of the latter, the galaxies themselves are not inherently in a transitory phase, but rather appear so due 
to heavy extinction of the stellar continuum coming from the dusty component in the galaxy (see, e.g., Cardamone et al.\ 
2010). While we could, in principle, correct the color and magnitude of each galaxy for extinction effects, the extinction 
values coming from the SED fitting process are not robust enough to do this in practice. In principle, we could also use
the FIR-derived extinction values, values that we derive for a subsets of the full SPIRE sample later in this paper (see \S\ref{AtoK})
to correct the colors and magnitudes of each SPIRE-detected galaxy. However, since we are comparing to an optically selected 
sample of galaxies undetected in SPIRE in each redshift bin, the colors and magnitudes of the comparison sample cannot be 
corrected in this manner, which limits the usefulness of the method. 

Instead, as is done in Kartaltepe 
et al.\ (2010b), we employ a color-color diagram (CCD) to investigate which of these two possibilities is more 
applicable to the full SPIRE sample. Plotted in Fig. \ref{fig:initialCCD} is the $NUV-r^{\prime}$ vs. $r^{\prime}-J$ CCD of
the full SPIRE sample against the backdrop of the SPIRE-undetected sample. As was done for both panels of Fig. \ref{fig:initialCMDnCSMD}, we require here 
that any galaxy plotted be detected in the $K_s$ band at a magnitude brighter than the completeness limit in order to ensure 
accurate rest-frame NIR magnitudes. Overplotted on the CCD is a dividing line between galaxies with quiescent colors and 
those consistent with young stellar populations (see Ilbert et al.\ 2013, Muzzin et al.\ 2013, and Arnouts et al.\ 2013, and discussions 
and references therein). The dividing lines given in Fig. \ref{fig:initialCCD} are slightly different than those presented in Ilbert 
et al.\ (2013) due to a different methodology used to determine the minima in color-color space between the two populations.
In this paper, the CCD was rotated to search for the optimal angle of approach for the diagonal line in each redshift 
bin, requiring that this angle must remain the same in each bin. The optimal angle of approach was defined as the angle 
which maximized the disparity between the local minimum and the local maxima of the colors projected on the new axes 
(i.e., the axis defined by the diagonal line and an axis perpendicular to this line). The diagonal line was truncated for each redshift bin 
at the point where it began to intersect the star-forming locus, i.e., where continuing the line diagonally began to decrease the disparity 
between the minimum and local maxima of the rotated color-color histograms. The histograms along the ordinate of each panel in
Fig. \ref{fig:initialCCD} no longer simply represent the distributions of $M_{NUV}-M_{r}$ colors. Rather an effective color is plotted, 
defined as the effective vertical offset of each galaxy from its projection onto the dividing line. The analysis of galaxies 
in this color-color space has the distinct advantage of almost completely eliminating extinction effects, as galaxies with higher extinction 
move parallel to the main discriminator nearly excluding the possibility of dust causing star-forming galaxies to appear quiescent. 

In this context, the interpretation of full SPIRE sample changes dramatically. While in color-magnitude and 
color-stellar-mass space a large fraction of SPIRE-detected galaxies had colors consistent with the red sequence or 
green valley ($30$\%), this effect is largely gone in the CCD. The galaxies of the full SPIRE sample have colors that 
place them on the star-forming sequence (i.e., below the dividing line), with only 8.0\%, 5.5\%, 6.5\%, and 0\% 
of SPIRE-detected galaxies having colors consistent with the quiescent part of this diagram in the $0<z\le0.5$, 
$0.5<z\le1$, $1<z\le2$, and $2<z\le4$ bins, respectively. This is in contrast to the SPIRE-undetected sample, in which 
41.3\%, 34.0\%, 26.8\%, and 3.8\% of the galaxies have colors consistent with quiescent galaxies in the same bins, 
respectively. Transition galaxies are also more apparent in this diagram, as galaxies between the quiescent region 
and the star-forming sequence are there as a result of either age or metallicity effects. While there still remains 
a small fraction of galaxies that appear to be undergoing a transition, especially in the redshift range $0.5<z\le2$, the main 
conclusion from this diagnostic is that $\sim95$\% of the galaxies in the full SPIRE sample have colors consistent with 
star-forming galaxies once extinction effects are minimized. 

\subsection{The star forming main sequence}
\label{sfmainseq}

\begin{figure*}
\plottwokindaspecialrot{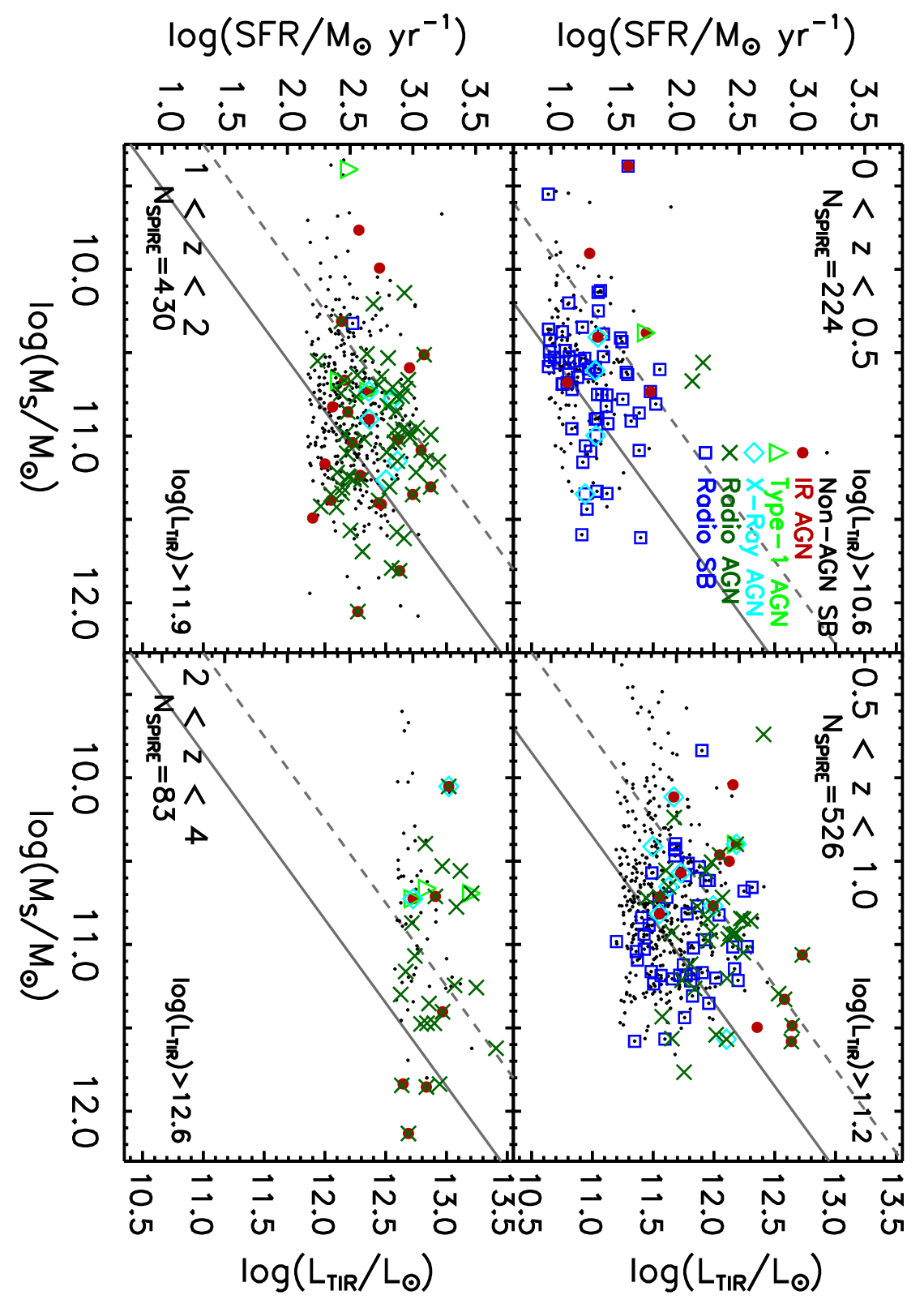}{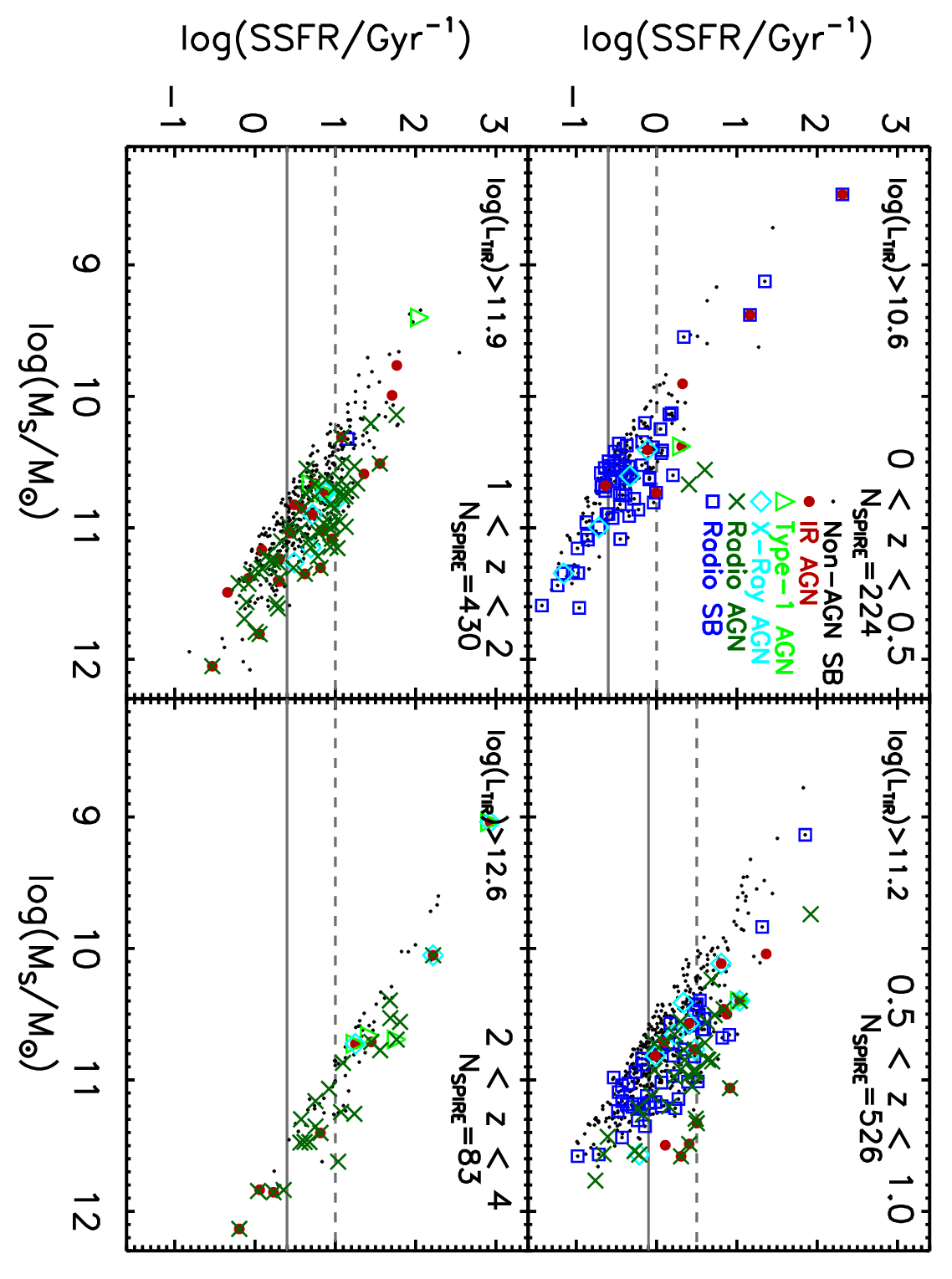}
\caption{\emph{Left:} Total infrared (TIR)-derived $\mathcal{SFR}$ plotted against stellar mass for volume-limited samples of all optically matched galaxies 
detected in at least one SPIRE band at $\ge3\sigma$ for four different redshift bins. $L_{TIR}$ is shown on the right ordinate for reference. 
Because the $\mathcal{SFR}$ conversion employs a Salpeter IMF, the stellar masses shown here are converted from a Chabrier to a Salpeter IMF. 
Black points represent galaxies detected in SPIRE, while galaxies detected in radio imaging and those galaxies hosting AGN are 
overplotted in larger symbols. The main sequence relationship for star-forming galaxies is adopted from E11 for galaxies at $0<z<0.5$, an average of
the relation in E11 and that relation of the lowest stellar mass bin of R11 for galaxies at $0.5<z<1$, and the relation from the lowest stellar mass 
bin in R11 for galaxies at $z>1$ and is plotted as a solid gray line. The dashed line denotes the transition point between galaxies classified as normal star forming to
those classified as starbursts (4$\times$ the main sequence $\mathcal{SSFR}$ threshold). There is no evidence in three of the four redshift bins of a correlation 
between $\mathcal{SFR}$ and $\mathcal{M}_{\ast}$ to the depth of our SPIRE imaging. \emph{Right:} Specific $\mathcal{SFR}$ plotted against stellar mass for the same samples as the
left panel. Symbols and lines are identical to the left panel. The dynamic range of \emph{Herschel} is small in any given redshift bin while the  
stellar mass range is not, which leads to the observed trend.} 
\label{fig:SFRnSSFRvsmass}
\end{figure*}

In this section we consider $L_{TIR}$ limited subsamples of the full SPIRE sample in the same four redshift bins as were used in the previous section. 
These $L_{TIR}$ cuts, which are shown in the right panel of Fig. \ref{fig:totlum}, allow us to consider subsamples which are volume limited, such that the 
effects of Malmquist bias are minimized. This is essential for the study that follows.  
Plotted in the left panel of Fig. \ref{fig:SFRnSSFRvsmass} are the stellar masses and TIR $\mathcal{SFR}$s of the volume-limited full SPIRE sample in four redshift
bins. Due to the method in which $\mathcal{SFR}$s are calculated, SED-fit stellar masses are converted from a Chabrier IMF to a Salpeter IMF 
for the remainder of this section using the relation of Ilbert et al.\ (2010). One of the most striking features of the left panel of Fig. \ref{fig:SFRnSSFRvsmass} 
plot is the high rate of star formation observed in galaxies hosting AGN relative to SPIRE-detected galaxies where no AGN is observed, a subject which will be covered extensively in \S\ref{sbstrength}. 
In this section, we make no distinction between galaxies hosting a powerful AGN and those without an active nuclei, and instead consider the question: is there a 
relationship between stellar mass and $\mathcal{SFR}$ in the volume-limited SPIRE sample, and, if so, does that relationship evolve as a function of redshift? 

In seminal \emph{Herschel} study, Elbaz et al.\ (2011; hereafter E11) used a combination of \emph{Spitzer}, \emph{AKARI}, 
and \emph{Herschel} observations to suggest the existence of an infrared main sequence for star-forming galaxies detected in the mid- and 
far-infrared. In this picture, stellar mass ($\mathcal{M}_{\ast}$) is intimately linked to the amount of star formation in such galaxies, 
such that, if a galaxy is star forming, the higher the stellar mass the more vigorous the star formation. Multiple studies over a large range of redshifts 
using a variety of different selection methods have found that, in star-forming galaxies, increasing $\mathcal{M}_{\ast}$ generally leads to an increase in $\mathcal{SFR}$ 
(e.g., Brinchmann et al.\ 2004; Daddi et al.\ 2007; Elbaz et al.\ 2007; Noeske et al.\ 2007a; Santini et al.\ 2009; Gonz$\rm{\acute{a}}$lez et al.\ 2011; 
Muzzin et al.\ 2012; Koyama et al.\ 2013), but this phenomenon was observed in E11 for the first time with a purely IR-selected sample of galaxies.
These observations have in turn lead to significant tension with semi-analytic models of galaxy formation, specifically in the evolution of this relationship 
as a function of redshift
and its overall normalization (for an excellent review of this topic, see Dav$\rm{\acute{e}}$ 2008). The uniqueness of the data presented in E11 was, however, that
the sample presented was strictly IR-selected, probed a large range of redshifts, and probed a dynamic range of stellar masses and TIR-derived $\mathcal{SFR}$s of greater than 
two orders of magnitude. Using the observed $\mathcal{M}_{\ast}$-$\mathcal{SFR}$ relationship and arguments based on the measured compactness of galaxies as measured in their
IR photometry, E11 created a new method for defining starbursts by relating the current $\mathcal{SFR}$ of a galaxy to its past averaged $\mathcal{SFR}$ (Kennicutt 1983). The past
averaged $\mathcal{SFR}$ was proxied by the $\mathcal{M}_{\ast}$-$\mathcal{SFR}$ infrared main sequence (MS), and therefore a galaxy significantly departing from the MS at a given 
redshift is considered a starburst galaxy (more specifically, when the ratio of the $\mathcal{SSFR}$ of a galaxy and a typical value of the $\mathcal{SSFR}$ for a MS galaxy 
at the same redshift exceeds two). This definition of starburst, which has been heavily used in the literature (see, e.g., Magnelli et al.\ 2012 and 
references therein), relies heavily both on the premise that the infrared MS exists and that
the typical value of the $\mathcal{SSFR}$ for an MS galaxy can be measured precisely as a function of redshift. The latter relies on both a precise quantification 
of the relationship between $\mathcal{M}_{\ast}$ and $\mathcal{SFR}$, the precision with which each parameter can be estimated, and the assumption that this relationship 
does not change based on the selection method of the sample nor the dynamic range of $\mathcal{SFR}$s or stellar masses probed in a given sample. 

While our observations do not reach the depth achieved by E11 in the GOODS-N/S fields, we probe an area which is more than order of magnitude 
greater than the observations of E11. At the expense of fainter sources, our observations are more sensitive to
rare galaxies which are undergoing extremely vigorous star-formation episodes. Galaxies with $\mathcal{SFR}$s in excess of 100 $\mathcal{M}_{\odot}$ yr$^{-1}$, a population
which is largely absent in the E11 sample, are found in abundance in our observations for all redshifts greater than $z\ga0.5$. To our knowledge, a statistical sample of this population
has been investigated by only one other study (Rodighiero et al.\ 2011; hereafter R11, see also Rodighiero et al.\ 2014 for a followup study).



In our sample, we see little evidence for an infrared MS for the $\mathcal{SFR}$/$L_{TIR}$ range probed by our SPIRE observations. 
In all redshift bins the correlation between $\mathcal{M}_{\ast}$ and $\mathcal{SFR}$ is weak: the Spearman rank correlation 
coefficient is 0.17 on average and only one bin ($0.5<z<1$) shows a $>3\sigma$ deviation from the null hypothesis that 
the two quantities are uncorrelated. The results of these tests remain essentially unchanged if we apply the stellar mass completeness limits from the previous section.
This lack of correlation argues against the existence of the $\mathcal{M}_{\ast}$-$\mathcal{SFR}$ MS for IR-selected galaxies at all redshifts that have $\mathcal{SFR}$s in excess of the $\mathcal{SFR}$ 
limits used to form the volume-limited samples here\footnote{If such galaxies were, by definition, starbursts this result would not be surprising, and perhaps would even be 
nonsensical. However, adopting the definition of starburst from R11, i.e., galaxies whose $\mathcal{SSFRs}$ are in excess of 4$\times$ the MS value as a function of stellar mass, the bulk of the full 
SPIRE sample is not comprised of starburst galaxies.}. While this lack of correlation observed in the full SPIRE sample cannot be due to differential effects
as a function of stellar mass, e.g., imposing a stellar mass and $\mathcal{SFR}$ limits have little effect on the lack of correlation, our data
do not probe deep enough at any redshift to claim that such a correlation does not exist for galaxies anything but the most prodigious star formers. However, as mentioned 
previously, the large area coverage of our SPIRE reduces cosmic variance and allows for the characterization of starburst 
galaxies over a large range of $\mathcal{SFR}$s by adopting the formalism of those studies which probe deep enough to observe such a correlation.

In order to determine the fraction of starburst galaxies in each redshift bin we now turn to the $\mathcal{SSFR}$s of the full SPIRE sample plotted in the 
right panel of Fig. \ref{fig:SFRnSSFRvsmass}. Again a Salpeter IMF is adopted in calculating stellar masses. Galaxies in our sample are observed over a 
large range of $\mathcal{SSFR}$s, spanning nearly three orders of magnitude in all redshift bins. This effect is largely driven by the stellar 
mass of the galaxies, since the \emph{Herschel} observations have a limited dynamic range for a given redshift range. 
We now adopt the definition of a starburst from a combination of those definitions in E11 and R11 for the remainder of this section. 
Applying these cuts to the full SPIRE sample at all redshift bins and imposing the stellar mass cuts from the previous section 
results in a fraction of starburst galaxies, so defined, of $20.3\pm1.4$\% over all redshift ranges. The fractional contribution of 
starburst galaxies to the global star formation rate density ($\mathcal{SFRD}$) in a given (stellar-mass and volume-limited) redshift bin varies immensely, 
ranging from a 28\% contribution in the lowest redshift bin to 61\% in the highest redshift bin. These numbers are, however, subject to the 
relatively shallow $\mathcal{SFR}$ limits in each bin and should be viewed as strict upper limits in each redshift bin. This exercise 
shows that starburst galaxies make up a non-negligible fraction of the SPIRE population and contribute significantly to the overall number of 
stars being formed at all redshifts, at least to the depth of our data.  

While the observed starburst fraction in the full SPIRE sample is considerably higher than that found in R11 for galaxies between $1.5<z<2.5$ (2-3\%), 
except for the lowest redshifts probed in our sample, the data presented here does not go deep enough to probe 
the BzK selected samples of R11 that make up the bulk of the lower $\mathcal{SFR}$ and lower stellar mass sample. Thus, due to the differing sample selections and various differential biases, 
these numbers can not be directly compared to those of R11 in anything other than a broad sense, and, given the limits of our data, 
are not in direct contradiction with their results. In an attempt to make a more rigorous comparison, a volume-limited sample was selected between $1.5<z<2.5$ from our data by
imposing $\mathcal{SFR}(L_{TIR})>200$ $\mathcal{M}_{\odot}$ yr$^{-1}$ and a stellar mass cut of $\log(\mathcal{M}_{\ast})>10.2$, which 
allows for a direct comparison with the sample of R11. In this volume-limited sample we find that 35.6$\pm5.7$\%
of the $\mathcal{SFRD}$ at $1.5 < z < 2.5$ for IR-bright galaxies with $\mathcal{SFR}$s in excess 200 $\mathcal{M}_{\odot}$ yr$^{-1}$ is contributed by galaxies
classified as starbursts. While this number is marginally inconsistent with the $\sim20-25$\% contribution reported by R11, within the
relative uncertainties of the two samples it is plausible that the two results are consistent. 

However, a larger issue remains. Wide variations in the starburst fraction and fractional contributions to the $\mathcal{SFRD}$ by starbursts of up to
nearly an order of magnitude \emph{in our own sample} result if we impose slightly different stellar mass completeness 
criteria, redshift ranges, or vary the definition of a starburst by as little as 0.3 dex. The fraction of starburst galaxies and their contribution to the global 
$\mathcal{SFRD}$ appear highly subject to the stellar mass completeness limits and the definition of a starburst. The latter is intimately linked both with the precision 
at which one can constrain the MS \emph{at a given redshift}, the method used and precision to which one can measure stellar masses and $\mathcal{SFR}$s 
(see Rodighiero et al.\ 2014 for a thorough disucssion of the latter), and the methodology used in determining galaxies are significantly offset from this
sequence. The extreme level of variance observed in our data clearly shows the necessity of precisely and accurately determining the MS at a given redshift and 
to properly and precisely quantify the stellar mass and $\mathcal{SFR}$ limits of the data in order to use such analysis to constrain galaxy models with any sort of reasonable accuracy.
The problem of constraining the contribution of starburst galaxies to the overall star formation rate density of the universe using this framework appears, in the true 
sense of the term, a chaotic one.
 
While this discussion may appear to be a matter of semantics, the operational definition of the term starburst is an 
extremely important one in the study of galaxy evolution as such galaxies are connected to a physical phase 
of galaxy formation, which is then used in turn to contextualize the evolution of galaxies across a wide variety of 
environments (e.g., Goto 2006; Moran et al.\ 2007; Oemler et al.\ 2009; Dressler et al.\ 2013). 
Given the large variation in results seen both in our own sample and from contrasting our sample with 
those results from the literature, it is clear that significant work is necessary to understand the relationship between
\emph{Herschel}-bright sources and galaxies undergoing normal star formation, to better understand the 
nature of the star-forming MS, and to find proper methodologies to robustly define starburst 
populations. It is outside the scope of this paper to attempt a precise definition of the term starburst, and, indeed, our data 
generally preclude such possibility. Thus, we instead apply the term starburst throughout this paper as a proxy for ``SPIRE-detected, 
strongly star-forming galaxy'' in an agnostic view.

\subsection{General properties of SPIRE galaxies at different TIR luminosities}
\label{generalTIR}

In \S\ref{colormagnSM} the color, magnitude, and stellar mass properties of the full SPIRE sample were investigated without differentiating galaxies  
based on their TIR luminosities. While doing this allowed us to determine, broadly, the properties of our sample, marginalizing over 
TIR luminosity limits our ability to discern the more subtle properties of the full SPIRE sample, as it has been observed that dusty starburst galaxies of differing 
TIR luminosities exhibit different dust temperatures and SEDs (Symeonidis et al.\ 2013; Hwang et al.\ 2010b), rest-frame colors (Kartaltepe et al.\ 2010b), and are found to lie in 
different environments (Kocevski et al.\ 2011a). In this section we move from considering the full SPIRE sample as a whole to galaxies in individual TIR bins. 
We define here three samples: a ``low-$L_{TIR}$'' sample ($11<\log(L_{TIR})<11.5$, low luminosity luminous infrared galaxy (LIRG) level emission), a ``moderate-$L_{TIR}$'' 
sample ($11.5<\log(L_{TIR})<11.9$, high luminosity LIRG level emission), and a ``high-$L_{TIR}$'' sample ($\log(L_{TIR})>11.9$, ultra-LIRG (ULIRG) level). We restrict 
the redshift range of each of these subsamples to redshifts $0.5<z<2$, which is the redshift range containing the bulk of our sample. The TIR levels and redshift 
range were chosen in such a way to minimize redshift-dependent effects in each bin (i.e., sampling a representative population over the chosen redshift range) 
while still maximizing the number of SPIRE-detected galaxies in each bin. Furthermore, we attempted to construct the bins such that roughly an equal number of 
SPIRE-detected galaxies with spectral redshifts were placed in each bin. There is some concern that galaxies of the same $L_{TIR}$ have redshift-dependent 
properties, such that performing this analysis over such a large redshift bin will confuse galaxy populations which are fundamentally different. However, for 
galaxies which we could compare over the entire $0.5<z<2$ redshift range (i.e., the high-$L_{TIR}$ sample), the spectral, photometric, color, and stellar mass 
properties were all nearly identical from low ($z<1.5$) to high ($z>1.5$) redshift, which likely precludes the possibility that redshift-dependent properties will 
confuse the analysis which follows. The sample properties of the galaxies in each $L_{TIR}$ bin are summarized in Table \ref{tab:lumsamples}. 

\begin{table*}
\caption{Properties of the luminosity subsamples of the full SPIRE sample \label{tab:lumsamples}}
\centering
\begin{tabular}{lccccccc}
\hline \hline
Sample & $\log(L_{TIR})$ Range & Equivalent $L_{TIR}$ Class & $N_{total}$ & $N_{spec}$\tablefootmark{a} & $\langle \log(L_{TIR}) \rangle$ & $\langle M_{r^\prime} \rangle$ & $\langle M_{NUV}-M_{r^{\prime}} \rangle$ \\[0.5pt] 
\hline
low-$L_{TIR}$ & $11<\log(L_{TIR})<11.5$ & low-lum LIRG & 175 & 42 & 11.35 & -21.90 & 3.74 \\[4pt]
moderate-$L_{TIR}$ & $11.5<\log(L_{TIR})<11.9$ & high-lum LIRG & 266 & 52 & 11.70 & -22.12 & 3.50 \\ [4pt]
high-$L_{TIR}$ & $\log(L_{TIR})>11.9$ & ULIRG & 358 & 43 & 12.30 & -22.67 & 3.14 \\ [0.5pt] 
\hline
\end{tabular}
\tablefoot{
\tablefoottext{a}{Only VVDS galaxies are considered as ORELSE galaxies comprised $\lsim5$\% of each subsample}
}
\end{table*}

\subsection{Spectral properties of SPIRE galaxies at different TIR luminosities}
\label{spectralTIR}

We begin by contrasting the VVDS rest-frame NUV/optical spectral properties of each of the three $L_{TIR}$ subclasses. The typical individual VVDS spectrum in our
sample has an median signal-to-noise (S/N) ratio of order $\sim5-10$ pix$^{-1}$, which is not sufficient for much of the analysis presented in this section. Thus, rather 
than study the spectra of individual galaxies we instead chose to create an inverse-variance weighted composite spectrum (referred to hereafter as ``coadded spectrum'' or ``coadd'') 
of the galaxies in each luminosity bin using the methodology described in Lemaux et al.\ (2009, 2012) adapted for the VVDS data. As was the case in Lemaux et al.\ (2012), 
the average flux density of the spectrum of each galaxy is normalized to unity prior to the stacking process. The normalization factor for 
each spectrum is determined over the range of wavelengths free of poorly subtracted airglow lines or other reduction artifacts. This 
stacking and weighting scheme has two main advantages: it allows to increase the S/N of our spectra significantly (the coadded spectrum 
in each of the three bins has S/N$\sim$50), while allowing us to retain the ability to study the 
\emph{average} galaxy in each luminosity bin (though subject to the caveats discussed in Lemaux et al.\ 2012). It is important to note that the subsample of galaxies
in each $L_{TIR}$ bin with available high-quality VVDS spectra fully spans the stellar mass, color, and absolute magnitude phase space of their full photometric parent 
samples, and, further, are distributed through this phase space in an approximately representative manner relative to their parent samples. In addition, the 
distributions of the subsample of those galaxies with high-quality VVDS spectra have a (scaled) $L_{TIR}$ distribution and a median $L_{TIR}$ that are essentially identical to the 
parent photometric sample. The distributions of the two samples in an $i^{\prime}-m_{250\mu m}$ CMD also appear as nearly scaled versions of each other. The observed similarity
in these two phase spaces precludes the possibility that a large population of TIR-bright, highly dust-extincted galaxies are lost in the spectroscopic sample.
Thus, we feel confident in applying the results of the spectral analysis presented in this and the following two sections to the full galaxy populations of each $L_{TIR}$ bin.

\begin{figure*}
\plottwo{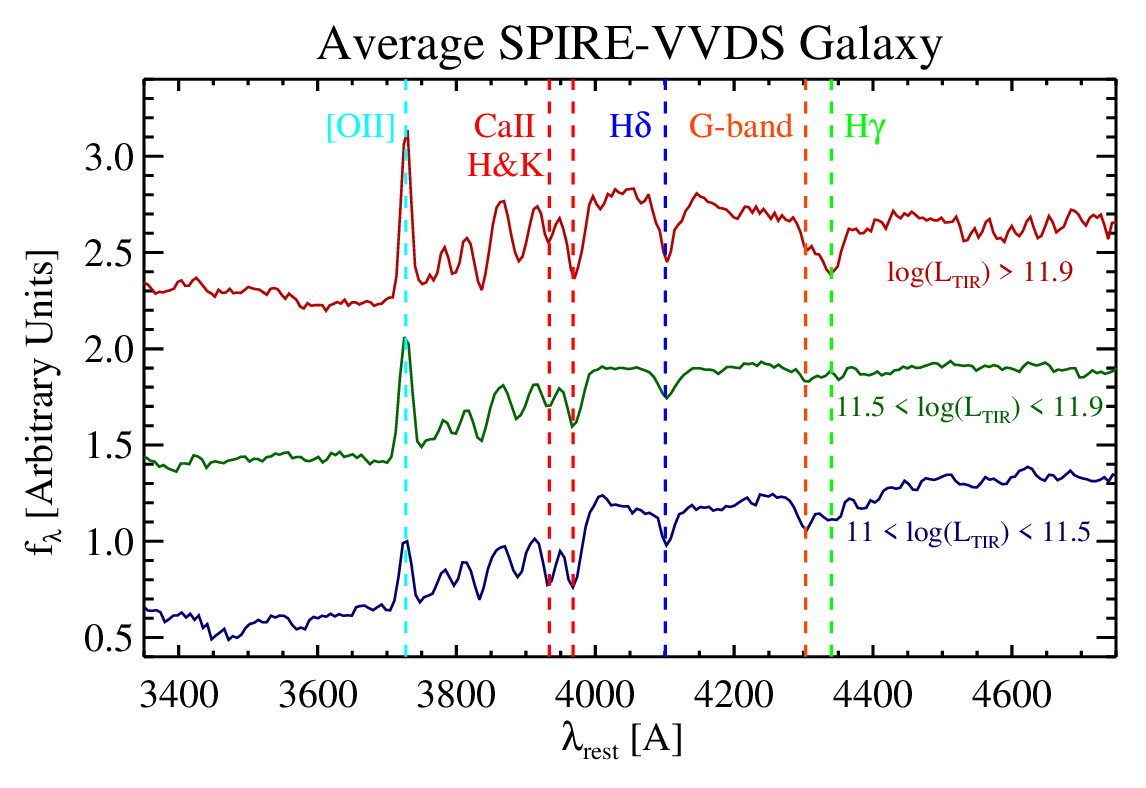}{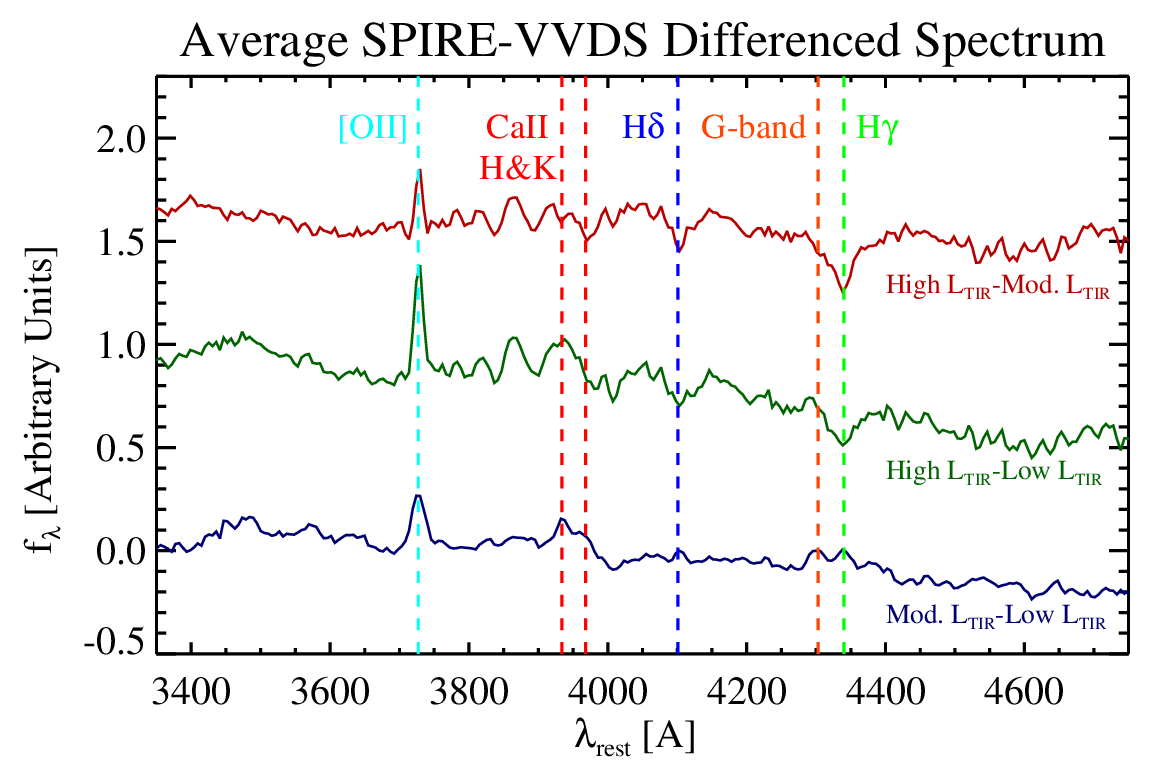}
\caption{\emph{Left:} Unit-weighted spectral coadditions of VVDS spectra of galaxies detected in at least one SPIRE band at $\ge3\sigma$ for three different
luminosity bins. Each coadded spectrum contains $\sim50$ galaxies, with only VVDS galaxies being used as ORELSE galaxies comprise only $\lsim5$\% of any given sample. 
Important spectral lines are overplotted and spectra are offset vertically for clarity. Though all $L_{TIR}$ subsamples of the full SPIRE sample show strong
[OII] emission and strong Balmer features, which indicate significant ongoing star formation, the subsamples also show marked differences. \emph{Right:} Differenced
coadded spectra of the three $L_{TIR}$ subsamples. The high-$L_{TIR}$ subsample appears to be comprised of galaxies which, on average, have stronger [OII] emission
and much stronger Balmer absorption features than either of their lower-$L_{TIR}$ counterparts. Also clearly visible are weaker features associated with older stellar 
populations (CaII H\&K and G-Band) in the average high-$L_{TIR}$ galaxy relative to the other two $L_{TIR}$ samples. The average galaxy in the moderate-$L_{TIR}$ 
subsample largely continues this trend relative to that of the low-$L_{TIR}$ subsample.}
\label{fig:spectralplot}
\end{figure*}

Plotted in Fig. \ref{fig:spectralplot} is the coadded spectra of those low-$L_{TIR}$, moderate-$L_{TIR}$, and high-$L_{TIR}$ with available VVDS spectra. 
Qualitatively, the average galaxy in each bin appears to exhibit considerably different spectral properties from the average galaxies in the other two bins. 
Considering the main features marked in Fig. \ref{fig:spectralplot}, it is apparent that the average high-$L_{TIR}$ galaxy has stronger $\lambda$3727\AA\
[OII] emission, stronger Balmer features (i.e., $\lambda$4340\AA\ H$\gamma$, $\lambda$4101\AA\ H$\delta$, and the higher order Balmer lines blueward of the
CaII H\&K features), and weaker spectral features typically associated with older stellar populations, i.e., CaII H\&K, $\lambda$4305\AA\ $G$-band, and the 
strength of the continuum break at 4000\AA\ known as $D_n(4000)$, than its lower TIR-luminosity counterparts. For the latter point, the one exception is the $\lambda$3969\AA\ 
CaII H line, which appears stronger in the average high-$L_{TIR}$ galaxy. However, the Balmer series absorption line $\lambda$3970\AA\ H$\epsilon$ is 
coincident with this feature at the resolution of the rest-frame VVDS spectrum. Given the strength of the Balmer features and the strong asymmetry in the strengths 
between the CaII H\&K lines (see Rose et al.\ 1985 and Rumbaugh et al.\ 2012 for a detailed discussion of this asymmetry), it is almost certainly the case that 
absorption coming from younger stellar populations, i.e., B- and A-type stars, is the dominant contribution to the observed feature. The Balmer series lines, particularly
that of H$\delta$, also appear progressively wider at progressively higher $L_{TIR}$ due to Stark broadening, an effect attributable to a larger fractional contribution of 
earlier A-type stars in the higher $L_{TIR}$ coadds (Struve 1926; Mihalas 1964). These qualitative 
observations are reinforced by analyzing the spectra quantitatively. For each spectrum, bandpass equivalent width ($EW$) measurements of the [OII]
and H$\delta$ features were made using a methodology almost identical to Lemaux et al.\ (2012). A slight modification was made to this method to accommodate the 
decreased resolution of the VVDS data by widening the bandpass used to measure the H$\delta$ feature by $\sim$6\AA\ on each side of the nominal Fisher et al.\ (1999) 
bandpasses. The strength of $D_n(4000)$, typically used to measure luminosity-weighted stellar 
ages (e.g., Kauffmann et al.\ 2003), was also measured for each coadd using the ratio of continua blueward and redward of 4000\AA, as defined by Balogh et al.\ (1999), 
using the methodology of Lemaux et al.\ (2012). These values are given in Table \ref{tab:specproperties}. 

In Fig. \ref{fig:spectralplot} we also plot differenced spectra for each of the three combinations of TIR luminosity 
bins. The conclusions reached previously by a simple qualitatively visual inspection of the individual coadds are broadly confirmed here: the contribution from young stellar populations  
is dominant in the average high-$L_{TIR}$ galaxy relative to the average galaxies in the other two bins. Taking the extremes of the distribution, features associated with
recently created stars in the high-$L_{TIR}$/low-$L_{TIR}$ differenced spectrum all appear strongly in absorption, meaning the amplitude of the absorption is significantly 
higher in the average high-$L_{TIR}$ galaxy. Conversely, the $\lambda$3933\AA\ CaII K feature and the $G$-band appear in emission in the differenced spectrum, implying 
a higher fractional contribution from late-type (i.e., older) stars in the average low-$L_{TIR}$ galaxy. Visual inspection of the high-$L_{TIR}$/moderate-$L_{TIR}$ differenced 
spectrum implies that the luminosity-weighted stellar population of the average moderate-$L_{TIR}$ galaxy is intermediate to that of the other two samples, with significantly 
less contribution from early-type stars (proxied by the Balmer absorption) than that of the high-$L_{TIR}$ sample, but also significantly less contribution from late-type 
stars than that of the low-$L_{TIR}$ sample. The range plotted in Fig. \ref{fig:spectralplot} was determined by requiring that at least half of the spectra in each 
luminosity subsample went into the coadded spectrum over the entire wavelength range. Restricting the wavelength of interest in this manner is necessary so that 
differences between the various populations can be attributed to properties inherent to the galaxy population in each bin and not to artifacts from the stacking process. 
This point is particularly important here as we begin to to discuss the differences in average stellar continua and spectral fits to stellar population synthesis models, 
as the loss of spectra over a particular wavelength range can artificially induce a significant change in the color of the spectra.

\begin{table*}
\caption{Composite spectral properties of the luminosity subsamples of the full SPIRE sample \label{tab:specproperties}}
\centering
\begin{tabular}{lccccc}
\hline \hline
Sample & $\langle EW(\rm{[OII]}) \rangle$ & $\langle EW(\rm{H}\delta) \rangle$\tablefootmark{a} & $\langle D_{n}(4000) \rangle$\tablefootmark{b} & $\langle A/K \rangle$ & $\langle E(B-V) \rangle$ \\ [0.5pt]
\hline \\[-1pt]
low-$L_{TIR}$ & -8.63$\pm$0.14\AA & 4.11 (4.30)$\pm$0.11\AA & 1.284$\pm$0.003 (1.224$\pm$0.004) &0.73 &0.661$\pm$0.005 \\[4pt]
moderate-$L_{TIR}$ & -14.47$\pm$0.13\AA & 3.80 (4.12)$\pm$0.10\AA & 1.137$\pm$0.002 (1.081$\pm$0.003) &1.65 &0.703$\pm$0.004 \\ [4pt]
high-$L_{TIR}$ & -14.99$\pm$0.15\AA & 6.77 (7.09)$\pm$0.14\AA & 1.115$\pm$0.003 (1.050$\pm$0.004) &9.00 &0.807$\pm$0.005 \\ [0.5pt]
\hline 
\end{tabular}
\tablefoot{
\tablefoottext{a}{The numbers given in parentheses are EW(H$\delta$) values corrected for emission infill (see text)}
\tablefoottext{b}{The numbers given in parentheses are $D_n(4000)$ values corrected for extinction (see text)} 
}
\end{table*}

\subsubsection{A/K Ratios and Continuum Measurements}
\label{AtoK}
Now that we are confident that the full spectral range can be used to probe properties inherent to each galaxy population,
we move from measuring single indices to considering the coadded spectra as a whole. To further quantify the difference in the average galaxy of each bin, two component stellar 
population synthesis models from Bruzual (2007; hereafter CB07) were fit to each coadded spectrum to determine the average ratio between the contribution of late-type and 
early-type stars commonly referred to as an A/K ratio. The CB07 models are modified versions of the original Bruzual \& Charlot (2003) models updated with a new prescription
to account for the mass loss of thermally pulsing asymptotic giant branch stars (Charlot, priv. comm.). For the purposes of the analysis presented here, the results obtained using
CB07 models are negligibly different from those obtained with those of Bruzual \& Charlot (2003). The method used to calculate A/K ratios for each
spectrum is similar to that used in Quintero et al.\ (2004) and Yan et al.\ (2006), though different models are used here that result in offsets in the absolute value of the A/K 
ratio. Since we are interested only in the relative difference between the average galaxy in each TIR luminosity bin, this absolute offset makes no difference to the results in
presented in this study. The A component of the spectrum was represented by a single burst model with a $\tau=0.01$ Gyr, ``observed'' 300 Myr after the starburst, and smoothed
to match the resolution of the VVDS coadded spectra. The time after the starburst was chosen to be consistent with those models used in the works mentioned above and is roughly the 
time at which the luminosity-weighted contribution of A-type stars to the synthetic spectrum is maximized (i.e., when $EW($H$\delta)$ is maximized, see Fig. \ref{fig:sbplot}). 
The K component of the spectrum also used a smoothed single burst $\tau=0.01$ Gyr model but observed 5 Gyr after the 
initial starburst, a time when late-type stars dominate the synthetic spectrum. For each coadd, the [OII] emission feature was interpolated over 
and fit a linear combination of the two components to each coadded spectrum over the entire wavelength range presented in Fig. \ref{fig:spectralplot} was performed, 
adopting the best-fit as that linear combination which minimized the $\chi_{\nu}^2$ statistic. The A/K ratio was then defined as the quotient of 
the amplitudes of the A and K components of the best-fit composite model. The results of these fits, given in Table \ref{tab:specproperties}, add to the growing body of 
evidence that the luminosity-weighted stellar populations in the average galaxy
of each of the three TIR luminosity bins is significantly different. The measured A/K ratio of the average high-$L_{TIR}$ galaxy shows that the continuum light of this galaxy 
is completely dominated by early-type stars (A/K=9.0). Conversely, the stellar continuum of the average low-$L_{TIR}$ galaxy has a K-star component which is more prominent  
than that of A-type stars (A/K=0.73), with the moderate-$L_{TIR}$ average galaxy again falling at an intermediate stage between the galaxies in 
the two other TIR luminosity bins (A/K=1.65). 

\begin{figure}
\epsscale{1.25}
\plotone{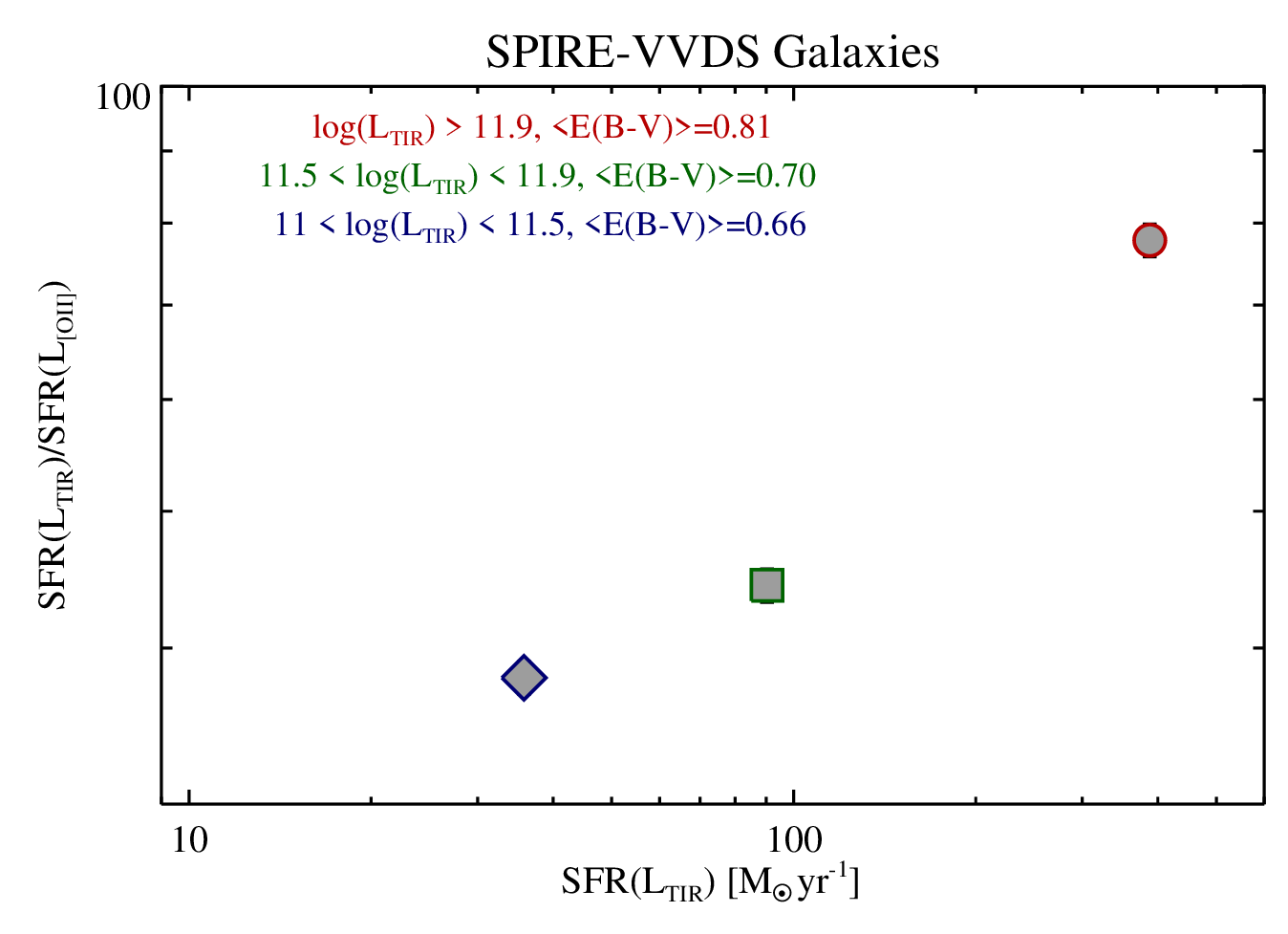}
\caption{Ratios of $\mathcal{SFR}$s determined from $L_{TIR}$ to those determined by [OII] emission for the average galaxies of the three $L_{TIR}$ subsamples
plotted against $L_{TIR}$. In the \emph{Herschel} subsamples, [OII] emission underestimates the true $\mathcal{SFR}$ by a factor of $\sim$40$-$80. Increasing
$L_{TIR}$ results in an increase in the $\mathcal{SFR}$($L_{TIR}$)/$\mathcal{SFR}$([OII]) ratio, presumably a consequence of the larger dust content in the average high-$L_{TIR}$ 
galaxy relative to its lower luminosity counterpart. Only random errors are used resulting in error bars the sizes of the symbols. Systematic errors related to 
$\mathcal{SFR}$ conversions and extinction are quite large, but are unnecessary to include here as the comparison is relative and any systematic 
offset would affect each ratio in the same manner. The $\mathcal{SFR}$($L_{TIR}$)/$\mathcal{SFR}$([OII]) ratio is used to determine the nebular extinction, $E(B-V)$, for the average 
galaxy in each subsample. These values are given at the top of the plot along with the definitions of each subsample.}
\label{fig:EBV}
\end{figure}

The color of the stellar continuum in the differenced coadded spectra also allows for insights on the average luminosity-weighted stellar population. Both the spectra which are
differenced with respect to the low-$L_{TIR}$ sample appear quite blue, which, in the absence of dust effects, also implies a larger fractional contribution of late-type stars in the
average low-$L_{TIR}$ galaxy relative to that of the high-$L_{TIR}$ and moderate-$L_{TIR}$ samples. Of course, given that our sample is selected, by definition, as galaxies containing 
a significant amount of obscuring dust, the interpretation of this color difference is not as straightforward. If the strength of the stellar extinction in the average low-$L_{TIR}$ galaxy 
(i.e., $E_s(B-V)$) is higher than that of the high-$L_{TIR}$ or moderate-$L_{TIR}$ samples, the blueness observed in the differenced spectra
could be attributable solely to dust. While it has been shown that the opposite trend is true, i.e., that galaxies with higher TIR luminosities have, on average, a higher value of 
$E(B-V)$, and by proxy, a higher value of $E_s(B-V)$, than those galaxies of lower TIR luminosities (Kocevski et al.\ 2011a), such statistical arguments may not hold when applied to 
the small number of galaxies in each spectral bin ($\sim50$). 

Instead, we calculated for each coadded spectrum an average nebular extinction value, $\langle E(B-V) \rangle$, which 
allows us to completely remove the effects of extinction when discussing average quantities of the various subsamples. In order to derive an $\langle E(B-V) \rangle$ for each
sample it is necessary to compare star formation rates coming from two different indicators, one that is significantly affected by dust and one that is largely dust-independent.
Because each of our coadded spectra has the [OII] emission feature within its usable range, the [OII]-derived $\mathcal{SFR}$ can be used in concert with the $\mathcal{SFR}$ derived from the
average TIR luminosity to determine $\langle E(B-V) \rangle$ for each of our subsamples. A calculation involving this method has two inherent assumptions: $i)$ that the $\mathcal{SFR}$ 
derived from the TIR luminosity is, for our sample, a good representation of the true amount of star formation in a given galaxy, and $ii)$ the star-formation properties of the galaxy as a whole 
is well represented by the portion of the galaxy covered by the slit. As stated in \S\ref{TIRnSFRs}, the TIR luminosity dominates the SED of these galaxies, meaning that we can safely ignore the 
first assumption. The second assumption is a bit more complex. In a recent study of galaxy clusters at $z\sim0.3-0.5$, Oemler et al.\ (2013) found little difference between the strength of 
the EW(H$\delta$) and EW([OII]) for galaxies in their sample when slit apertures of two different sizes were used. The covering fraction of these two slit sizes
was disparate enough for them to conclude that a spectrum observed with a $1\arcsec$ slit is sufficiently representative of the galaxy as a whole for a large
fraction of their sample. The sample we consider here is bounded on the low-redshift end by the high-redshift galaxies in the sample of Oemler et al.\ (2013). As such, the 
covering fraction for the galaxies in our sample, given that the VVDS galaxies were observed with a $1\arcsec$, is equivalent or higher than that of Oemler et al.\ (2013) and
approaches unity at $z\gsim1$. Thus, we ignore all considerations of slit effects for the remainder of this discussion and equate the properties observed in each coadded spectrum
with the properties of the galaxies as a whole.

Because of the way the coaddition process was performed, absolute flux calibration is not preserved. Thus, rather than using an average [OII] line flux it is necessary to rely on 
the $\langle$EW([OII])$\rangle$ value for each bin to derive the dust-dependent $\mathcal{SFR}$. For each bin, the average [OII]-derived $\mathcal{SFR}$ was calculated by

\begin{equation}
\begin{split}
\mathcal{SFR}(L_{[OII]}) = -(7.87\pm1.97)\times10^{-10}(10^{\frac{-(\langle M_{u^{\prime}} \rangle +48.6)}{2.5}})\\
(\frac{c}{\lambda_{eff}^2}) \langle EW([OII]) \rangle, 
\end{split}
\label{eqn:SFROII}
\end{equation}

\noindent where $\langle M_{u^{\prime}} \rangle$ is the mean rest-frame $u^{\prime}$ absolute magnitude of a given bin, chosen because it provides a fair sampling of the rest-frame continuum 
emission surrounding the [OII] feature, $\lambda_{eff}$ is the effective wavelength of the $u^{\prime}$
filter curve $3543\times10^{-8}$ cm, and c is the speed of light in cm s$^{-1}$. The constant of proportionality is adopted from a combination of 4$\pi d_L^2$, where $d_L$ is the luminosity
distance to each source, 10 pc in units of cm for the case of an absolute magnitude, conversion of the $EW$ from \AA\ to cm, and the [OII] $\mathcal{SFR}$ formula of Kewley et al.\ (2004). The 
negative sign is a consequence of the $EW$ convention adopted in this paper. The IMF is assumed to be Salpeter, the same that was used for the TIR $\mathcal{SFR}$ derived in \S\ref{TIRnSFRs}. The 
average nebular extinction, $\langle E(B-V) \rangle$, for each coadd is then

\begin{equation}
\langle E(B-V) \rangle = \frac{2.5\log(\frac{\mathcal{SFR}(\langle L_{TIR} \rangle)}{\mathcal{SFR}(L_{[OII]})})}{k^{\prime}_{[OII]}},
\label{eqn:EBV}
\end{equation}

\noindent where $\langle$TIR$\rangle$ is the mean TIR luminosity in a given bin and $k^{\prime}_{[OII]}$ is the Calzetti et al.\ (2000) reddening curve evaluated at the wavelength of
[OII]. The ratio of the $\mathcal{SFR}$ indicators as a function of $L_{TIR}$ for the three bins are given in Fig. \ref{fig:EBV} and the $\langle E(B-V) \rangle$ values for each bin 
are given in Table \ref{tab:specproperties} and Fig. \ref{fig:EBV}. The values found for the first two subsamples are similar to those found in Choi et al.\ (2006), who applied 
almost identical methodologies to a near- (2.2$\mu$m) and mid- (24$\mu$m) infrared selected spectroscopic sample of ULIRGs spanning a slightly lower redshift range than the current
sample. The mean $L_{TIR}$-derived $\mathcal{SFR}$ of the high-$L_{TIR}$ subsample is prohibitvely high to make a similar comparison. However, the $E(B-V)$ value found for the high-$L_{TIR}$ 
subsample is similar to those values derived for a \emph{Herschel}/PACS-selected ULIRG sample with roughly equivalent median $L_{TIR}$ and redshifts 
(Buat et al.\ 2011), albeit with an extinction law which is slightly modified from the one chosen here.

\begin{figure}
\plotonespecial{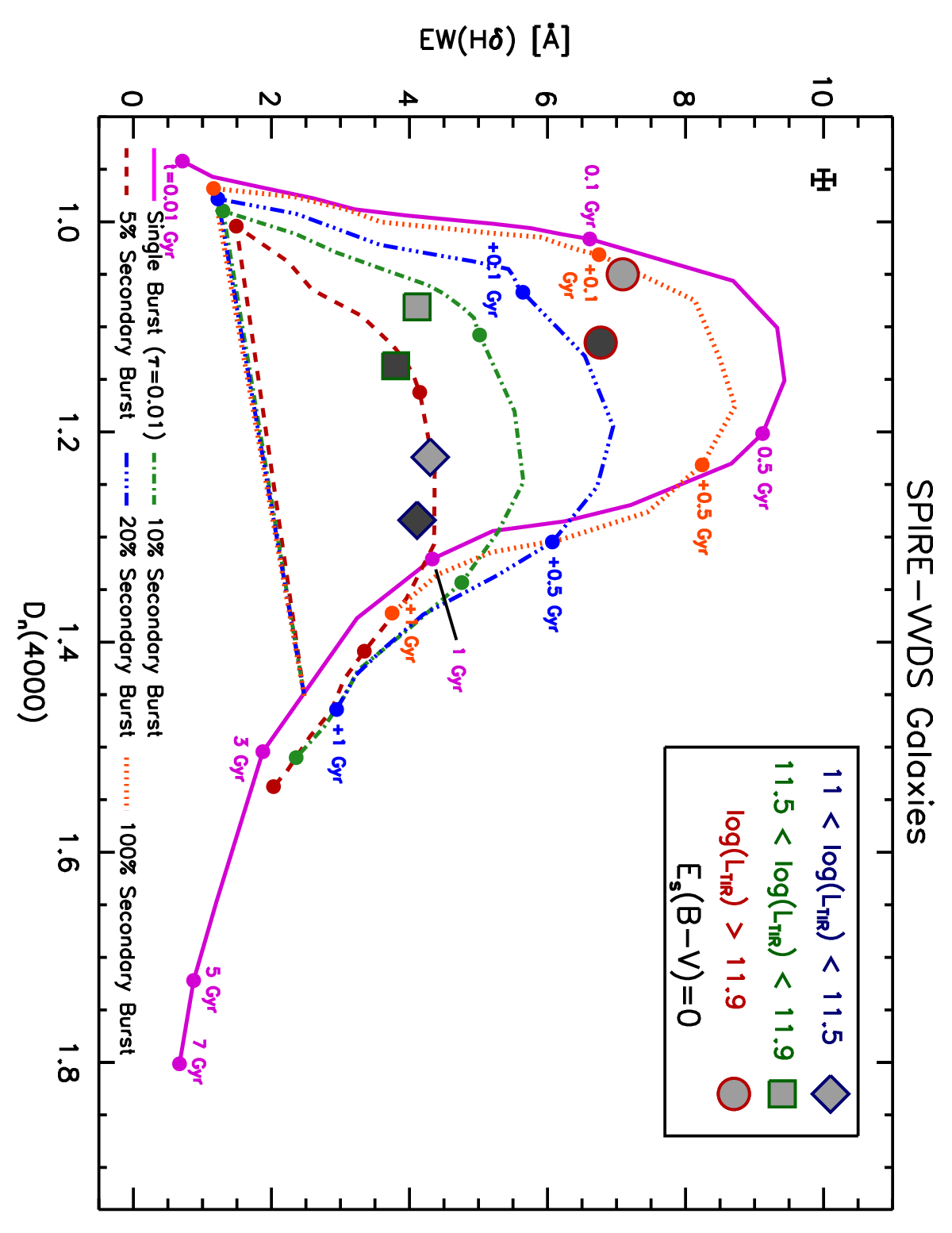}
\caption{Measurements of EW(H$\delta$) and $D_n(4000)$ of the coadded spectra of the high-$L_{TIR}$ (circles), moderate-$L_{TIR}$ (diamonds), and low-$L_{TIR}$ subsamples 
corrected (filled light gray symbols) and uncorrected (filled dark gray symbols) for the effects of extinction. These measurements are plotted against the backdrop of 
equivalently measured quantities from CB07 synthetic spectra for various time steps using various models. For models with secondary bursts, these bursts occur 2 Gyr 
after the initial burst. No extinction is applied to the CB07 models and the models employ a Salpeter (1955) IMF, identical to the IMF used for all $\mathcal{SFR}$ calculations. 
The decreasing starburst strength relative to the already established stellar mass for galaxies of decreasing $L_{TIR}$ suggest that 
the LIRG population is experiencing rejuvenated star formation at moderate levels while the average ULIRG is undergoing either its initial starburst or an 
extremely vigorous (i.e., 100\% by mass) rejuvenated star-formation event.}
\label{fig:sbplot}
\end{figure}

\subsubsection{Mean Stellar and Starburst Ages as a Function of TIR Luminosity} 
\label{stellarnsbages}

As stated earlier, for the trend observed in the differenced continua observed in the previous section to be,
at least in part, due to differential dust effects rather than to the difference in the mean stellar populations of each $L_{TIR}$ subsample, it would be necessary for the average
low-$L_{TIR}$ galaxy to have a higher extinction than that of the high-$L_{TIR}$ subsample. What we found in the previous section is the opposite, namely that the nebular extinction of the 
average low-$L_{TIR}$ galaxy ($\langle E(B-V) \rangle=0.66$) is significantly \emph{lower} than that of the average high-$L_{TIR}$ galaxy ($\langle E(B-V) \rangle=0.81$). The average
moderate-$L_{TIR}$ galaxy again falls in between that of the two other subsamples, with a $\langle E(B-V) \rangle=0.70$. These results are consistent with those of Kocevski 
et al.\ (2011a), in which it was observed that, for \emph{Spitzer}/MIPS selected galaxies in a large-scale structure at $z\sim0.9$, the nebular extinction increased with increasing
$L_{TIR}$. These results are also broadly consistent with those found in the field by Roseboom et al.\ (2012) in a spectroscopic sample of MIPS- and SPIRE-selected galaxies 
at redshifts similar to the ones probed in our sample ($0.8<z<1.8$). However, the difference in nebular extinction between the subsamples does not necessarily translate in a 
straightforward manner to a difference in the extinction of the stellar continuum. We assume here that the nebular extinction is related to the extinction of the stellar continuum, 
the quantity of interest in this discussion, via 
$E(B-V)_s=0.44E(B-V)$ (Calzetti 1997; Calzetti et al.\ 2000). In principle it is possible or even likely that a variety of dust geometries for a given galaxy would change the constant of 
proportionality relating the two quantities, such that we could not draw a definite conclusion based on only nebular extinction values. However, in this analysis we are averaging 
over a large population of galaxies in each subsample and thus mitigating the effects of varying dust geometries when comparing the various subsamples. Given this and given the 
small variance in the constant of proportionality observed for individual galaxies in Calzetti et al.\ (2000), it is likely that the variation in the constant of proportionality 
is minimal for the average galaxy of the three $L_{TIR}$ subsamples. Therefore, the intrinsic properties of the stellar population of the mean galaxy of each subsample must
be the cause of the observed color difference in the various subsample, as an extinction correction of the coadded spectra would only serve to increase the blue color of the 
differenced spectra. Thus, there remains two main effects that could cause a increasingly redder stellar continuum with decreasing TIR luminosity: metallicity and age of the
mean stellar populations. While differing metallicities between the various subsamples could account for some of the observed color difference between the various subsamples, 
the effect of metallicity on the color of galaxies over the wavelength range considered in the coadded spectra is too weak to explain the observed difference (see, e.g., Cooper
et al.\ 2008; Lemaux et al.\ 2012). Thus, these results, along with those obtained earlier on the analysis of the spectral lines and the A/K ratios, strongly implies that 
\emph{the mean luminosity-weighted age of the stellar populations of an average (U)LIRG decreases with increasing TIR luminosity.}

This result has one of two implications for the galaxies studied here. One way to induce this correlation is if the average galaxy in the low-$L_{TIR}$ subsample is observed 
later in the development of the starburst than the average galaxy in either of the other subsamples, thereby decreasing the contribution of A-type stars to the 
observed spectral features and stellar continuum color. An alternative scenario is to appeal to a well-established older stellar population in the average low-$L_{TIR}$ 
galaxy, and to a lesser extent in the average moderate-$L_{TIR}$ galaxy, embedded into the galaxy from star-formation events that occurred prior to the onset of the currently observed 
burst. In the diagnostics used thus far, the two
scenarios appear equally probable, as the two scenarios would manifest themselves in an identical manner. In order to disentangle these scenarios we revisit the 
EW(H$\delta$) and $D_n(4000)$ measurements made of the coadded spectra of each subsample \S\ref{spectralTIR}. To begin, in Fig. \ref{fig:sbplot} we plot the measured 
EW(H$\delta$) and $D_n(4000)$ values in dark filled symbols against the backdrop of measurements made of solar-metallicity synthetic spectra from CB07 
using a variety of different models using the method described in Lemaux et al.\ (2012). Unlike the method of Lemaux et al.\ (2012) we do 
not, however, redden the models here as noted by the $E_s(B-V)$ value in the legend of Fig. \ref{fig:sbplot}. 

The values of EW(H$\delta$) and $D_n(4000)$ plotted in dark symbols (and given outside the parentheses in Table \ref{tab:specproperties}) represent the measurement taken directly 
from each coadded spectrum. In order to make these measurements meaningful in the context of the CB07 models plotted in Fig. \ref{fig:sbplot} 
corrections need to be made to both EW(H$\delta$) and $D_n(4000)$. The correction to EW(H$\delta$) is straightforward. Because EW measurements are generally
insensitive to the effects of dust (see discussion in Lemaux et al.\ 2010), it is only necessary to make a correction for the infill to the stellar H$\delta$ absorption from  
nebular emission. This is necessary because the CB07 models are only stellar and do not contain a component which models emission from HII regions. This correction was 
based on the method of Kocevski et al.\ (2011a) and uses the EW([OII]) of each coadded spectrum. The emission infill corrected EW(H$\delta$) values are given in parentheses 
in Table \ref{tab:specproperties}. The correction to $D_n(4000)$ is also straightforward now that we have already derived the average extinction for each $L_{TIR}$ subsample. 
The average $E(B-V)_s$ values for each coadd were used to apply a Calzetti reddening curve to each coadded spectrum and a de-reddened $D_n(4000)$ 
measurement was made of each coadded spectrum. These values are given inside the parentheses
in Table \ref{tab:specproperties}. The corrected values of EW(H$\delta$) and $D_n(4000)$ for each $L_{TIR}$ subsample should be completely independent of dust, nebular emission, 
and, given that metallicity effects are not strong in the age range considered here (Kauffmann et al.\ 2003), metallicity effects, and can thus be compared directly to the BC07
models to determine the age since the onset of the starburst for each $L_{TIR}$ subsample. The corrected values are plotted in Fig. \ref{fig:sbplot} as lightly filled symbols.

\begin{figure*}
\plottwospecial{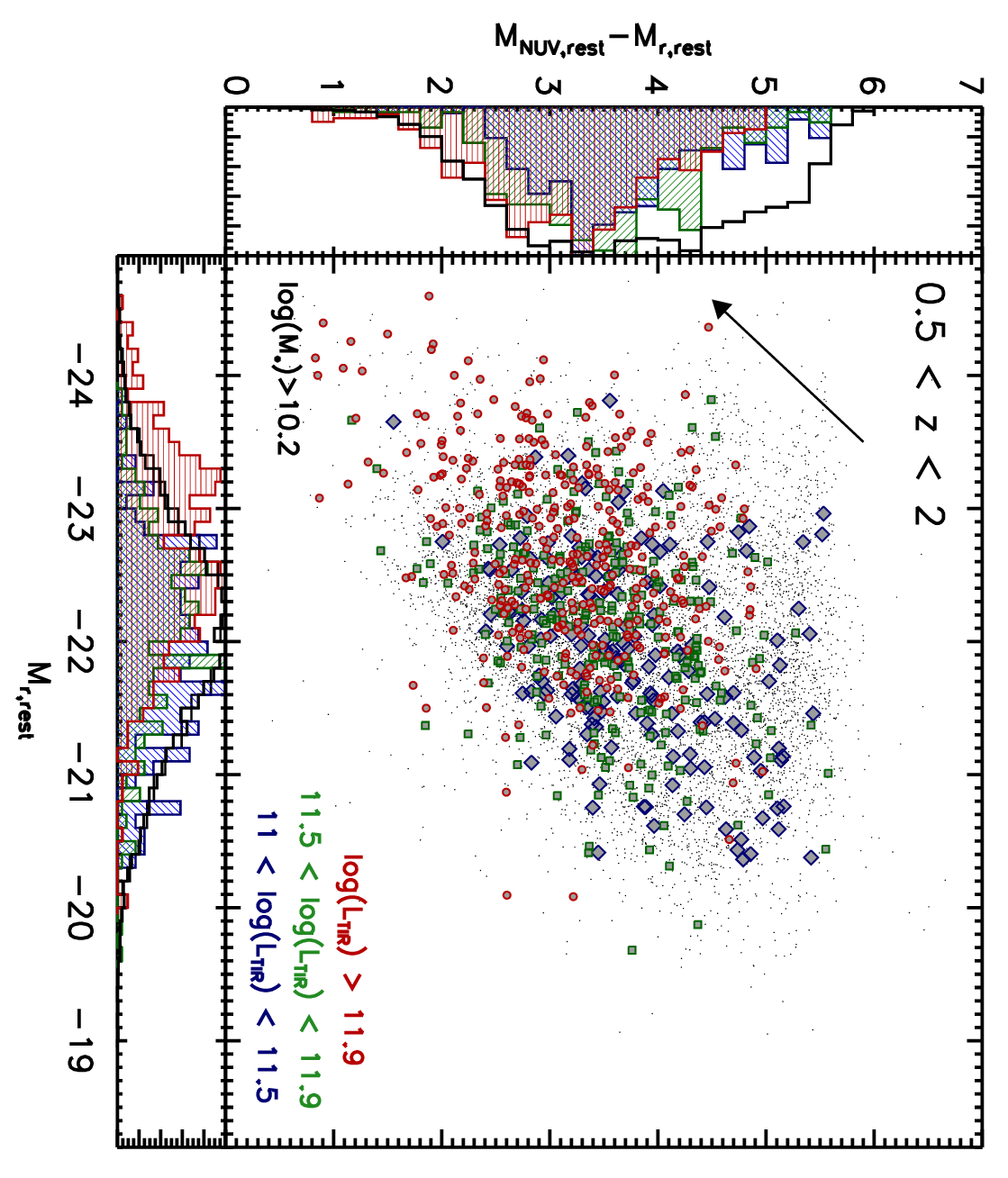}{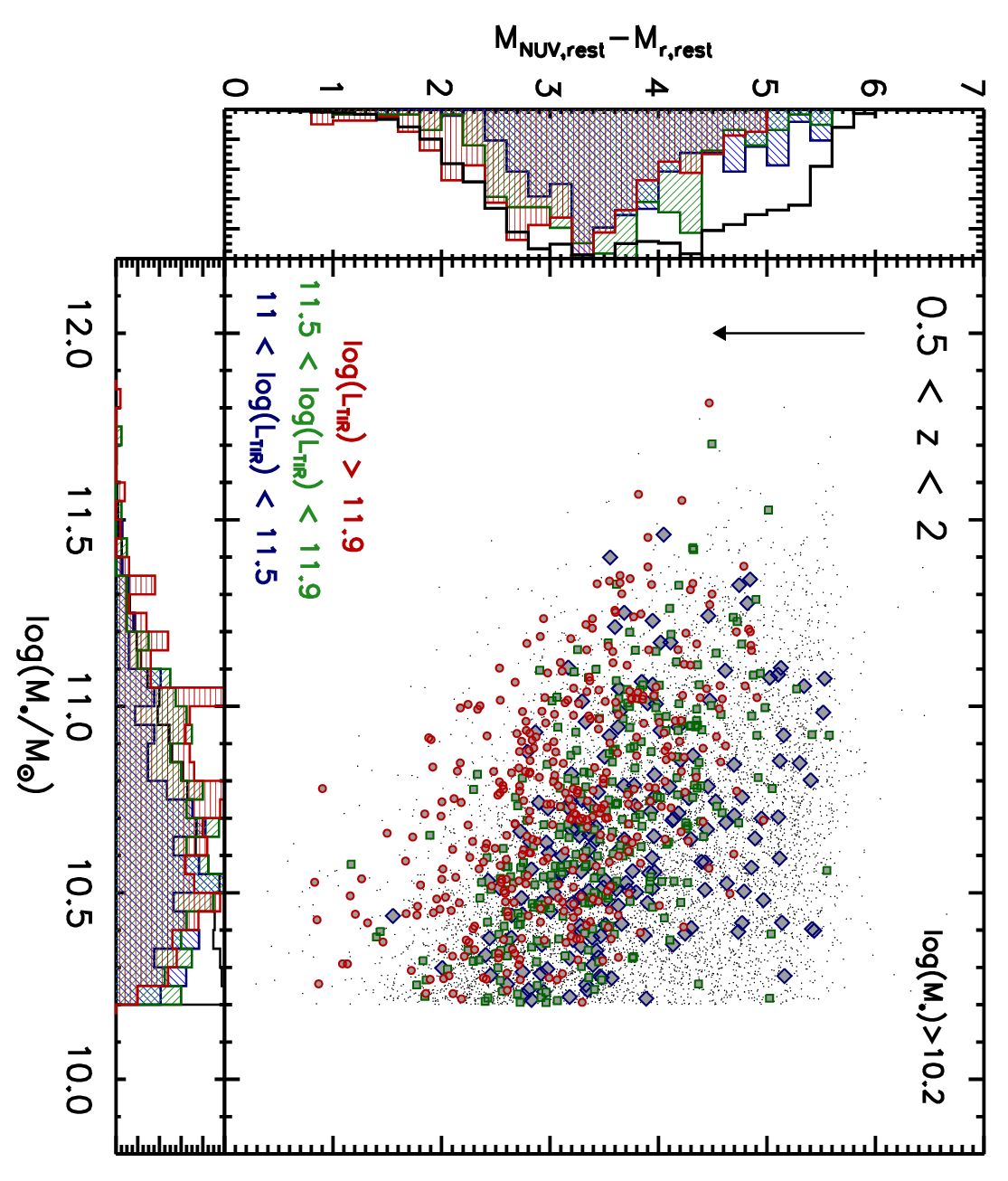}
\caption{Similar to Fig. \ref{fig:initialCMDnCSMD} except that only a single redshift range is plotted ($0.5<z<2$), a redshift range which comprises the bulk
of our sample. SPIRE-detected galaxies are broken into high-$L_{TIR}$ (red filled circles), moderate-$L_{TIR}$ (green filled squares), and
low-$L_{TIR}$ (blue filled diamonds) subsamples. Again in both the CMD (\emph{left}) and CSMD (\emph{right}), galaxies which went undetected or were not detected
at high enough significance in the SPIRE imaging are shown as small black points. Histograms are normalized in the same way as Fig. \ref{fig:initialCMDnCSMD} and
arrows again represent the movement in each phase space when an stellar continuum extinction of $E_{s}(B-V)=0.3$ is applied.
Only those galaxies with $K_s<24.0$ and stellar masses larger than the completeness limit ($\log(\mathcal{M}_{\ast})>10.2$) are plotted. While the LIRG subsamples show 
a high level of similarity in their $M_{NUV}-M_{r}$ colors and $M_{r}$ absolute magnitudes (\emph{left}), the ULIRG population appears both bluer and brighter optically. The range of stellar masses spanned by the 
ULIRGs at \emph{any given color} is generally larger than that of the LIRG samples (\emph{right}) suggesting that the ULIRG population is inhomogenous, comprised of nascent galaxies undergoing their initial
starbursts and older galaxies undergoing intense rejuvinated starburst events.}
\label{fig:singlepanelCMDnCSMD}
\end{figure*}

By comparing the corrected EW(H$\delta$) and $D_n(4000)$ values of the three $L_{TIR}$ subsamples to the closest model track in Fig. \ref{fig:sbplot} the first scenario is 
almost entirely ruled out, as the age of the starburst in the average galaxy of each of the three subsamples is nearly the same (between 50-140 Myr). Indeed, the populations
that showed the most marked difference in the other diagnostics, the high- and low-$L_{TIR}$ subsamples, appear here to be nearly identical in terms of the starburst age of 
their average galaxies. The major difference between the three subsamples appears to be, rather, the strength of the burst. While this is not unexpected given that the populations
were binned in terms of $L_{TIR}$, which is directly proportional to the $\mathcal{SFR}$, the strength of the burst probed here is not the absolute strength (which is probed by $L_{TIR}$), but
rather the strength of the burst relative to the \emph{already established stellar population in that galaxy}. The results in Fig. \ref{fig:sbplot} 
appear to directly confirm the second scenario proposed above. Both the average galaxy of the moderate-$L_{TIR}$ and low-$L_{TIR}$ are consistent with galaxies having already 
undergone significant star formation in the past, which are now undergoing a moderate (by mass) burst of star formation. Conversely, the average galaxy in the high-$L_{TIR}$
is inconsistent with either of these models. Instead, the measurements of the high-$L_{TIR}$ spectrum indicate a population of incipient galaxies undergoing their initial phase of
star formation. However, caution must be used when interpreting this population, as a secondary-burst model that doubles the mass (a 100\% burst in the language of Fig. 
\ref{fig:sbplot}) and occurs several Gyr after the initial star-formation event is essentially indistinguishable in this phase space. These two scenarios will be discussed
further in the next section. Combining the results of all the spectral analysis here and in the previous sections clearly indicates that, amongst the full SPIRE sample, galaxies 
with progressively lower $L_{TIR}$ have a progressively higher abundance of established older stellar populations relative to the strength of the starburst. In other words,
the LIRGs\footnote{The term LIRG(s) is used for the remainder of \S\ref{fullSPIRE} to refer to the combined low- and moderate-$L_{TIR}$ subsamples and ULIRG to refer 
to the high-$L_{TIR}$ subsample, see Table \ref{tab:lumsamples}} in the full SPIRE sample appear to be galaxies that have a high fractional abundance of older stellar 
populations born of previous star-formation events and are currently undergoing moderate rejuvenated star formation. The ULIRGs in the full SPIRE sample appear to be a 
decidedly different population, being largely comprised of galaxies that are either undergoing a primordial starburst or a massive secondary starburst event which is 
considerably increasing the amount of stellar mass in these galaxies. These results will have important consequences for the interpretation of SPIRE-detected galaxies 
hosting AGN discussed later in this paper.  

\subsubsection{Photometric and Stellar Mass Properties} 
\label{photnmassbyTIR}

\begin{figure}
\epsscale{1.8}
\plotonespecial{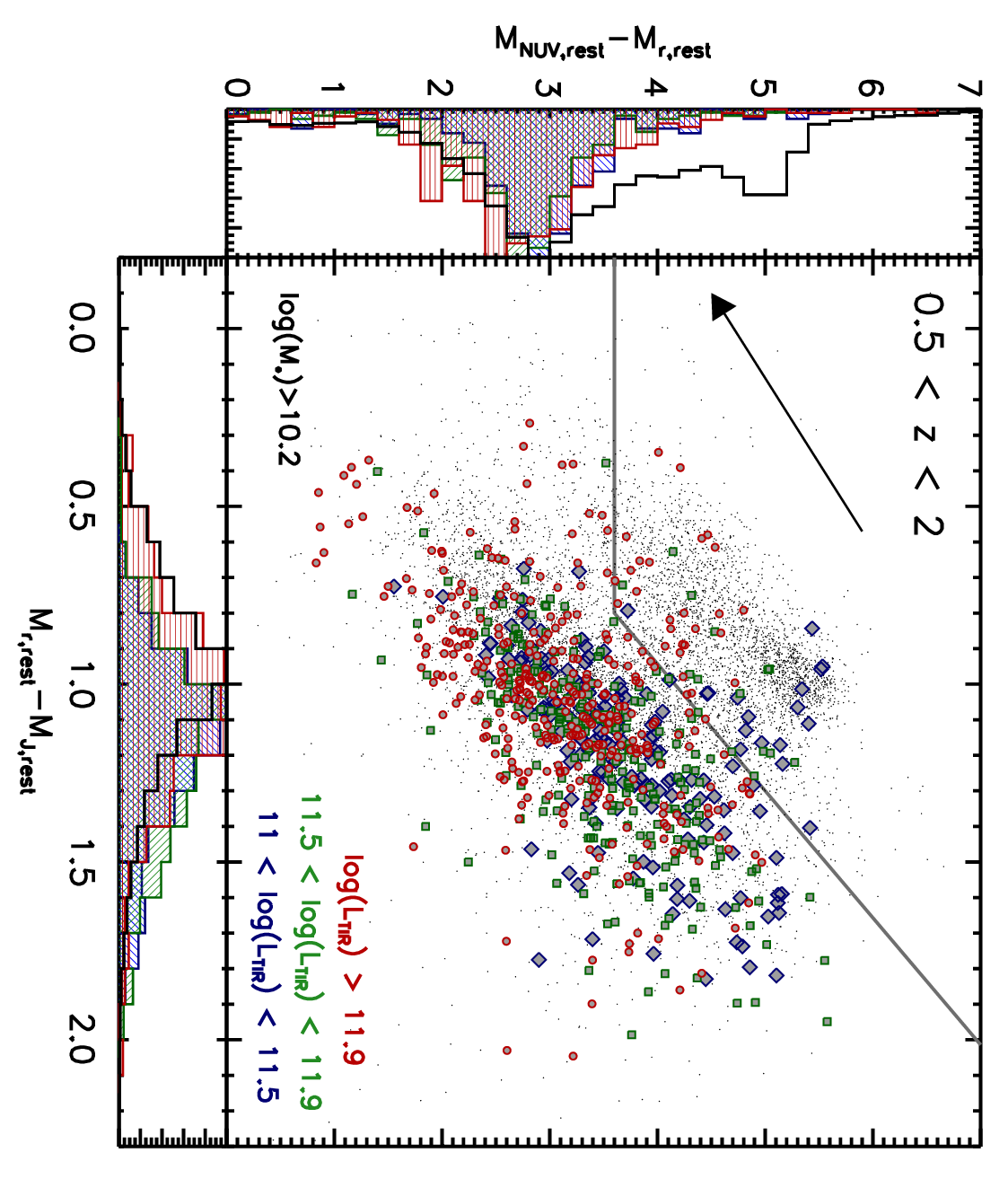}
\caption{Similar to Fig. \ref{fig:initialCCD} but plotting only those galaxies in the redshift range 0.5$<z<$2 broken into $L_{TIR}$ subsamples. The symbols are identical to 
those of Fig. \ref{fig:singlepanelCMDnCSMD}. Again only those galaxies brighter than $K_s<24.0$ and with stellar masses $\log(\mathcal{M}_{\ast})>10.2$ are plotted.
Histograms are normalized and defined in the same way as Fig. \ref{fig:initialCCD} (i.e., effective colors) and arrows again represent the effect of applying $E_{s}(B-V)=0.3$.
The fraction of galaxies in the quiescent region (above the line) of the diagram is roughly the same ($\sim$5-8\%) for all $L_{TIR}$ subsamples. However, the
ULIRG galaxies in the quiescent region can easily be moved to the lower left part of the diagram by applying the average extinction found in \S\ref{AtoK}. On
the contrary, the presence of LIRGs in the quiescent part of the diagram cannot be explained by extinction effects alone which hints that the optical colors of 
these galaxies is \emph{dominated} by light from older stellar populations.}
\label{fig:singlepanelCCD}
\end{figure}

Having established the nature of the LIRGs in the full SPIRE sample, we now turn from the spectral properties of the galaxies of different TIR luminosities to their 
color, magnitude, and stellar mass properties. This is done both to contextualize the results of the previous section and to investigate more fully the nature of the 
SPIRE-detected ULIRGs. As was mentioned in the beginning of \S\ref{spectralTIR}, neither the color, absolute magnitude, or stellar mass distributions of the galaxies with 
spectra in each $L_{TIR}$ subsample used in the previous section are appreciably different than their parent population such that the conclusions drawn from the spectral
analysis in the previous section can be directly combined with conclusions drawn from the full $L_{TIR}$ subsamples. With this in mind, plotted in
the left panel of Fig. \ref{fig:singlepanelCMDnCSMD} is the $NUV-r^{\prime}$ CMD of the full galaxy populations of each of the $L_{TIR}$ subsamples 
along with SPIRE-undetected galaxies in the redshift range $0.5<z<2$ that satisfy the criteria given in \S\ref{colormagnSM}. While the color histograms 
of all three $L_{TIR}$ subsamples peak at roughly the same color, the galaxies in the full 
LIRG subsample have significantly redder colors than the SPIRE-detected ULIRGs. This is confirmed by a KS test, which rejects the hypothesis that the galaxies 
of either of the LIRG subsamples are drawn from a parent galaxy population with the same color properties as the ULIRG subsample at $>>99.99$\% CL. 
As was established in the previous section, the redder colors of the LIRGs relative to that of the ULIRGs cannot be the result of differential extinction effects and is,
in fact, only enhanced if the rest-frame colors are corrected by the $\langle E_s(B-V) \rangle$ of each subsample. 
A comparison of the rest-frame $M_{r}$ magnitudes of the three subsamples finds an even more marked difference between the populations. SPIRE-detected ULIRGs appear significantly 
brighter in the rest-frame $r^{\prime}$ band than SPIRE-detected LIRGs, with ULIRGs representing some of the brightest galaxies \emph{at any color} in this redshift range.  
The magnitude distributions reflect this, as the magnitude distribution of the ULIRGs peaks $\sim$1.5 mags brighter than either of the LIRG 
populations, a difference that is also enhanced if the rest-frame magnitudes are corrected for the effects of extinction. A similar picture emerges from
the CCD plotted in Fig. \ref{fig:singlepanelCCD}. Of the galaxies safely in the region of the diagram populated by galaxies with quiescent colors (i.e., the area
where even large extinction effects could not move the galaxies out of the quiescent region), almost all are made up of low-luminosity LIRGs. The ULIRG galaxies that
are in the quiescent region are easily moved out when the $\langle E_s(B-V) \rangle$ derived in the previous section is applied to the observed colors (i.e., one arrow
length). These properties are 
broadly consistent with the results of the previous sections: the average stellar continuum of a LIRG in our sample has a larger fractional contribution from 
older stellar populations that that of a ULIRG. The blue colors and extremely bright rest-frame $M_{r}$ magnitudes of the ULIRGs show that this is a population
largely undergoing a violent burst of star-formation in which its optical colors are completely dominated by the newly formed young stellar populations.

However, the ambiguity related to the ULIRG population revealed in the previous section still remains. In star-forming galaxies, the optical luminosity 
(i.e., absolute magnitude) is a confusing quantity to interpret, as a galaxy with a well-established stellar population undergoing a moderate strength
starburst can appear equally luminous as a nascent galaxy undergoing a massive starburst event. While it is tempting to interpret those ULIRGs which 
are extremely blue, extremely (optically) luminous as being comprised of the latter population, it is necessary to examine the stellar masses of these 
galaxies in order to contextualize their luminosities. Plotted in the right panel of Fig. \ref{fig:singlepanelCMDnCSMD} is the CSMD of the three $L_{TIR}$
subsamples along with all SPIRE-undetected galaxies in the same redshift range. What is immediately noticeable is that the distributions of the three subsamples,
which were extremely distinct in absolute magnitude space, have now become essentially indistinguishable in stellar mass space. A KS test of the various
distributions confirms this, with only the low-$L_{TIR}$ and high-$L_{TIR}$ subsamples having even moderate levels of incompatibility (i.e., a rejection of
the null hypothesis at the $\sim3\sigma$ level). 

The most striking change between the two panels of Fig. \ref{fig:singlepanelCMDnCSMD} takes place for the extremely blue, 
extremely luminous ULIRGs. Nearly all of the ULIRGs which appeared at the highest luminosities in the CMD are found in 
galaxies with only moderate stellar masses. Even so, some of these ULIRGs have large stellar masses. In Fig. 
\ref{fig:singlepanelCMDnCSMD} there is a distinct envelope of ULIRGs, comprised entirely of the extremely bright 
and blue ULIRGs observed in the CMD, which essentially binds the blue cloud at the high stellar mass end. These galaxies 
appear at the highest stellar mass for a given color and are likely massive late-type galaxies 
undergoing strong star-formation events. The ULIRG phenomenon also appears prevalent all the way to the stellar mass 
limit of this redshift range, appearing prominently in galaxies of moderate stellar masses (i.e., $10.4\le \log(\mathcal{M}_{\ast}) 
\le10.8$). For the typical strength of the starburst in the ULIRG sample, $\sim350\mathcal{M}_{\odot}$ yr$^{-1}$, 
these are stellar masses that are built up in only 50-100 Myr, a time roughly equal to the typical length of the peak of 
star-formation in a starburst event (Swinbank et al.\ 2006; Hopkins et al.\ 2008, 2012; McQuinn et al.\ 2009; Wild et al.\ 2010) 
and roughly equal to the age since the beginning of the burst derived in the previous section. This scenario does not, 
however, work for even for the least massive LIRGs in our sample as a burst at the mean $\mathcal{SFR}$ of the LIRG sample 
($\sim50\mathcal{M}_{\odot}$ yr$^{-1}$) would have to be sustained at a constant level for the previous 
500 Myr$-$Gyr to build up the necessary stellar mass in the observed burst. Similarly, for galaxies at the high mass 
end of the ULIRG stellar mass function, the stellar masses are prohibitively high to allow the interpretation that a majority 
of their stellar mass was created in the extant starburst event. Even at the $\mathcal{SFR}$s of the typical ULIRG in our sample, these 
galaxies require sustained star formation at this level for 200-500 Myr to create the requisite stellar mass in a single event, 
a timescale that is unphysical for a starburst of constant strength. Because of the strength of the starburst, the
spectrum of such galaxies will be dominated by features of the young stars recently formed in the burst. This allows
higher-mass ULIRGs to be consistent with the results of \S\ref{stellarnsbages}, in which young luminosity-weighted stellar ages were 
observed for the average ULIRG. Indeed, we attempted to analyze massive and lower-mass ULIRGs separately using the spectral analysis
 presented in the previous section and found no significant differences between the two samples within the framework of the analysis 
of this paper. Thus, it appears the ULIRG population is comprised both of newly formed galaxies undergoing their first major 
star-formation event and massive galaxies undergoing star-formation events that are adding significantly to an already established 
stellar population.

\section{The Role and Prevalence of Active Galactic Nuclei in Starbursting Galaxies}
\label{AGNfull}

In this section we no longer consider the SPIRE sample as a whole, but rather divide the 1753 galaxies in our full SPIRE 
sample into two distinct subsamples: the $\sim$10\% of SPIRE-detected galaxies that host an active galactic nuclei and 
the $\sim$90\% of SPIRE-detected galaxies that, within the limitations of our data, have galactic nuclei which appear 
dormant. This section, unlike the previous section, will almost entirely focus on the multiwavelength imaging properties
of the two subsamples. The spectra of the AGN hosts, as well as those galaxies not hosting powerful AGN, are used here only to constrain 
redshifts and, where appropriate, the results of the spectral analyses of \S\ref{fullSPIRE} will be invoked. A study comparing 
the spectral properties of radio AGN hosts detected by SPIRE, a population which comprises the bulk of our SPIRE-detected AGN host 
sample, and AGN hosts undetected by SPIRE will be presented in a companion paper (Lemaux et al.\ 2015). 
Many of the numbers and results derived in this section are a function of our data quality and imaging depth, 
both of which vary immensely over the datasets used in this paper. The global trends that will be presented in this 
section are primarily internal comparisons and, thus, are largely independent of the details of the data.
However, caution must be exercised when comparing the raw numbers presented in this section to other observational datasets,
though we attempt as much as possible to make comparisons between datasets free of bias. With this warning, a warning to which 
we will return frequently throughout this section, we continue on to investigate the nature of coeval starbursts and AGN.

\subsection{The Prevalence of Active Galactic Nuclei in Starbursting Galaxies}
\label{SBfrac}

In total 230 IR, X-ray, spectral Type-1, and radio AGN were selected from our sample using the methods described in \S\ref{AGNselection} and subject to the criteria
given in \S\ref{TIRnSFRs}. Accounting for AGN detected using multiple diagnostics, AGN are found in 181 unique galaxies in the full SPIRE sample.
This corresponds an AGN rate of 10.3$\pm$0.8\% in the full SPIRE sample, a value broadly similar to preliminary studies investigating the prevalence of AGN 
in sub-mm-selected samples at similar redshifts (e.g., Pope et al.\ 2008; Serjeant et al.\ 2010) and one which only slightly exceeds the detection rates of 
AGN in surveys of optically-selected galaxies at a variety of different redshifts (e.g., Kauffmann et al.\ 2003; Martini et al.\ 2009, 2013; Montero-Dorta et al.\ 2009; 
Sarajedini et al.\ 2011; Klesman et al.\ 2012). While this comparison suggests that the presence of a starburst in a galaxy does not increase the chances that its 
nucleus will be undergoing a powerful active phase, these comparisons are subject to a slew of observational and selection biases. Of the former, one of the most 
crucial is that our data only allow us to observe this relationship (or lack thereof) for the hosts of powerful AGN, and, indeed, different trends have been observed 
for less powerful type-2 AGN (see, e.g., Symeonidis et al.\ 2013). Regardless, when viewed in this manner the phenomena of dusty starbursts and powerful AGN (see \S\ref{AGNselection}
for the meaning of powerful in this context) do not appear at first glance to be especially linked, as a large majority of dusty starbursts selected by SPIRE 
do not host a powerful AGN in contrast to initial investigations of the link between the two phenomena (e.g., Sanders et al.\ 1988).

\begin{table}
\caption{Fraction of galaxies and AGN hosts detected in SPIRE\label{tab:AGNfractions}}
\centering
\begin{tabular}{lcc}
\hline \hline
Sample & $N$ & $f_{SB}$ \\[0.5pt]
\hline
All Optical\tablefootmark{a} & $\langle32\rangle$ & $\langle5.8\rangle \pm$1.2\% \\[4pt]
Radio-Starburst & 146 & 58.2$\pm$4.8\% \\[4pt]
4-Channel IRAC & 498 & 45.0$\pm$2.0\% \\[4pt]
IR AGN & 62 & 36.1$\pm$4.6\% \\[4pt]
VVDS Type-1 AGN & 10 & 13.0$\pm$4.1\% \\[4pt]
X-ray AGN & 26 & 20.0$\pm$3.9\% \\[4pt]
Radio AGN & 132 & 41.8$\pm$3.6\% \\[4pt]
All AGN\tablefootmark{b} & 181 & 32.3$\pm$2.4\% \\[0.5pt]
\hline
\end{tabular}
\tablefoot{
\tablefoottext{a}{The numbers here reflect the mean number and rate of SPIRE detections for hosts of (largely) dormant nuclei that were matched in sample size, $\mathcal{M}_{\ast}$, and $z$ to the ``All AGN" sample, see Appendix B} 
\tablefoottext{b}{This subsample is comprised of unique galaxies in which an AGN is hosted, i.e., AGNs detected using multiple diagnostics are counted only once}
}
\end{table}

\begin{figure}
\epsscale{1.2}
\plotone{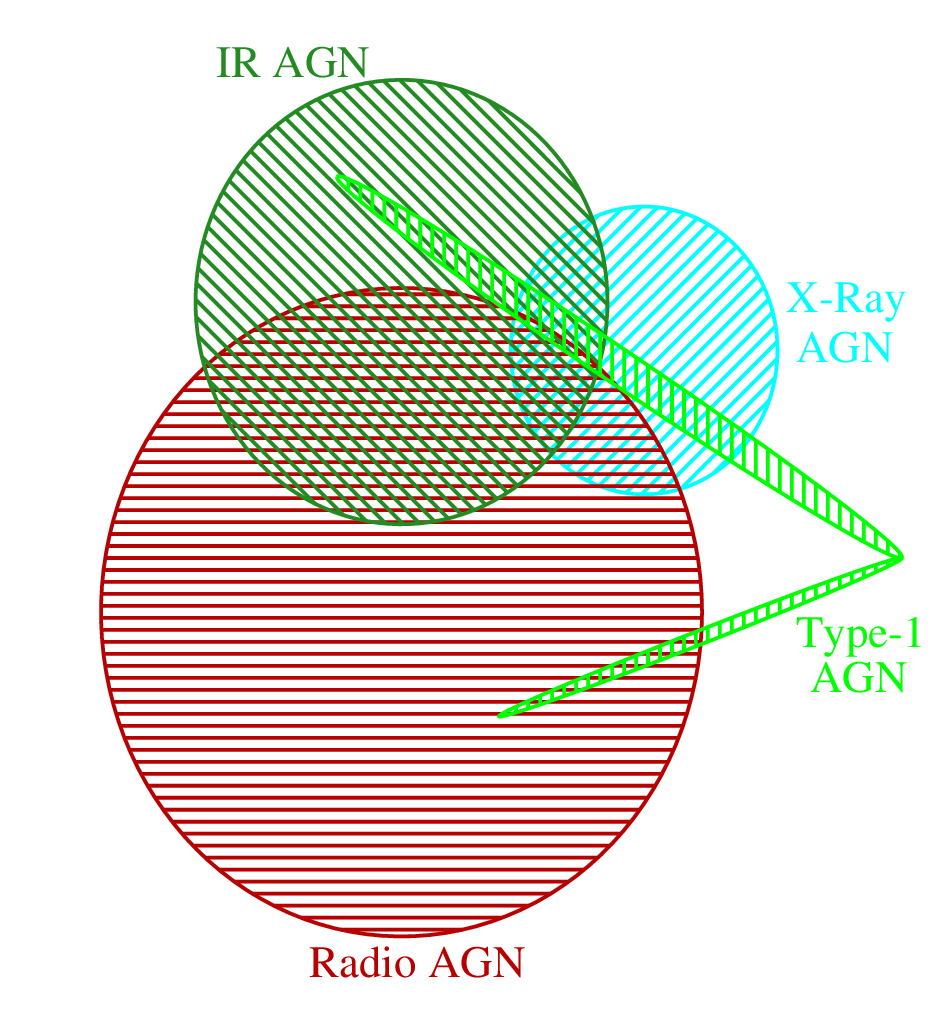}
\caption{Venn diagram of the 181 unique galaxies significantly detected in SPIRE which are host to a powerful AGN selected
by the various methodologies presented in \S\ref{AGNselection}. While some galaxies are selected as hosting a powerful AGN in with multiple methods, only one galaxy is selected
by all four methods, and a large fraction of the AGN selected are selected in only one of the methodologies, underscoring the importance of selecting AGN at multiple wavelengths 
using multiple selection techniques. Less powerful AGN are (generally) not separable or, in some cases, detectable in our dataset and are thus not represented in this diagram.}
\label{fig:spunkyvenn}
\end{figure}

However, a different approach can be taken. In the previous view, the question of whether the presence of a starburst in a galaxy increases the chance that an AGN
is present was addressed. This is perhaps not a fundamental probe of the link between the phenomena. There is evidence to suggest that an AGN operates on a different timescales 
than a starburst (e.g., Hopkins et al.\ 2008) and that AGN also lag appreciably behind the peak of the star-formation activity (e.g., Wild et al.\ 2010; Hopkins 2012). Under these premises, 
the presence of an AGN in any particular starburst depends completely on the time at which the starburst is observed. If observed near inception, the lack of an AGN in a 
starburst galaxy is perhaps not surprising. Similarly, if the starburst is viewed 100-200 Myr after the onset of the burst, the lag time between the onset of significant AGN 
activity and the initial burst is spanned, and the presence of an AGN, if one is to form, is expected on this timescale. To eliminate this temporal selection effect 
we instead selected galaxies whose nuclei are undergoing an active phase and address the question of whether or not the presence of an AGN significantly increases the 
occurrence of a starburst relative to galaxies with dormant hosts. 


Listed in Table \ref{tab:AGNfractions} are the fractions of each of the host galaxies of the full AGN samples detected with SPIRE at a 
$\ge3\sigma$ level in at least one of the three SPIRE bands (hereafter $f_{SB}$). Also given in Table \ref{tab:AGNfractions} 
is this fraction for a stellar-mass/redshift-matched samples of optical galaxies that are generally not host to 
a powerful AGN (i.e., all AGN selected in \S\ref{AGNselection}). For more details on the generation of the matched optical samples see 
Appendix B. To the depth of our SPIRE imaging, which is limited at $\sim25 \mathcal{M}_{\odot}$ yr$^{-1}$ for the bulk of our sample, 
it appears that LIRG-level starbursts are rare amongst galaxies with dormant galactic nuclei. In the 500 realizations of a matched
optical sample with dormant AGN hosts, an average of only 32 of the 561 galaxies in the sample ($f_{SB}$=$5.8\pm1.2$\%) were detected 
at $\ge3\sigma$ significance in at least one of the three SPIRE bands. For galaxies hosting an AGN of any type, this fraction was 
significantly higher. Combining the samples of AGN selected by the variety of methods presented in \S\ref{AGNselection} and 
counting those AGN detected in multiple diagnostics only once (see Fig. \ref{fig:spunkyvenn}), the fraction of unique galaxies hosting an 
AGN that also host a SPIRE-detected starburst $f_{SB}=32.3\pm$2.4\%. This fraction of AGN hosts observed by starburst galaxies is 
similar to those fractions calculated over a similar redshift range in a PACS survey of X-ray selected AGN using extremely deep 
X-ray data in the two Chandra Deep Fields (Rosario et al.\ 2013), though these numbers can only be compared in an extremely broad 
sense given the different AGN selection methods and imaging depths of the two studies. Still, the two studies indicate a strong 
connection between the two phenonmena, observing similarly drastic differences in starburst fractions amongst AGN hosts and 
matched samples of galaxies.

It is, however, possible that this dramatic difference in starburst rate observed in our own sample is
due to observational bias rather than astrophysical reasons. In Appendix B we address this issue extensively and show that 
observational bias cannot account for the differences in $f_{SB}$ of galaxies with dormant and active galactic nuclei. Thus, it 
appears that \emph{the presence of a powerful active galactic nuclei in a galaxy dramatically increases the 
likelihood that the galaxy will be undergoing a starburst}. In the standard paradigm, where an AGN is responsible for quenching the star formation of its host galaxy, 
this is a counter-intuitive result, as the presence of an AGN, coupled with the relatively short timescale of the starburst, should generally preclude the possibility of the 
two phenomena being observed coevally for large numbers of galaxies. We will return to this point later in the paper as the nature of SPIRE-detected AGN hosts becomes 
clearer.

While the largest discrepancy is observed between the starburst fraction of galaxies with powerful AGN and those without, there is significant variance between the starburst fraction
of galaxies hosting different types of AGN. Hosts of X-ray and unobscured type-1 AGN have the smallest starburst frequency of all the AGN types, with $f_{SB}$ of 
20.0$\pm3.9$\% and 13.0$\pm4.1$\%, respectively. These fractions are broadly consistent with similar fractions found by other studies (e.g., Hatziminaoglou et al.\ 2010; Page et al.\ 2012). The 
host galaxies of IR-AGN, while exhibiting a significantly higher fraction of starbursts than either the X-ray or type-1 AGN hosts, are a peculiar case. In order to detect an 
IR-AGN, we require that a galaxy be detected in all four \emph{Spitzer}/IRAC bands. Because of the SPIRE matching process (see \S\ref{herschelmatch}), this results in such galaxies 
being more likely to be detected in SPIRE by virtue of their increased $24\mu m$ detections and bright $3.6\mu m$ magnitudes. In fact, the peak of the distribution of the 
$3.6\mu m$ and $4.5\mu m$ magnitudes are brighter than that of the full SPIRE sample by 1.1 and 1.5 magnitudes, respectively, the latter especially suggesting widespread 
obscured AGN activity (e.g., Hickox et al.\ 2009). Because of the increased likelihood of a SPIRE detection for this population, it is necessary to compare the starburst 
fraction observed in the IR-AGN hosts with a control sample of galaxies detected in all four IRAC bands. The fraction of the control sample exhibiting starbursts is, in fact, 
higher than that of the IR-AGN hosts, suggesting that, for the IR-AGN hosts, observational effects are biasing the starburst fraction higher relative to galaxies hosting other 
flavors of AGN. 

Perhaps the most surprising result to come from this analysis is the starburst fraction of galaxies hosting a radio AGN. Numerous studies have shown that 
galaxies hosting radio AGN are predominantly massive galaxies, containing older stellar populations and low or non-existent levels of star formation (e.g., Hickox et al.\ 
2009; Kauffmann et al. 2009 and references therein). Indeed, radio-mode AGN feedback is commonly invoked in both observational studies and in semi-analytic models as an efficient mechanism to 
inject energy into the hot gas halos observed surrounding massive ellipticals, thereby preventing further star formation from occurring (e.g., Croton et al.\ 2006, 
Smol$\rm{\check{c}}$i$\rm{\acute{c}}$ et al.\ 2009a and references therein). In this context, the fact that the hosts of radio AGNs are galaxies with the highest incidence of starbursts, with
$f_{SB}=41.8\pm3.6$, is extremely puzzling. Though we will continue to investigate this population as part of the full AGN sample throughout the remainder of this paper, 
a detailed investigation of those radio AGN hosts both detected and undetected in SPIRE is given in an accompanying paper (Lemaux et al.\ 2015). 

\subsection{Starburst Strengths of Active Galactic Nuclei Hosts} 
\label{sbstrength}

In this section we move from considering a binary quantity, namely whether an SPIRE-detected starburst is or is not present in a given galaxy hosting an AGN (or vice versa), 
to contrasting the properties of those starbursts occurring in AGN hosts with the full SPIRE sample. Because the main metric used for comparison in this section 
is the strength of the starburst as proxied by $L_{TIR}$, we move for the remainder of the section back to a SPIRE volume limited sample for each redshift bin considered 
(as was done in \S\ref{sfmainseq}) to remove the effects of Malmquist bias. 

\begin{figure*}
\epsscale{0.7}
\plotonesuperspecial{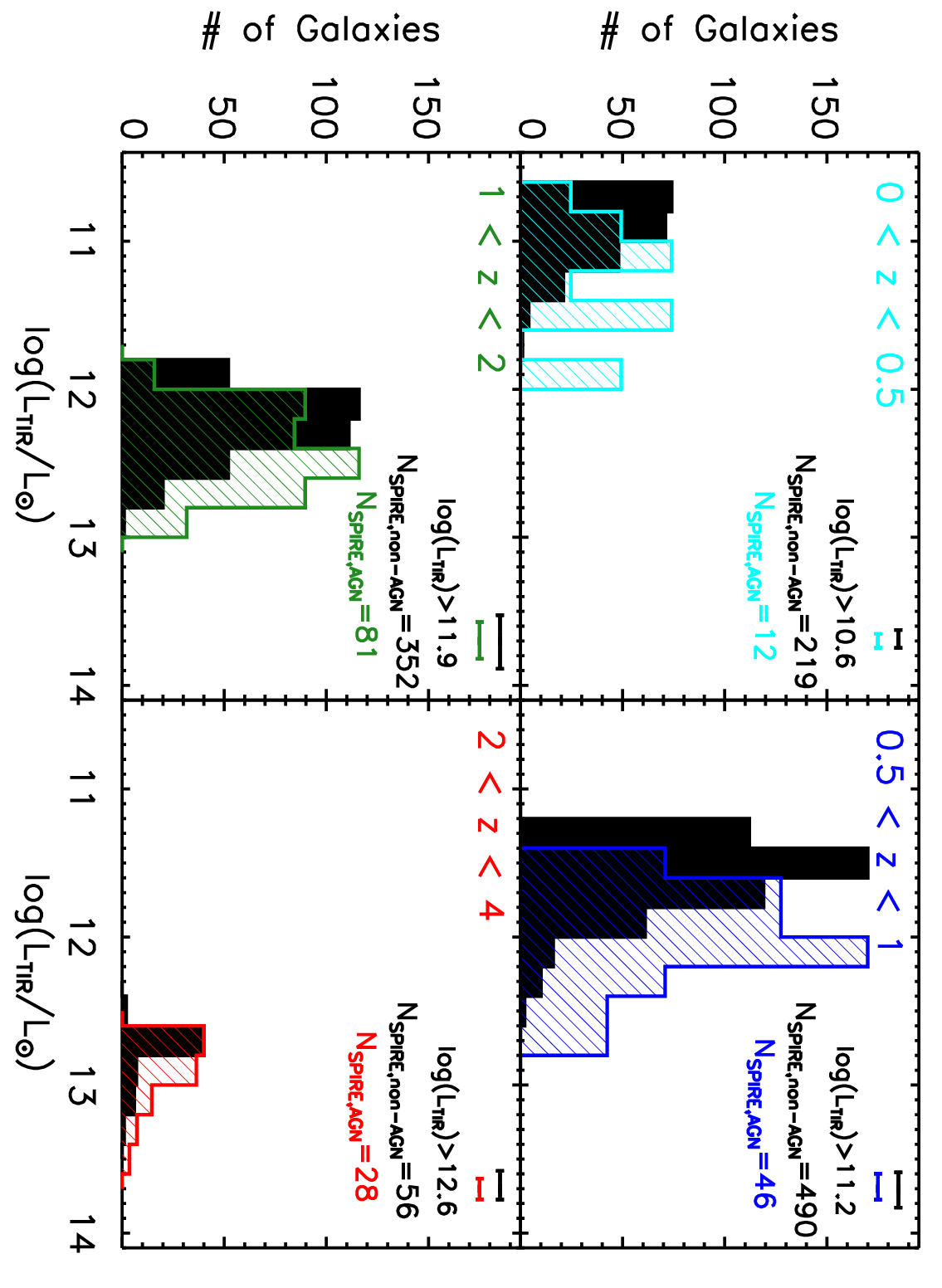}
\caption{Volume-limited $L_{TIR}$ histograms of all galaxies detected at a significance of $>3\sigma$ in at least one of the three
SPIRE bands which are not host to a powerful AGN (black filled histograms) plotted against $L_{TIR}$ histograms of those SPIRE 
galaxies hosting a powerful AGN (dashed color histograms) for four different redshift bins. The histograms are normalized in such
a way that the two histograms in each panel have the same maximum value. The $L_{TIR}$ limit used to construct the volume-limited 
sample for each redshift bin is shown in each panel along with average errors on $L_{TIR}$ for each subsample.
The hosts of powerful AGN are shifted to higher $L_{TIR}$ values than their counterparts whose nuclei 
are not undergoing a powerful active phase. The difference in the $L_{TIR}$ distributions of the two samples is statistically 
significant at $>3\sigma$ for all redshift bins, implying a relationship between the presence of a powerful AGN and a stronger 
than average starburst event.}
\label{fig:TIRAGNvscontrol}
\end{figure*}

In \S\ref{sfmainseq} it was mentioned that one of the most striking and perhaps surprising results of the analysis of the relationship between $\mathcal{M}_{\ast}$ 
and $\mathcal{SFR}(L_{TIR})$ was the large number of AGN hosts observed at extremely high $\mathcal{SFR}$s. The fact that, in the left panel of Fig. \ref{fig:SFRnSSFRvsmass}, it was 
observed that nearly all galaxies undergoing the most vigorous starbursts at a given redshift are host to an AGN strongly suggested that there is a relationship 
between the strength of a starburst in a galaxy and the presence of an AGN. This relationship becomes more apparent in Fig. \ref{fig:TIRAGNvscontrol}, 
where we plot a normalized $L_{TIR}$ histogram of AGN host galaxies in the volume-limited SPIRE sample the backdrop of the full volume-limited SPIRE sample for four redshift bins. 
While we did not explicitly impose this condition (like was done in \S\ref{SBfrac}), the average SPIRE-detected AGN host in each redshift bin has a stellar mass 
which does not differ significantly from the comparison sample (i.e., the average stellar mass difference between the two samples is $\sim0.1$ dex, roughly comparable to
the median stellar mass errors drawn from the SED-fitting process). Thus, the SPIRE-detected AGN host sample can be thought of as, again, matched in redshift and stellar mass to the 
galaxy sample it is being compared to, thereby eliminating any additional complications on $L_{TIR}$, or equivalently $\mathcal{SFR}$, from those two quantities. This will remain true throught 
the entirety of \S\ref{AGNfull}. The hosts of AGN are clearly shifted to higher values of $L_{TIR}$ at all redshifts, a trend confirmed at $>3\sigma$ in all redshift bins by a KS test 
and at $>>3\sigma$ for the two bins with the largest number of AGN (i.e., $0.5<z<2$). Considering galaxies at redshifts $z>0.5$, a redshift range where a majority of our AGN host 
galaxies lie, an overwhelming fraction of AGN host galaxies belong to the high-$L_{TIR}$ sample defined in \S\ref{generalTIR}, suggesting that \emph{the average AGN host galaxy 
detected by SPIRE is experiencing ULIRG-level star-formation activity}. This statement holds if different methodologies are chosen for selecting AGN hosts (in particular 
radio AGN hosts, see \S\ref{AGNselection}). It is also indifferent to our choice of sample selection and remains significant if a (SPIRE) magnitude-limited 
or stellar mass-limited sample is instead considered. In the reverse view, we calculated the global fraction of \emph{all types} of AGN selected in our sample for ULIRGs in this redshift range,
finding that $18.4\pm1.5$\% of all ULIRGs between the redshift range of $0.5<z<2$ host some type of powerful AGN. While this fraction is almost double the $\sim10$\% global 
AGN fraction of the full SPIRE sample, it is considerably less than the 50\% of similarly bright ULIRGs found to host AGN of all types at all luminosities in 
a recent study at similar redshifts ($0.3<z<1$) by Juneau et al.\ (2013). The difference between these numbers strongly suggest that lower power AGN 
selected by a variety of multiwavelength methods play a significant role in ULIRGs and their evolution.

\begin{figure}
\plotonespecial{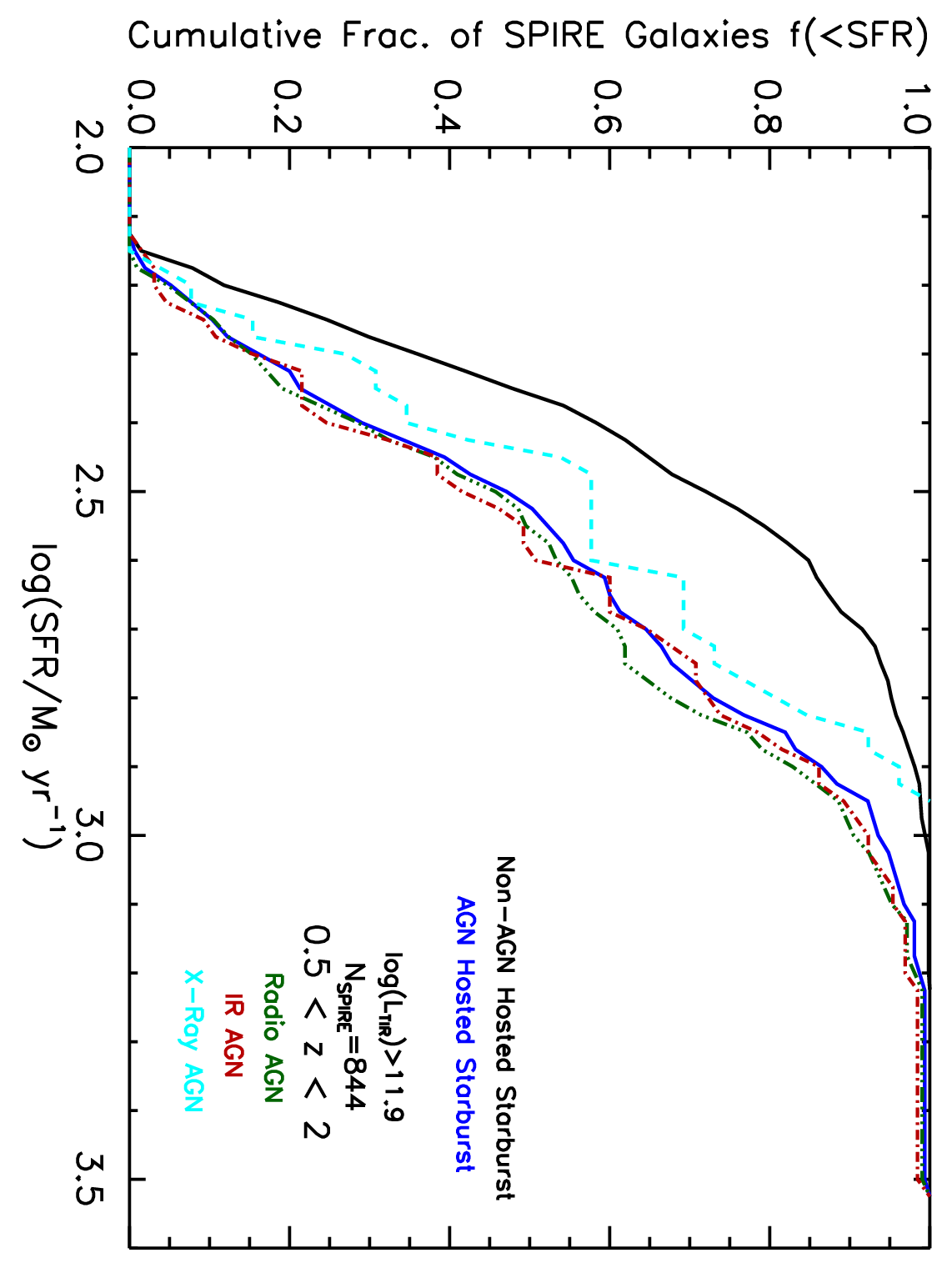}
\caption{Cumulative $L_{TIR}$-derived $\mathcal{SFR}$ fraction for a volume-limited SPIRE sample between $0.5<z<2$, the redshift range containing
the bulk of our SPIRE sample and the bulk of the galaxies in our sample which host powerful AGN. Galaxies are divided in those hosting 
a powerful AGN (solid blue line) and those not (solid black line). Only those galaxies detected in at least one of the three SPIRE 
bands at $>3\sigma$ are plotted. The cumulative $\mathcal{SFR}$ fraction of galaxies hosting various flavors of powerful AGN are represented by 
dashed (X-ray AGN), dot-dashed (IR AGN), and dot-dot-dot-dashed lines (Radio AGN). The sample of VVDS Type-1 AGN are not shown 
individually due to their small number, but they are included in the full AGN sample. The cumulative $\mathcal{SFR}$ fraction of hosts of AGN types 
of all flavors as well as the combined AGN host sample differs appreciably and significantly from galaxies that do not host a powerful 
AGN. Galaxies hosting a powerful AGN are both more likely to be undergoing a starburst event than galaxies not hosting powerful AGN 
and, furthermore, galaxies hosting powerful AGN are more likely to be undergoing more vigorous starbursting events than the average 
starburst galaxy at these redshifts.}
\label{fig:cumfrac}
\end{figure}

The relationship between AGN hosts and the strength of the starburst
is further demonstrated in Fig. \ref{fig:cumfrac}, in which we plot the cumulative fraction of galaxies at a given $\mathcal{SFR}$ for the host galaxies of the 
three main AGN types and a combined AGN sample and contrast it to the full SPIRE sample in the redshift range $0.5<z<2$. This redshift range was
chosen to encompass both the bulk of our full SPIRE sample and the bulk of the full SPIRE sample hosting a powerful AGN. Though the $L_{TIR}$ limit
imposed on this sample necessary to ensure a volume-limited sample ($\log(L_{TIR}$)$>$11.9) is quite high (i.e., ULIRG-level), the results we present
here are unchanged if we instead choose a sample at lower redshifts which allows for a less-restrictive $L_{TIR}$ cut. The cumulative fraction distribution 
of the hosts of the full AGN sample are again clearly biased towards much higher $\mathcal{SFR}$s than the overall SPIRE sample, with a KS rejecting that the two 
distributions are similar at a significance of $>3\sigma$. Since only galaxies that comprise the high-$L_{TIR}$ subsample are plotted in Fig. 
\ref{fig:cumfrac}, all galaxies are necessarily forming stars at high rates. However, at nearly all $\mathcal{SFR}$s less than that of the mean $\mathcal{SFR}$ of the high-$L_{TIR}$ 
sample (i.e., $\log(\mathcal{SFR})<2.5$), the cumulative fraction of galaxies with dormant nuclei is essentially double that of galaxies hosting powerful AGN. 
In other words, most ($\sim$75\%) ULIRG galaxies are forming stars at a rate below the mean $\mathcal{SFR}$ of the high-$L_{TIR}$ sample. For galaxies hosting 
powerful AGN, this fraction drops precipitously, as only $\sim45$\% of ULIRGs hosting powerful AGN are forming stars below the same limit.
Combining these results with the results of the previous section, it becomes apparent that \emph{not only are starbursts more common in those galaxies 
that host all types of AGN, but the starbursts occurring in those galaxies are considerably more powerful than the average starburst at all redshifts}.
Such a trend is also seen in AGN hosts selected by a variety of other methods (e.g., Santini et al.\ 2012; Gruppioni et al.\ 2013), which suggests
this phenomenon is common among nearly all types of AGN. 

This, again, is an extremely puzzling result if one considers the standard picture of interplay between an AGN and a coeval starburst in its 
host galaxy. In this picture a feedback process associated with the AGN, primarily thought to be AGN-driven outflows which inject large amounts of 
kinetic energy into the ISM and which can rival the bolometric luminosity of the AGN itself, efficiently suppress the starburst in the galaxy over short timescales (e.g., 
Springel et al.\ 2005; Somerville et al.\ 2008; Harrison et al.\ 2012a). This picture is thought to be especially pertinent for exactly the type of powerful AGN 
considered in this paper. The resulting population predicted by this picture are galaxies with signatures of post-starburst or post-quenching features, either 
in their optical spectroscopy, colors, or morphologies, and residual moderate-level AGN activity. The latter is thought to occur due to the aforementioned 
lag time of the AGN activity relative to the onset of the starburst coupled with the relatively long lifetime of such AGN (i.e., several 100s of Myr; e.g., 
Marconi et al.\ 2004). Indeed, such populations have been observed across a wide range of environments and redshifts (e.g., Kauffmann et al.\ 2003; 
Yan et al.\ 2006; Georgakakis et al.\ 2008; Kocevski et al.\ 2009b; Schawinski et al.\ 2009; Rumbaugh et al.\ 2012), suggesting that this picture is,
at least in part, correct. When considering whether our results are in direct conflict with this picture, or, alternatively, how to reconcile the two pieces of
observational evidence, caution must be exercised. As discussed extensively in the introduction, any inferred causal relationships between an AGN and its host
galaxy based on observational data is highly subject to the timescales of both phenomena. In the remaining section we will investigate what causal relationships
can be inferred from our data and discuss the implications and limitations of those inferences.

\subsubsection{The Role of Active Galactic Nuclei in Starbursting Galaxies}
\label{agnrole}

Because of the large number of seemingly disconnected results presented in this paper, before going further in this investigation we pause to review the 
pertinent results of this study that relate to the connection between AGN and their parent galaxies in the full SPIRE sample. 

\begin{figure*}
\plottwospecial{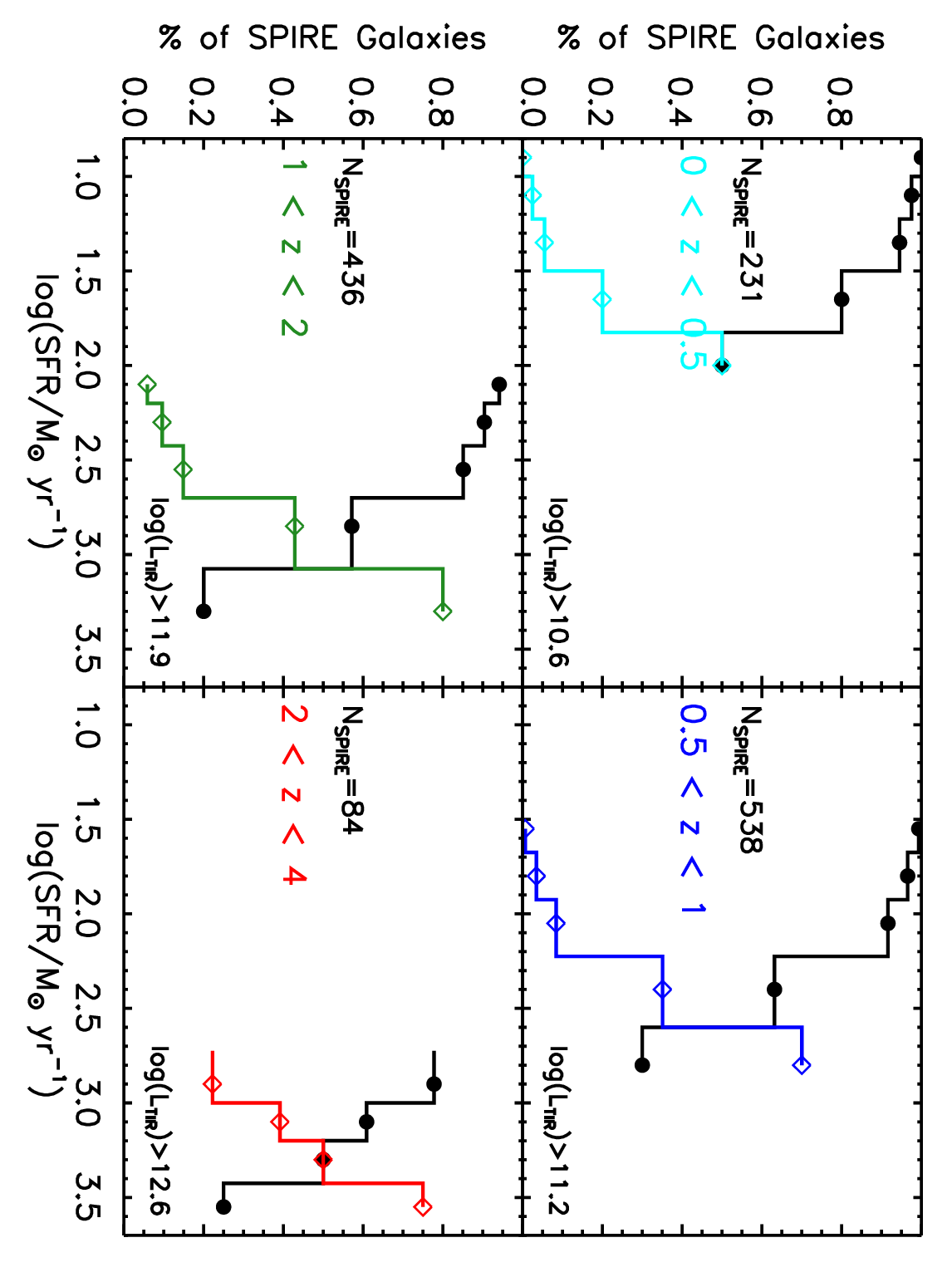}{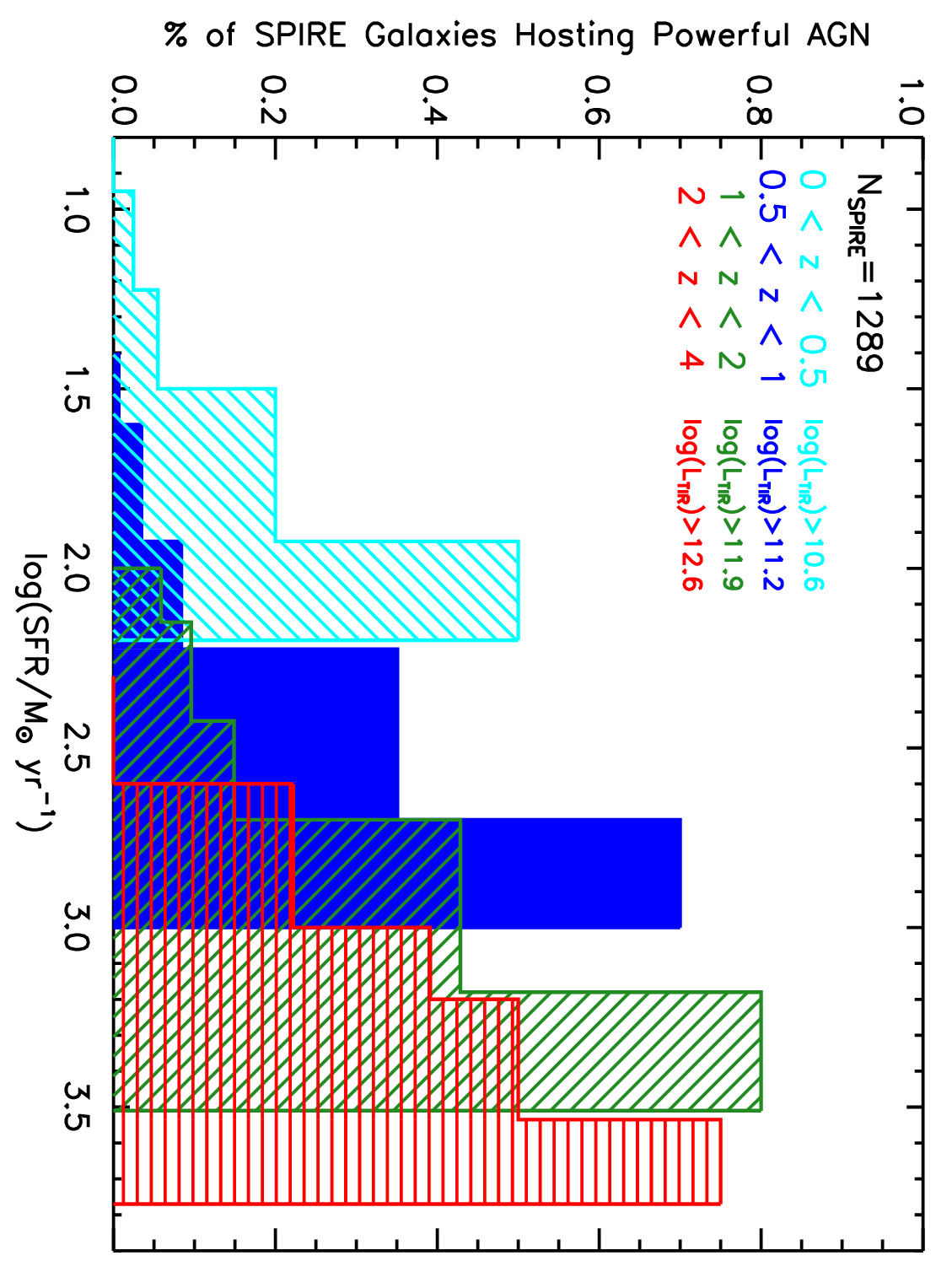}
\caption{\emph{Left:} Fraction of galaxies hosting a powerful AGN as a function of $L_{TIR}$-derived $\mathcal{SFR}$ in volume-limited samples of SPIRE 
galaxies in four different redshift bins. Only those galaxies detected at $>3\sigma$ in at least one of the three SPIRE bands 
are plotted. Solid colored lines show the fraction of galaxies at any given $\mathcal{SFR}$ ($L_{TIR}$) bin that host a powerful AGN, solid
black lines show the fraction of SPIRE galaxies with that $\mathcal{SFR}$ which do not host a powerful AGN. For all redshifts, the fraction
of powerful AGN increases monotonically with increasing $\mathcal{SFR}$. Galaxies at the highest $\mathcal{SFR}$ must quench rapidly in order 
to avoid violating the observed evolution in the stellar mass function (see \S\ref{agnrole}). The presence of an 
extremely powerful AGN in the majority of galaxies undergoing extreme starbursts provides a natural and plausible candidate 
for quenching. \emph{Right:} Similar to the left panels but employing SPIRE volume-limited samples at all redshifts.
Starting from high redshift, the most vigorous star-forming galaxies at any given redshift are unobserved in samples at subsequently 
lower redshifts. Such galaxies are also the galaxies with the highest fraction of AGN in any given redshift bin. The fraction of 
powerful AGN hosted by moderately powerful ULIRGs (i.e., $\log(\mathcal{SFR}_{TIR})\sim2.8$) increases by a factor of $\sim3$ from the highest to lowest redshifts, 
a phenomenon which is used to argue for the presence of downsizing amongst dusty starburst galaxies driven, at least in part, by powerful AGN activity.} 
\label{fig:fracSFRAGNvsnon}
\end{figure*}

\begin{itemize}
\renewcommand{\labelitemi}{--}

\item The AGN selected in this sample are of the powerful variety. The X-ray AGN have a luminosity limit of $L_{X,\;2-10 keV}\ga
10^{43}-10^{44}$ ergs s$^{-1}$ (or, equivalently, $L_{X,\;bol}\gsim10^{44.5}$ ergs s$^{-1}$) for the bulk of the redshifts considered here. The 
IR-selected AGN have significantly brighter $4.5\mu m$ magnitudes than the bulk of full SPIRE sample, suggesting significant enshrouded 
AGN activity. The radio AGN were selected such that only powerful ($\log(P_{\nu,1.4GHz})>23.8$) radio are included in our sample. 
The type-1 AGN selected spectroscopically typically dominate the continuum of the host galaxy and have broad, luminous emission, typical
for a powerful unobscured active nucleus. 

\item Considering the color-magnitude, color-stellar-mass, and spectral properties of galaxies in three different luminosity bins, we
found that galaxies undergoing LIRG-level starbursts had significantly different properties than galaxies undergoing a ULIRG-level 
starburst. Galaxies in a ULIRG phase had a smaller fractional contribution from older stellar populations than LIRGs, suggesting
that the latter were galaxies undergoing lower-level starbursts which added only moderately to the established stellar populations 
in those galaxies. The ULIRGs were comprised of galaxies undergoing primeval bursts and galaxies undergoing star-formation events
which were contributing significantly to the already established stellar matter in those galaxies.

\item Within the uncertainties of the models used, we found that the average starburst ages were roughly the same for galaxies of all 
$L_{TIR}$ in our sample. We are typically observing these galaxies $\sim100$ Myr after the inception of the burst.

\item While SPIRE-detected starbursts were not commonly associated with AGN activity ($\sim10$\% of the time) galaxies with AGN activity
selected independent of SPIRE were, generally, much more commonly associated with starbursts than similarly selected optical samples of
galaxies with dormant nuclei. 

\item Those SPIRE-detected galaxies hosting AGN are typically ULIRGs. In a SPIRE volume-limited sample between $0.5<z<2$ we found that 
the majority of powerful AGN hosts \emph{of all types} were forming stars at a significantly higher rate than the average ULIRG in the 
same redshift range. 

\end{itemize}

In the left panels of Fig. \ref{fig:fracSFRAGNvsnon} we plot the fraction of galaxies in the full volume-limited SPIRE sample that are host to powerful
AGN as a function of four different redshift bins. Again in this section we marginalize over AGN selection method, combining all AGN hosts
into a single sample for each redshift. As noted in the previous section, the mean stellar mass of the AGN hosts in this sample is 
comparable to the average SPIRE detected galaxy in the volume-limited sample at each redshift. 
Given our previous results it is perhaps not surprising that at all redshifts we observe that the fraction of galaxies hosting AGN is a 
strong function of $\mathcal{SFR}$ (or, equivalently, $L_{TIR}$). This is not the first time this trend has been observed. In two 
different studies covering different fields, Kartaltepe et al.\ (2010a) and Juneau et al.\ (2013; hereafter J13) observed that the 
fraction of galaxies which host AGN, AGN selected by a wide variety of methods, was a strong function of $L_{TIR}$ (see also the review of this trend for galaxies hosting X-ray AGN in Lapi et al.\ 2014). 
These samples were, however, selected at shorter wavelengths, meaning that the contamination to $L_{TIR}$ from the AGN was more problematic
than it is in a SPIRE-selected sample. In addition, the majority of these samples were distributed over a relatively
narrow redshift range, which limited the ability of these studies to investigate the redshift evolution of this fraction. 
Still, the redshift evolution of these fractions was investigated in J13 by comparing their observed $70\mu m$-selected sample to an
equivalently selected sample drawn from Sloan Digital Sky Survey (SDSS), finding essentially no evolution in the AGN fraction of IR-selected sources 
as a function of $L_{TIR}$ between the redshifts of $0.05<z<1$. 

Here, we broadly reaffirm the results found by Kartaltepe et al.\ (2010a) and J13 using a sample limited 
to the host galaxies of more powerful AGN. Considering only a SPIRE volume-limited sample between $0<z<1.5$, a redshift
range in which a volume-limited sample does not restrict us only to ULIRGs, the fraction of galaxies hosting any type of AGN 
at the highest $\mathcal{SFR}$s (equal to $\log(L_{TIR})\sim12.5$) is $\sim60$\%, roughly equivalent to that found at the same $L_{TIR}$ in J13 and
identical within the errors to the fraction computed for low-redshift IR-selected sources. Interestingly, we find a significantly
lower fraction of galaxies hosting AGN in lower-$L_{TIR}$ galaxies than was found in J13, again suggesting that lower power
AGN, especially those selected using the methods of J13, are important to the evolution of LIRGs and lower-luminosity ULIRGs (see also Netzer 2009).
The similarity between the AGN host fraction at the highest $L_{TIR}$ luminosities between the various samples suggests, 
on the other hand, that this population is dominated by high-power AGN. It is tempting to interpret the temporal invariance of 
the AGN fraction of high-$L_{TIR}$ galaxies over two-thirds of the Hubble time ($0.05<z<1.5$) as a lack of evolution. However, 
the value of this fraction is subject to a large variety of systematic and observational biases, which we do not 
attempt to control here. 

While complications arise from differing volumes and the $L_{TIR}$ that are necessary to impose to 
make the samples volume-limited, these biases are not present when comparing internally between different redshift bins in our 
sample. It is clear from the right panel of Fig. \ref{fig:fracSFRAGNvsnon} that, while the overall trend of increasing AGN fraction with 
increasing $L_{TIR}$ holds for all redshift bins, the fraction at a given $\mathcal{SFR}$ ($L_{TIR}$) is a strong function of redshift. 
There are two bins which have a large number of AGN hosts in all redshift bins. For LIRGs at $\log(\mathcal{SFR}_{TIR})\sim2.0$ 
($\log(L_{TIR})\sim11.8$) the fraction of galaxies hosting an AGN increases precipitously with decreasing redshift, rising from $\sim8$\% at 
redshift $z\sim1.5$ to roughly $50$\% at redshift $z\sim0.25$. Similarly, for galaxies undergoing ULIRG-level $\mathcal{SFR}$s of $\log(\mathcal{SFR}_{TIR})\sim2.8$ 
($\log(L_{TIR})\sim12.5$), this fraction increases from just over $20$\% to $70$\% from redshifts $z\sim3$ to $z\sim0.75$. This trend is 
strikingly similar to the trend of downsizing observed in galaxy evolution (Cowie et al.\ 1991, 1996), in which galaxies 
at fainter magnitudes are found to transition to redder colors at later epochs than their brighter counterparts. In 
our study, the most powerful AGN are observed to increase in frequency in galaxies at a given $L_{TIR}$ with decreasing redshift. In other
words, if the picture of downsizing is correct and there does indeed exist a relationship between the optical magnitude of
a galaxy and the epoch in which it quenches its star formation, one would expect that the galaxies at highest-$L_{TIR}$ 
values (i.e., ULIRGs), galaxies which are generally observed at the brightest absolute optical magnitudes for a given color, to shut down their 
star-formation activity at high redshifts and gradually transition to redder colors. This process would then occur at
progressively lower redshifts for galaxies with lower-$L_{TIR}$ luminosities, galaxies which are also generally fainter in the 
rest-frame optical (see \S\ref{photnmassbyTIR}). The dearth of extremely high-$L_{TIR}$ sources at lower redshifts 
is exactly what we observe, as the $\mathcal{SFR}$ of the brightest $L_{TIR}$ galaxies in a given redshift bin drops by an order of 
magnitude between redshift $z\sim3$ and $z\sim0.75$ (see the right panel of Fig. \ref{fig:fracSFRAGNvsnon}).  Such a picture is qualitatively similar 
to the observed redshift evolution of the space density of dusty starbursts as a function of $L_{TIR}$ detected in the CFHTLS-D1 and other fields 
(see, e.g., Rodighiero et al.\ 2010a and references therein). The fact that extremely bright ULIRGs, and indeed
the galaxies with the brightest $L_{TIR}$ at every epoch, are observed with the highest fraction of powerful AGN strongly suggests
that the presence of the AGN is involved in the observed trend.

There are also other reasons to suspect the rapid truncation of the star-formation events amongst the ULIRGs. From \S\ref{photnmassbyTIR}, ULIRGs comprise 
some of the most massive galaxies at \emph{every optical color}. Given the high rate of star formation in these galaxies (i.e., $\sim$350 $\mathcal{M}_{\odot}$ 
yr$^{-1}$) and the general lack of galaxies at a given color more massive than the ULIRGs, one of two things must occur over 
a short time period. Either the ULIRG must cease or considerably slow down its buildup of stellar mass, causing either a rapid
or gradual transition to redder optical colors, or the galaxy can move to redder optical colors through some other method 
and continue forming stars at the same rate. The effect of increasing metallicity is not enough to cause large enough change 
in optical colors to power the latter scenario (see \S\ref{AtoK}), and extinction effects are equally implausible given that
we observe no relationship between the optical colors of ULIRGs and their $\mathcal{SFR}$s (which should be, in turn, linked to the 
dust content of the galaxies, and, by extension, the amount of extinction). There is further evidence to suggest the latter scenario is
not plausible. If ULIRGs were to continue to form stars at their current rates for even $\sim100$ Myr, the \emph{average} ULIRG would
increase in stellar mass by 31$\pm$6\%\footnote{Since this quantity is determined from the median $\mathcal{SSFR}$ of ULIRGs in our sample, stellar masses are again converted to a 
Salpeter IMF}. This would have the effect of enormously increasing the number of galaxies with extremely high stellar masses ($\log(\mathcal{M}_{\ast})>11$) 
over the redshift range $0.5<z<2$, a phenomenon which is inconsistent with observations of the stellar mass function of galaxies
(e.g., Peng et al.\ 2010; Ilbert et al.\ 2013). Because of these arguments and because the most prodigious star-forming galaxies 
in a given redshift bin in Fig. \ref{fig:fracSFRAGNvsnon}, galaxies which are typically ULIRGs, are unobserved in each 
subsequently lower redshift bin and because these are the same galaxies with the highest fraction of powerful AGN at any given redshift, 
it seems easy to appeal to a scenario in which a galaxy undergoes a strong burst, is quenched by the powerful AGN associated 
with that burst, and subsequently transitions to a red, quiescent galaxy. Such a scenario is identical to that advocated by
Gruppioni et al.\ (2013) for the population most resembling the ULIRGs we consider here, in which the number counts of various types of \emph{Herschel}
sources across four fields were used, along with a variety of other evidence, to deduce the evolutionary scenario for galaxies with powerful 
coeval starburst and AGN activity. It is also similar to the model adopted to fit a compilation of \emph{Herschel} observations of X-ray AGN 
hosts in Lapi et al.\ (2014). The above scenario does not preclude the possibility of 
additional quenching occuring due to processes associated with the star formation itself (see, e.g., Cen 2012), processes which can be quite strong 
in galaxies with the typical $\mathcal{SFR}$s of the ULIRGs in the full SPIRE sample. However, a large fraction of SPIRE sources ($\sim90$\%) do not show 
powerful AGN activity, but the reverse view does not hold, i.e., a large fraction of galaxies with powerful AGN activity do exhibit a starburst.
These observations strongly indicate that the starburst persists at least until the point at which a powerful AGN can be ignited, meaning quenching processes 
related to star formation are not effective to quench the starburst over a timescale equal to the lag time between the initial burst activity and the onset of the
AGN. Additionally, everything else being equal, the effectiveness of such processes at quenching the starburst will diminish as redshift decreases 
and the typical $L_{TIR}$ (or, equivalently, $\mathcal{SFR}$) of the galaxies with the highest fraction of AGN drops precipitously (see Fig. \ref{fig:fracSFRAGNvsnon}).  

As noted in the previous section, our results, when viewed in a different way, seem, at the surface, to contradict this
picture, since a large number of powerful AGN are observed to be coeval with powerful starbursts. However, this is perhaps not
the case. The observed trend of increasing AGN fraction with increasing $L_{TIR}$ is a trend that would not be observed if the 
truncation process in the starburst took place over a long timescale. Such gradual quenching is also disfavored by modeling 
of \emph{Herschel}-bright X-ray AGN hosts, in which the fraction of IR-bright starbursts amongst powerful X-ray AGN hosts is shown to be 
highly sensitive to the star-formation quenching timescale (Lapi et al.\ 2014; compare, e.g., Figs. 10 and 12). If the starburst is allowed to die out gradually, 
such powerful AGN would be associated not only with the strongest starbursts at a given redshift, but would rather be equally 
distributed over all values of $L_{TIR}$. This argument does not, however, constrain the lifetime that the two phenomena 
can be observed coevally, nor does it say definitively that it is the AGN which causes the quenching. For the former point
we can appeal to two different pieces of evidence. The first is the typical starburst age of the galaxies which are hosting the
AGN. According to \S\ref{stellarnsbages}, roughly 100 Myr has passed since the inception of the burst in a typical galaxy 
hosting an AGN. This is roughly the timescale of the lag time of the peak of the AGN activity inferred from both observations 
and simulations. Starbursts, excepting extreme cases, are typically not thought to live for more than twice this timescale,
with the powerful activity of an AGN dying out on roughly the same timescale, though with large uncertainties. The fact that,
for the bulk of our sample, there are few galaxies at the highest TIR luminosities which \emph{do not host a powerful AGN},
strongly suggests that the powerful AGN activity does not die out before the starburst is significantly truncated. If the 
truncation timescale must be rapid and efficient, as suggested above, this means that \emph{in ULIRGs the truncation of the 
starburst occurs during the peak of the AGN activity}. Given the efficiency of feedback mechanisms associated with
luminous AGN at the peak of their activity (e.g., Harrison et al.\ 2012a), it is natural to appeal to the AGN as
the mechanism causing the cessation of star-formation activity in their ULIRG host galaxies. While a seemingly equally 
plausible scenario is one in which the AGN ceases its activity and, at the same time, the starburst strength drops to a 
lower level and continues to form stars in the absence of an AGN, the large number of both low-luminosity and powerful 
AGN observed with post-starburst features in their spectra or with transitional colors (Schawinski et al.\ 2007; Georgakakis et al.\ 2008; Hickox et al.\ 2009; 
Kocevski et al.\ 2009b, 2011b; Rumbaugh et al.\ 2012) and the small fraction of X-ray AGN hosts detected in the FIR (see Lapi et al.\ 2014 and references therein)
strongly argues against this scenario. It still remains a possibility, however, that the power of the 
starburst and AGN decrease in tandem, with the AGN continuing to accrete matter, but emitting at a level below our detection limits. If this 
scenario were generally correct, however, one would expect to see a flattening of the AGN host fraction as a function
of $L_{TIR}$ in samples that probe both powerful and lower-luminosity AGN, something that is not observed (J13). It 
seems, then, that \emph{our data favor a scenario where both massive galaxies undergoing a ULIRG-level burst of star-formation 
and nascent galaxies undergoing an initial ULIRG-level starbursting event are commonly associated with powerful AGN activity, which
quickly and efficiently truncates the star formation in that galaxy, and that the typical luminosity of those galaxies
undergoing this process is a strong function of redshift.} The large populations of \emph{Herschel} sources observed in
this sample undergoing secondary, lower-level starbursting events largely in the absence of powerful AGN, suggest that
the picture we have drawn here is incomplete and only applies to galaxies which are undergoing extremely strong bursts of 
star formation. Future work will be necessary to investigate whether or not the nascent ULIRGs observed in this sample
can be connected to the population of older, more massive LIRGs through cyclical star formation induced in a gentler 
way so as to not re-excite powerful AGN activity.

\section{Summary \& Conclusions}
\label{conclusions}

In this paper we have studied the properties of star-forming \emph{Herschel}/SPIRE detected galaxies over a broad  
redshift range, a study which has taken a variety of forms along the way. Here we briefly outline the most important results 
and consequences of this study. 

\begin{itemize}
\renewcommand{\labelitemi}{$\bullet$}

\item We have assembled and matched a massive collection of data over 0.8 deg$^2$ the CFHTLS-D1 field that includes deep spectroscopy from 
the VVDS and ORELSE surveys and associated X-ray, radio, optical, and near-infrared imaging. To these data we added three-band 
\emph{Herschel}/SPIRE imaging from the HerMES survey complete to a $3\sigma$ depth of $f_{\nu,250\mu m}>12$ mJy ($m_{250\mu m} < 13.7$). 
In total, 1753 galaxies detected significantly in SPIRE were matched to counterparts with well-defined redshifts (referred to as the
full SPIRE sample), including 263 with spectroscopic redshifts. The photometric and spectroscopic redshifts of the full SPIRE sample 
span the range of $0<z<4$ and have a median redshift of $\langle z\rangle =0.85$. 

\item All galaxies detected significantly in SPIRE and matched to optical counterparts
were initially compared to SPIRE-undetected optically-selected samples at a variety of redshifts. In this
comparison we found that the galaxies comprising the full SPIRE sample spanned a large range of color, absolute magnitudes, and
stellar masses, ranges that were equivalent to those spanned by the optically-selected samples at all redshifts. Though a large
number of SPIRE galaxies appeared to have NUV-optical colors consistent with quiescent or transitional populations, an analysis
using a combination of NUV-optical and optical-IR colors revealed that the colors of these galaxies were largely driven by 
extinction effects rather than properties intrinsic to their stellar populations. In volume-limited samples of SPIRE galaxies
spanning a large redshift range we found little or no correlation between the stellar mass of a galaxy and its $\mathcal{SFR}$. While 
we argued against the existence of a star forming main sequence in our data, the data do not probe deep enough
to rule out the phenomenon for all FIR-selected samples of galaxies. Instead, relationships were adopted from various other
studies to explore the issues inherent in the classification and interpretation of starburst galaxies. Large variance
of up to an order of magnitude in starburst fractions and their fractional contribution to the overall $\mathcal{SFRD}$ of the universe 
(for galaxies above a certain $\mathcal{SFR}$ limit) were observed in our own sample if the definition of starburst or the stellar mass,
$\mathcal{SFR}$, or redshift limits were changed slightly. This variance was used to argue against attempting to contextualize starburst
in this framework unless the properties of the data and the properties of the star-forming main sequence are known extremely 
precisely. 

\item SPIRE-detected galaxies were divided into bins of $L_{TIR}$ and compared internally. With a variety of analyses primarily 
involving the use of composite VVDS spectra, we found that the mean luminosity-weighted age of stellar populations of SPIRE galaxies increased 
with decreasing $L_{TIR}$. The age since the inception of the starburst, however, was not correlated with $L_{TIR}$, and was found to be
$\sim100$ Myrs for all $L_{TIR}$ subsets. This evidence was used to argue that the LIRGs in our sample were, on average, undergoing rejuvenated 
starbursts and were building up a smaller fraction of their stellar mass in the observed burst than ULIRGs. Combining these spectral results 
with analysis of the color, absolute magnitude, and stellar mass properties of both LIRGs and ULIRGs it was shown that the 
ULIRG population was inhomogeneous, comprised of both newly-formed galaxies undergoing an initial violent starburst event and 
galaxies that had already built up considerable stellar mass undergoing a vigorous rejuvenation event. 

\item In the second part of this paper, the full SPIRE sample was divided into galaxies hosting powerful X-ray, radio, IR, and 
spectrally-selected AGN and those galaxies where such phenomena were absent. The hosts of powerful AGN made up only a small 
fraction of the full SPIRE sample ($\sim10$\%), considerably less than other studies of FIR-selected populations of galaxies
with targeted deep multiwavelength observations. This suggested that low-luminosity AGN play a large role in these populations. However,
galaxies hosting powerful AGN \emph{of all types} were observed to be more commonly associated with a starbursting event than 
a stellar mass and redshift matched sample of galaxies without powerful AGN. Surprisingly, the most common association was between 
radio-selected AGN and SPIRE-selected starbursts, as 41.8$\pm$3.6\% of all radio AGN hosts were detected in SPIRE at 
$\ge3\sigma$ significance. Though we included radio AGN hosts in the All AGN sample presented in this work, 
the properties of SPIRE-detected and undetected galaxies are covered explicitly in a companion paper (Lemaux et al.\ 2015). 

\item At all redshifts covered by our sample the AGN host populations were undergoing more vigorous starbursts than the average SPIRE-selected
starburst at those epochs, with the AGN hosts undergoing, on average, ULIRG-level starbursts. The fraction of galaxies hosting a powerful
AGN was observed, at all redshifts, to be a monotonically increasing function of $L_{TIR}$ (i.e., $\mathcal{SFR}$), with the most powerful starbursts 
at every redshift hosting a powerful AGN, on average, $\sim$80\% of the time.  This trend was not coupled only with $L_{TIR}$, as the 
fraction of galaxies hosting a powerful AGN at the same $L_{TIR}$ was found to change considerably over cosmic time. Rather, the AGN 
fraction was tied to both $L_{TIR}$ and redshift; similar AGN fractions were found for the most vigorous star-forming galaxies at low
($0<z<0.5$) and high ($2<z<4$) redshift despite a difference of a factor of $\sim50$ in their $\mathcal{SFR}$s. Combining the results of the two parts 
of this paper along  with evidence from other observational and theoretical works, we argued for the existence of downsizing in the full 
SPIRE sample, a phenomenon driven by quenching from a powerful AGN in the most vigorous star-forming galaxies at each epoch.  

\end{itemize}

While this study constrains the relationship between extreme starburst galaxies and powerful AGN, due to the limits of a variety of data 
presented in this study, the relationship between lower power starbursts and lower power AGN is less clear. While several works have
been mentioned throughout the course of this paper studying exactly the link between the latter two, we have demonstrated clearly the 
usefulness of SPIRE observations to unambiguously decouple starburst and AGN phenomena. Areas of the sky with either existing or planned
deep multiwavelength data have already been imaged by \emph{Herschel}/SPIRE. Though few have the rich spectroscopy dataset available
on the CFHTLS-D1 field, future studies using the data in these fields, especially high resolution imaging, will help to further 
understand what relationships exist between lower power AGN and the galaxies which host them.  

\begin{acknowledgements}
B.C.L. thanks St\'ephanie Juneau, S\'ebastien Heinis, and Nicholas Rumbaugh for several useful and enlightening discussions. B.C.L. also thanks Rebecca Leigh Polich
for being supportive during this journey even though, sometimes, it can be hard for some kids. Some of the work presented herein was based on data obtained with 
the European Southern Observatory Very Large Telescope, Paranal, Chile, under Large Programs 070.A-9007 and 177.A-0837. In addition, this work utilized 
observations obtained with MegaPrime/MegaCam, a joint project of CFHT and CEA/DAPNIA, at the Canada-France-Hawaii Telescope (CFHT) which is operated 
by the National Research Council (NRC) of Canada, the Institut National des Sciences de l'Univers of the Centre National de la Recherche Scientifique (CNRS) 
of France, and the University of Hawaii. This work is partially based in part on data products produced at TERAPIX and the Canadian Astronomy Data Centre 
as part of the Canada-France-Hawaii Telescope Legacy Survey, a collaborative project of NRC and CNRS. Funding for this work came from the European 
Research Council Advanced Grant ERC-2010-AdG-268107-EARLY. SPIRE has been developed by a consortium of institutes led by Cardiff University (UK) and including 
Univ. Lethbridge (Canada); NAOC (China); CEA, LAM (France); IFSI, Univ. Padua (Italy); IAC (Spain); Stockholm Observatory (Sweden); Imperial College London, 
RAL, UCL-MSSL, UKATC, Univ. Sussex (UK); and Caltech, JPL, NHSC, Univ. Colorado (USA). This development has been supported by national funding agencies: 
CSA (Canada); NAOC (China); CEA, CNES, CNRS (France); ASI (Italy); MCINN (Spain); SNSB (Sweden); STFC, UKSA (UK); and NASA (USA). Portions of this work are also based 
on observations made with the \emph{Spitzer Space Telescope}, which is operated by the Jet Propulsion Laboratory, California Institute of Technology under a 
contract with NASA. Portions of this work were also based on observations obtained with \emph{XMM-Newton}, an ESA science mission with instruments and 
contributions directly funded by ESA Member States and NASA. Additional observations were obtained from the Very Large Array run by the National Radio Astronomy 
Observatory, a facility of the National Science Foundation operated under cooperative agreement by Associated Universities, Inc. The spectrographic data presented herein were partially 
obtained at the W.M. Keck Observatory, which is operated as a scientific partnership among the California Institute of Technology, the University of California, and the 
National Aeronautics and Space Administration. The Observatory was made possible by the generous financial support of the W.M. Keck Foundation. We wish to thank the indigenous
Hawaiian community for allowing us to be guests on their sacred mountain; we are most fortunate to be able to conduct observations from this site.
\end{acknowledgements}


\appendix
\section{\normalsize{Near-infrared, x-ray, \& radio matching and \lowercase{$k$}-corrections}}

Source lists from the SWIRE NIR imaging were produced by running Source Extractor (Bertin \& Arnouts 1996)
on each the individual mosaics in the 3.6, 4.5, 5.8, 8.0, \& 24$\mu$m bands using the method described
in Surace et al.\ (2005). These catalogs were combined using the bandmerge software developed by the Spitzer Science
Center\footnote{http://irsa.ipac.caltech.edu/data/SPITZER/docs/dataanalysistools/ tools/bandmerge/},
a probabilistic matching technique described in detail in Surace et al.\ (2005). In our catalogs, we
required that an object be detected in the $3.6\mu$m band to be included in the SWIRE catalog. The $3.6\mu$m
source position was then used for matching to images in other wavelengths. The full SWIRE catalog was matched to a previous version of
our CFHTLS-D1 photometric catalog in Arnouts et al.\ (2007) using a nearest-neighbor matching technique
described in that paper. This catalog was in turn matched to the most recent version of our CFHTLS-D1
catalog using a known mapping between the two catalogs. The final CFHTLS-SWIRE band-merged catalog
was then cross-correlated with the the current generation of the VVDS and ORELSE spectral catalogs, again
using known mappings between the catalogs.

Initial radio-optical matching of the VLA 1.4 GHz was provided by Ciliegi et al.\ (2005) using a likelihood
ratio method similar to the method used to determine some of the \emph{Herschel} counterparts in this
study (see \S\ref{herschelmatch}). As described in \S\ref{ancillary}, the radio portion of this catalog
was updated to include both new fluxes and errors of the 1.4 GHz sources as well as to incorporate the
610 MHz GMRT imaging. This matching, which was done to the original CFHT12K VVDS imaging, was mapped to
our band-merged CFHTLS-SWIRE catalog by applying a known bulk astrometric shift between the two catalogs
followed by nearest-neighbor matching.

For those radio objects with optical counterparts having either a reliable photo-$z$ (see section
\S\ref{SEDfitting}) or a reliable spec-$z$ (see section \S\ref{spectra}), rest-frame power densities
were calculated by applying a \emph{k}-correction at the median redshift of the sample assuming
a spectral index of $P_{\nu}\propto \nu^{-\alpha}$. An $\alpha$ of 0.68 was chosen as this is the
median spectral index derived from those sources which were detected at $>5\sigma$ in both the VLA
and GMRT images. This value is also reasonably consistent with values used in other studies
(e.g., Prandoni et al.\ 2006; Hickox et al.\ 2009; Smol$\rm{\check{c}}$i$\rm{\acute{c}}$ 2009b) and is consistent with
values found in other sub-millimeter galaxy (SMG) studies (e.g., Ibar et al.\ 2010). In practice, our choice of $\alpha$ used
for the bulk \emph{k}-correction makes little difference on our results. If we instead
chose an $\alpha=0.5$, the difference in the derived power densities is, on average,
10\%, which results in only minor changes in our AGN sample (see \S\ref{AGNselection}). While
we could, in principle, use the spectral index derived for each individual object from their
610 MHz and 1.4 GHz fluxes and their individual redshift, we chose not to use this method due to the large uncertainties on
individual $\alpha$ measurements, opting instead for a statistical treatment.

Initial X-ray-optical matching was performed using a combination of modified nearest-neighbor
matching and visual inspection described in detail in Chiappetti et al.\ (2005). In addition,
a probability threshold using the methodology of Downes et al.\ (1986) was applied to
discard likely chance projections and to assign ranks to those matches that had a high
likelihood of being genuine (i.e., $p < 0.01$, which corresponds to the probability of finding
an $I=21$ object within 2$\arcsec$ of the X-ray source, see Chiappetti et al.\ 2005). Only
those matches considered genuine and unambiguous were used. The only exception were those
cases where multiple optical counterparts were assigned to a single X-ray object (rank 2
objects) where we were able to definitively assign an optical counterpart through visual
inspection. The XMDS matching was also done to the original CFHT12K VVDS imaging and was
mapped to our band-merged CFHTLS-SWIRE photometric catalog in the same manner as the
radio catalog.

\begin{figure}
\epsscale{0.6}
\plotonetiny{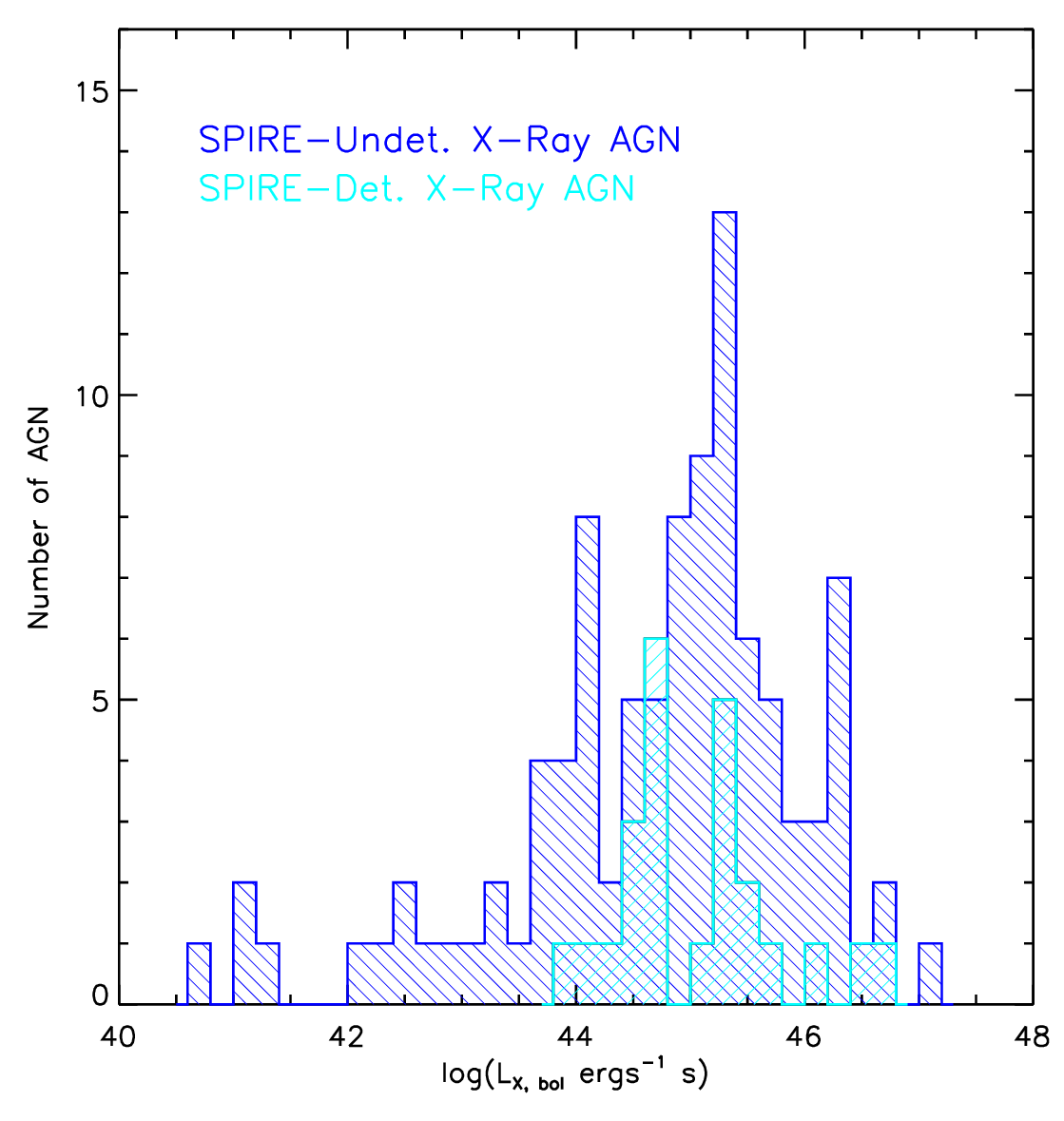}
\caption{Bolometric X-ray luminosity ($L_{X,\;bol}$) histogram of SPIRE-detected (dashed cyan histogram) and SPIRE-undetected (dashed blue 
histogram) X-ray AGN hosts. Though the conversion to $L_{X,\;bol}$ is extremely uncertain and though we do not explicitly use this quantity
in this work, the histogram is used to show the rough range and completeness limit of the X-ray AGN in our sample. $L_{X,\;bol}$ is
derived from the spectroscopic or photometric redshift and $XMM$/EPIC hard band X-ray luminosities (i.e., 2-10 keV) using the coversion
from Vasudevan \& Fabian (2007; see text). The rough completeness of the XMM imaging is $L_{X,\;bol}\gsim3\times10^{44}$ ergs s$^{-1}$, 
corresponding to a moderately powerful Seyfert. Note that most of the SPIRE-detected hosts of X-ray AGN lie at the more powerful
end of the $L_{X,\;bol}$ distribution.}
\label{fig:Xraybol}
\end{figure}

Observed-frame X-ray luminosities in the soft and hard band were derived for all
objects with a reliable redshift. These luminosities were transformed to the rest-frame
assuming a photon index of $\Gamma=1.4$, where $\nu F_{\nu}=\nu^{\Gamma}$ (Kocevski et al.\
2009a; Rumbaugh et al.\ 2012)\footnote{For clarity, this is equivalent to assuming an $\alpha$ of -0.4,
where $F_{\nu}\propto \nu^{-\alpha}$.}. This value corresponds to the approximate slope of the X-ray
background in both the 1-8 keV and 2-10 keV bands (Tozzi et al.\ 2001; Kushino et al.\ 2002). Bolometric
X-ray luminosities were calculated using a fourth-order polynomial fit to an emperical relationship
between the rest-frame hard band luminosity and the bolometric X-ray luminosity presented in Vasudevan
\& Fabian (2007) using the data of Marconi et al.\ (2004). The relationship is defined as

\begin{equation}
\begin{split}
L_{X,\;bol} = 1.36\times10^6 - 1.30\times10^5\log(L_{X}) + 4.65\times10^3 \log(L_{X})^2 \\
- 73.9\log(L_{X})^3 + 0.44\log(L_{X})^4)L_{X},
\end{split}
\label{eqn:bolometriclum}
\end{equation}

\noindent where $L_{X}$ is defined as the X-ray luminosity in the hard band (i.e., 2-10 keV).
As is discussed extensively in Vasudevan \& Fabian (2007), the relationship between these two
quantities is subject of much uncertainty. The relationship is especially questionable for certain types of
AGN which are pervasive in our sample. In the absence of a better method to calculate
bolometric X-ray luminosities this method is adopted, but we do so with some degree of trepidation.
While most of the results presented in this paper do not rely on the magnitude of the X-ray
luminosity, bolometric or otherwise, but are rather reliant solely on the binary function of
whether or not a galaxy is detected in the X-ray, we urge the reader to keep these reservations
in mind when comparing the $L_{X,\;bol}$ properties of the X-ray AGN in this sample with other studies.
In Fig. \ref{fig:Xraybol} we plot $L_{X,\;bol}$ for the AGN of galaxies detected in SPIRE and those 
which went undetected in the FIR.

\section{\normalsize{Herschel/SPIRE fraction matched samples}}

In order to determine if the high fraction of galaxies with powerful AGN detected in \emph{Herschel}/SPIRE is significant it is necessary to define a proper 
control sample of galaxies with which to compare the AGN hosts. In order to generate this control sample it is essential to identify the properties
of the galaxy population that affect their detectability in SPIRE. These properties can be astrophysical (e.g., higher likelihood of a starburst for 
reasons unrelated to the presence of an AGN) or they can be a result of the matching and selection scheme employed in this study. For the latter point, 
the main selection bias enters through the use of \emph{Spitzer} priors used in the matching process in such a way that an optical galaxy is much 
more likely be matched to a SPIRE detection if it has been detected at 24$\mu$m or is apparently brighter in the observed-frame 3.6$\mu$m band. However,
except for the host of the IRAC-selected AGN (a population for which a different control sample was generated, see \S\ref{SBfrac}), the hosts of powerful AGN
are not special in either of these regards relative to the overall SPIRE population: the fraction of the full AGN host sample with $24\mu m$ priors and the 
$3.6\mu m$ magnitude distribution of these galaxies are statistically indistinguishable from galaxies in the full SPIRE sample that do not host an AGN. 
Thus, we can safely ignore any effects of selection bias when generating the control sample and focus solely on biases entering through the physical 
properties of the AGN host population. 

While many physical properties of a galaxy can affect its likelihood of being detected in SPIRE, the primary drivers of differential bias in SPIRE detectability
are the redshift and the stellar mass of the galaxy. The former enters through the well-known increase in both the $\mathcal{SFRD}$ and the fraction of star-forming galaxies 
at higher redshifts. The latter is slightly more complicated. While one obvious effect of stellar mass on the SPIRE detectability enters through the dependency of 
the quiescent fraction on stellar mass, there are other, more subtle concerns. For star-forming galaxies, the rate at which a galaxy forms stars may monotonically 
increase with increasing stellar mass (though see \S\ref{sfmainseq}). Additionally, the typical environment in which a galaxy resides is also somewhat dependent on 
stellar mass, with environment having a complex relationship with $\mathcal{SFR}$ that is dependent on redshift and a variety of other factors. Thus, a sample of galaxies that do 
not host powerful AGN that are which are matched in \emph{both} stellar mass and redshift to the AGN host sample allows us to fairly assess whether the fraction of 
SPIRE detected galaxies ($f_{SB}$) of galaxies hosting a powerful active nucleus is significantly higher than galaxies devoid of such a phenomenon.

This control sample was generated in the following manner. All galaxies hosting powerful AGN, i.e., all galaxies selected 
in \S\ref{AGNselection}, were combined into a single sample. Counting host galaxies of AGN selected by multiple techniques only 
once resulted in a AGN host sample of 561 unique galaxies. This sample included both the SPIRE-detected AGN hosts that comprised the 
All AGN sample described in \S\ref{sfmainseq} and those galaxies hosting AGN which went undetected in SPIRE. A grid of stellar 
mass and redshift bins were generated spanning $\Delta\log(\mathcal{M}_{\ast})$ and $\Delta z$ of 0.2. For each two-dimensional 
bin in this phase space, both the number of galaxies hosting powerful AGN and the number of galaxies in the full optical sample 
that satisfied the redshift and stellar mass criteria of the bin were determined. A random sample of galaxies were selected 
from the full optical sample that satisfied the stellar mass and redshift criteria of each bin, 
with a sample size identical to the number of AGN hosts in that bin such that the final redshift and stellar mass bins of the control 
sample exactly matched those of the AGN hosts. The number of galaxies detected in SPIRE was then determined for the 561 galaxies of the 
control sample. Because the number of galaxies in the full optical sample in each bin, generally,
far exceeded the number of galaxies hosting a powerful AGN, this process was repeated 500 times to determine the effect of random 
sample variance. As much as was possible, galaxies known to be hosting a powerful AGN were removed from the full optical sample. 
Though a few stellar mass/redshift bins contained only AGN hosts, the effect of including these galaxies was negligible as the 
mean AGN host fraction of the 500 control samples was $<0.5$\%. The resulting $f_{SB}$ distribution of the 
500 realizations of the control sample was surprisingly Gaussian, with $\langle f_{SB}\rangle=5.8\pm1.2$ for a galaxies detected 
in SPIRE at a significance of $\ge3\sigma$ in one of the three SPIRE bands. This fraction is significantly lower than both 
the $f_{SB}$ of the full AGN host sample and that of the hosts of any particular flavor of AGN.  

\end{document}